\documentclass{aa}

\usepackage{graphicx}
\usepackage{txfonts}
\usepackage{url}
\usepackage{natbib}
\usepackage{color}
\bibpunct{(}{)}{;}{a}{}{,}
\begin{document}
\title{Dust variations in the diffuse interstellar medium: constraints on Milky Way dust from Planck-HFI observations}

\author{N. Ysard
\and M. K\"ohler
\and A. Jones
\and M.-A. Miville-Desch\^enes
\and A. Abergel
\and L. Fanciullo}

\institute{IAS, CNRS (UMR8617), Universit\'e Paris-Sud 11, B\^atiment 121, F-91400 Orsay, France\label{inst1}, \email{nathalie.ysard@ias.u-psud.fr}}

\abstract{The Planck-HFI all-sky survey from 353 to 857~GHz combined with the IRAS data at 100~$\mu$m (3\,000~GHz, IRIS version of the data) show that the dust properties vary from line of sight to line of sight in the diffuse interstellar medium (ISM) at high Galactic latitude ($10^{19} \leqslant N_{\rm H} \leqslant 2.5 \times 10^{20}$~H/cm$^2$, for a sky coverage of $\sim 12$\%).}
{These observations contradict the usual thinking of uniform dust properties, even in the most diffuse areas of the sky. Thus, our aim is to explain these variations with changes in the ISM properties and with evolution of the grain properties.}
{Our starting point is the latest core-mantle dust model. This model consists of small aromatic-rich carbon grains, larger amorphous carbonaceous grains with an aliphatic-rich core and an aromatic-rich mantle, and amorphous silicates (mixture of olivine and pyroxene types) with Fe/FeS nano-inclusions covered by aromatic-rich carbon mantles. We explore whether variations in the radiation field or in the gas density distribution in the diffuse ISM could explain the observed variations. The dust properties are also varied in terms of their mantle thickness, metallic nano-inclusions, carbon abundance locked in the grains, and size distributions.}
{We show that variations in the radiation field intensity and gas density distribution cannot explain variations observed with Planck-HFI but that radiation fields harder than the standard ISRF may participate in creating part of the observed variations. We further show that variations in the mantle thickness on the grains coupled with changes in their size distributions can reproduce most of the observations. We concurrently put a limit on the mantle thickness of the silicates, which should not exceed $\sim 10$ to 15~nm, and find that aromatic-rich mantles are definitely needed for the carbonaceous grain population with a thickness of at least 5 to 7.5~nm. We also find that changes in the carbon cosmic abundance included in the grains could explain part of the variations in dust observations. Finally, we show that varying the composition of metallic nano-inclusions in the silicates cannot account for the variations, at the same time showing that the amount of FeS they contain cannot be $> 50$\% by volume.}
{With small variations in the dust properties, we are able to explain most of the variations in the dust emission observed by Planck-HFI in the diffuse ISM. We also find that the small realistic changes in the dust properties that we consider almost perfectly match the anti-correlation and scatter in the observed $\beta-T$ relation.}

   \authorrunning{}
\titlerunning{Dust variations in the diffuse interstellar medium}

\maketitle
%

\section{Introduction}
\label{introduction}

Dust is seen everywhere in the Milky Way and it is known that dust properties evolve from the diffuse interstellar medium (ISM) to denser regions \citep{Bernard1999, Ridderstad2006, Schnee2008, Koehler2012, Ysard2013}. Except in a handful of studies (see for instance Bot et al. 2009 and Veneziani et al. 2010 for studies in emission, and Fitzpatrick \& Massa 2005, 2007, and 2009 for studies in extinction), before Planck-HFI \citep{PlanckProducts}, it has generally been assumed that dust properties were the same everywhere in the diffuse ISM (see for instance Boulanger et al. 1996, for $N_{\rm H} \lesssim 5 \times 10^{20}$~H/cm$^2$). However, Planck observations seem to contradict this assumption. Indeed, variations in dust spectral energy distribution (SED) were observed by Planck-HFI, which suggest that dust properties may vary within the diffuse ISM \citep{PCXXIV, PlanckBoulanger, PlanckMAMD, PlanckAniano}.
\nocite{Boulanger1996}
\nocite{Bot2009}
\nocite{Veneziani2010}
\nocite{Fitzpatrick2005}
\nocite{Fitzpatrick2007}
\nocite{Fitzpatrick2009}

Using a pixel-to-pixel modified black-body $\chi^2$ fit, Planck Collaboration XI (2014, hereafter PCXI) modelled the all-sky emission from 353 to 3\,000\,GHz: $I_{\nu} = \tau_{\nu_0} B_{\nu}(T) \left(\nu / \nu_0\right)^{\beta}$, where $\tau_{\nu_0}$ is the optical depth at $\nu_0 = 353$~GHz, $T$ is the dust colour temperature, and $\beta$ is the spectral index, assumed to be constant from 100~$\mu$m to 353~GHz. Besides, PCXI computed the dust radiance $\cal{R}$ $= \int_{\nu} I_{\nu} d\nu$, luminosity $L_{\rm H} = 4\pi \cal{R}$ $/ N_{\rm H}$, and opacity $\sigma_{\nu_0} = \tau_{\nu_0}/N_{\rm H}$.  For the most diffuse areas of the sky\footnote{These areas correspond to the two areas described as ``Low $N_{\rm H}$'' and ``Lowest 1\%'' in the last two lines of Tabs. 3 and 4 in PCXI and have a sky coverage of $\sim 12$\%.} ($10^{19} \leqslant N_{\rm H} \leqslant 2.5\times 10^{20}$~H/cm$^2$), the opacity is found to decrease with increasing temperature while the luminosity remains almost constant ($L_{\rm H} \sim 3.5\times 10^{-31}$~W/H). This strongly suggests variations in dust properties since variations in the interstellar radiation field (ISRF) would lead to an increased luminosity. Combining Planck-HFI and Sloan Digital Sky Survey (SDSS) data, PCXI derived the dust visible extinction for diffuse ISM at high Galactic latitude, $E(B-V)$, to far-IR emission ratio to be $E(B-V)/\cal{R}$ $\sim 5.4\times 10^5$. Additionally, PCXI found that the variations in $T$ and $\beta$ are globally anti-correlated on the sky. These authors used a two-step fit to limit the effect of instrumental noise on the estimate of $T$ and $\beta$ and showed that most of the variations measured at high Galactic latitude are truly related to variations in the dust emission, even if a significant but subdominant fraction of the observed anti-correlation comes from cosmic infrared background anisotropies. The uncertainties on $T$ and $\beta$ due to these anisotropies are $\sim 0.4$~K and $\sim 0.1$, respectively.

The goal of the present analysis is to explain the variations in the dust SED at high Galactic latitude as presented by PCXI, using a physical dust model \citep{Jones2013}. These variations may be explained by two phenomena. Either the dust environment is varying (density distribution, radiation field intensity and hardness) or dust properties are changing/evolving. In the following, we explore these two options: firstly, through building models of the ISM density distribution and varying the radiation field ; and secondly, through varying the properties of the \citet{Jones2013} dust model. Sect.~\ref{models} describes the models and tools that we use to describe dust emission and extinction in the ISM. Sect.~\ref{environment} shows the influence of varying the environment properties on the dust observables, whereas Sect.~\ref{dust_properties} details the influence of dust property evolution. The counterpart of all these variations in terms of extinction is also compared to observations. Concluding remarks are given in Sect.~\ref{conclusions}.

\section{Dust model and tools}
\label{models}

\subsection{Dust model}
\label{dust_model}

Our starting point is the \citet{Jones2013} dust model as updated by \citet{Koehler2014}. It assumes that dust in the diffuse ISM consists of small grains of aromatic-rich carbon (radius $a \leqslant 20$~nm) and larger core-mantle grains, for which the cores are composed either of amorphous aliphatic-rich carbon or of a mixture of amorphous silicates with the normative chemical compositions of olivine and pyroxene (with half of the total mass of silicate grains in each type). For all types of grains, the mantles are made of aromatic-rich carbons with a thickness of 20~nm for the carbon cores and 5~nm for the silicate cores. Moreover, iron and sulphur are incorporated into the silicate cores in the form of metallic nano-inclusions of Fe and FeS. They represent 10\% of the total core volume, of which 30\% is FeS and 70\% is pure Fe. The corresponding elemental abundances and dust-to-gas mass ratios are listed in Tab. 1.

\begin{table*}
\label{abundances_table}
\centering
\caption{Dust model abundances and dust-to-gas mass ratios. The first column lists the aromatic-rich carbon mantle thickness, the second column the grain density averaged over the size distribution, and the other columns the elemental abundances (in ppm) of C, Si, Mg, O, Fe, and S within the grains. For the carbonaceous grains, we distinguish between carbon in aromatic-rich (mantle) and aliphatic-rich (core) forms. The carbons included in the silicates are those of the aromatic-rich mantles.}
\begin{tabular}{ccccccccc}
\hline
\hline
\multicolumn{9}{c}{Small amorphous carbon grains ($a < 100$~nm)} \\
\hline
Mantle thickness (nm) & $\rho$ (g/cm$^3$) & $M_d / M_{{\rm H}}$ & \multicolumn{2}{c}{$\left[\frac{{\rm C}}{{\rm H}}\right]_{{\rm aromatic}}$} & \multicolumn{2}{c}{$\left[\frac{{\rm C}}{{\rm H}}\right]_{{\rm aliphatic}}$} & \multicolumn{2}{c}{$\left[\frac{{\rm C}}{{\rm H}}\right]_{{\rm total}}$} \\
\hline
0   & 1.6 & $0.17 \times 10^{-2}$ & \multicolumn{2}{c}{0}     & \multicolumn{2}{c}{143.5} & \multicolumn{2}{c}{143.5} \\
5   & 1.3 & $0.14 \times 10^{-2}$ & \multicolumn{2}{c}{111.7} & \multicolumn{2}{c}{1.7}   & \multicolumn{2}{c}{113.4} \\
7.5 & 1.3 & $0.14 \times 10^{-2}$ & \multicolumn{2}{c}{114.6} & \multicolumn{2}{c}{0.9}   & \multicolumn{2}{c}{115.5} \\
10  & 1.3 & $0.14 \times 10^{-2}$ & \multicolumn{2}{c}{115.4} & \multicolumn{2}{c}{0.5}   & \multicolumn{2}{c}{115.9} \\
20  & 1.3 & $0.14 \times 10^{-2}$ & \multicolumn{2}{c}{117.4} & \multicolumn{2}{c}{0.2}   & \multicolumn{2}{c}{117.6} \\
30  & 1.3 & $0.14 \times 10^{-2}$ & \multicolumn{2}{c}{118.1} & \multicolumn{2}{c}{0.1}   & \multicolumn{2}{c}{118.2} \\
\hline
\hline
\multicolumn{9}{c}{Big amorphous carbon grains ($100 \lesssim a \lesssim 200$~nm)} \\
\hline
Mantle thickness (nm) & $\rho$ (g/cm$^3$) & $M_d / M_{{\rm H}}$ & \multicolumn{2}{c}{$\left[\frac{{\rm C}}{{\rm H}}\right]_{{\rm aromatic}}$} & \multicolumn{2}{c}{$\left[\frac{{\rm C}}{{\rm H}}\right]_{{\rm aliphatic}}$} & \multicolumn{2}{c}{$\left[\frac{{\rm C}}{{\rm H}}\right]_{{\rm total}}$} \\
\hline
0   & 1.60 & $0.64 \times 10^{-3}$ & \multicolumn{2}{c}{0}    & \multicolumn{2}{c}{53.1} & \multicolumn{2}{c}{53.1} \\
5   & 1.35 & $0.54 \times 10^{-3}$ & \multicolumn{2}{c}{6.6}  & \multicolumn{2}{c}{45.0} & \multicolumn{2}{c}{51.6} \\
7.5 & 1.33 & $0.53 \times 10^{-3}$ & \multicolumn{2}{c}{9.3}  & \multicolumn{2}{c}{41.7} & \multicolumn{2}{c}{51.0} \\
10  & 1.32 & $0.53 \times 10^{-3}$ & \multicolumn{2}{c}{11.7} & \multicolumn{2}{c}{38.7} & \multicolumn{2}{c}{50.4} \\
20  & 1.31 & $0.52 \times 10^{-3}$ & \multicolumn{2}{c}{22.1} & \multicolumn{2}{c}{25.9} & \multicolumn{2}{c}{48.0} \\
30  & 1.30 & $0.52 \times 10^{-3}$ & \multicolumn{2}{c}{39.6} & \multicolumn{2}{c}{4.4}  & \multicolumn{2}{c}{44.0} \\
\hline
\hline
\multicolumn{9}{c}{Amorphous silicate grains ($100 \lesssim a \lesssim 200$~nm)} \\
\hline
Mantle thickness (nm) & $\rho$ (g/cm$^3$) & $M_d / M_{{\rm H}}$ & $\left[\frac{{\rm Si}}{{\rm H}}\right]$ & $\left[\frac{{\rm Mg}}{{\rm H}}\right]$ & $\left[\frac{{\rm O}}{{\rm H}}\right]$ & $\left[\frac{{\rm Fe}}{{\rm H}}\right]$  & $\left[\frac{{\rm S}}{{\rm H}}\right]$  & $\left[\frac{{\rm C}}{{\rm H}}\right]_{{\rm mantle}}$ \\
\hline
0   & 2.95 & $0.68 \times 10^{-2}$ & 44.5 & 63.1 & 152.1 & 29.1 & 6.2 & 0     \\
5   & 1.60 & $0.37 \times 10^{-2}$ & 38.5 & 54.5 & 131.5 & 25.2 & 5.4 & 33.9  \\
7.5 & 1.48 & $0.34 \times 10^{-2}$ & 36.0 & 51.0 & 122.9 & 23.5 & 5.0 & 48.1  \\
10  & 1.42 & $0.33 \times 10^{-2}$ & 33.7 & 47.7 & 115.1 & 22.0 & 4.7 & 61.1  \\
15  & 1.36 & $0.32 \times 10^{-2}$ & 29.3 & 41.6 & 100.3 & 19.2 & 4.1 & 85.5  \\
20  & 1.33 & $0.31 \times 10^{-2}$ & 24.8 & 35.1 & 84.7  & 16.2 & 3.5 & 111.2 \\
\hline
\end{tabular}
\end{table*}

In Sect.~\ref{dust_properties}, we deviate from this original composition by changing the mantle thickness on the carbon and silicate grains and the volume of the FeS inclusions in the silicates. We assume that 30\%, 50\% or 100\% of the metallic nano-inclusion volume is made of FeS. For the carbonaceous grains, we consider mantle thicknesses of 0, 5, 7.5, 10, 20, and 30~nm, and for the silicates of 0, 5, 7.5, 10, 15, and 20~nm. In all cases, we assume a fixed size distribution, meaning that when the mantle thickness increases (decreases), compared to the standard \citet{Jones2013} model, the size of the core decreases (increases) accordingly. The corresponding elemental abundances and dust-to-gas mass ratios are listed in Tab. 1. 

\subsection{Methodology}
\label{methodology}

The optical properties of the core-mantle grains are derived with the BHCOAT routine \citep{Bohren1983} and for the variations in the metallic inclusions, we use the Maxwell-Garnett effective medium theory as described in \citet{Koehler2014}. Once the grain optical properties are known, the corresponding emission and extinction are computed with the DustEM\footnote{The DustEM code and associated plotting routines can be downloaded at \url{http://www.ias.u-psud.fr/DUSTEM/}.} code described in \citet{Compiegne2011}. DustEM is also coupled to the 3D radiative transfer code CRT \citep{Juvela2005, Ysard2012}. This allows us to calculate the dust emission and extinction for various density distributions and a range of hydrogen column densities, $N_{\rm H}$, similar to those measured by PCXI.

The variations in the dust observables presented by PCXI are based on a pixel-by-pixel modified black-body $\chi^2$ fit. These fits were performed for the all-sky IRIS 100~$\mu$m band (3\,000~GHz, Miville-Desch\^enes \& Lagache 2005) and Planck-HFI 857, 545, and 353~GHz bands (350, 550, and 850~$\mu$m). To compare the results of PCXI to the dust model directly, we fit the modelled dust SED in the exact same way\footnote{N.B. As shown in Fig. 14 of \citet{Jones2013}, the spectral index $\beta$ actually varies across this entire wavelength range. However, fixing the $\beta$-value is the best one can do with only four observational bands.}, i.e. integration in the IRAS and Planck bands, followed by a modified black-body fit.

The results for the diffuse ISM ($10^{19} \leqslant N_{\rm H} \leqslant 2.5\times 10^{20}$~H/cm$^2$) presented in PCXI after computing $T$, $\beta$ and $\tau_{\nu_0}$, can be efficiently summarised with the following few parameters:\\
$-$ the radiance: $\cal{R}$ $= \int_{\nu} I_{\nu} d\nu$ ;\\
$-$ the luminosity: $L_{\rm H} = 4\pi \cal{R}$ $/ N_{\rm H}$ ;\\
$-$ the opacity: $\sigma_{\nu_0} = \tau_{\nu_0}/N_{\rm H}$. \\
From these parameters, six figures show the variations and relations among dust observables (see for instance Fig. \ref{Fig1}):\\
$-$ Panel {\it a}: the $\beta-T$ relation (Fig.~16 in PCXI, for which we apply the diffuse ISM mask corresponding to $10^{19} \leqslant N_{\rm H} \leqslant 2.5\times 10^{20}$~H/cm$^2$) ;\\
$-$ Panel {\it b}: the $L_{\rm H}$ vs. $T$ relation (their Fig.~20), in which $L_{\rm H}$ is found to be almost constant ; \\
$-$ Panel {\it c}: the $\sigma_{\nu_0}$ vs. $T$ anti-correlation (their Fig.~20), also observed by \citet{PlanckBoulanger} and \citet{PlanckAniano} ; \\
$-$ Panel {\it d}: the $\tau_{\nu_0}$ to $N_{\rm H}$ ratio (their Fig.~22) ;\\
$-$ Panel {\it e}: the $E(B-V)$ to $\tau_{\nu_0}$ ratio (their Fig.~23) ; and\\
$-$ Panel {\it f}: the $E(B-V)$ to $\cal{R}$ ratio (their Fig.~23).\\

In the following sections, we compare all our results to the observational results of PCXI using these six figures (see for instance Fig.~\ref{Fig1}). In panels~{\it a} (top left), {\it b} (top middle), and {\it c} (top right), PCXI estimated the hydrogen column density used to compute $L_{\rm H}$ and $\sigma_{\nu_0}$ from the Leiden/Argentine/Bonn (LAB) survey of Galactic HI for the atomic part \citep{Kalberla2005}, and from the Planck $^{12}$CO J=0$\rightarrow$1 map (type 3) for the molecular part, which has a high signal-to-noise ratio and is the best suited to detect faint CO regions at high Galactic latitudes \citep{PlanckCO}. Consequently, it does not account for the ionised hydrogen and molecular gas not detected in CO that can be present along the lines of sight. As explained in PCXI, their estimate of $N_{\rm H}$ for panels {\it b} and {\it c} is thus a lower limit. For these first three panels ({\it a}, {\it b}, and {\it c}), the observational results are the density of points maps, on which we overplotted yellow contours: the central contour encloses 50\% of the observed pixels and the external contour 75\%. Additionally, in panels~{\it b} and {\it c}, the median values of the luminosity and opacity, in equally spaced bins of temperature, are overplotted as white circles with ``error bars'', which are their $1\sigma$-dispersion.

In panel~{\it d} (bottom left), the opacity is plotted as a function of $N_{\rm H}$, also estimated from the LAB survey and the Planck type 3 CO map, but smoothed to 30' resolution. The observational results are represented by black circles, which are the median values of the opacity in each bin of column density (equally spaced in log), and the black ``error bars'' are the $1\sigma$-dispersion of $\sigma_{\nu_0}$ in each $N_{\rm H}$ beam.

The $E(B-V)$ values used in panels~{\it e} (bottom middle) and {\it f} (bottom right) come from the Sloan Digital Sky Survey (SDSS) on quasars \citep{Schneider2010}. These SDSS data cover a larger column density range ($N_{\rm H} \leqslant 6.25 \times 10^{21}$~H/cm$^2$) than shown in the four previous figures. These measures were performed along 53\,399 separated lines of sight. In panels~{\it e} and {\it f}, we only show the PCXI results corresponding to the same column density range as in previous figures, $10^{19} \leqslant N_{\rm H} \leqslant 2.5\times 10^{20}$~H/cm$^2$, in order to make all of them comparable. The observational results are represented by black circles, each one being the average of the $E(B-V)$ values in a $\tau_{\nu_0}$ (Fig.~{\it e}) or $\cal{R}$ (Fig.~{\it f}) bin. As explained by PCXI, the bin sizes are variable in order to keep the same number of samples per bin ($N = 1\,000$). The ``error bars'' are the standard deviation of the $E(B-V)$ for each bin, divided by $\sqrt{N}$. The solid black lines present the best linear fits computed by PCXI for the visible extiction to submm optical depth and radiance ratios, respectively.

Finally, in all panels, the thick coloured lines show our modelling results for a physically reasonable set of dust model parameters, without attempting to fit the data.

\section{Dust environment}
\label{environment}

\begin{figure*}[!th]
\centerline{
\begin{tabular}{ccc}
\includegraphics[width=6cm,height=4.8cm]{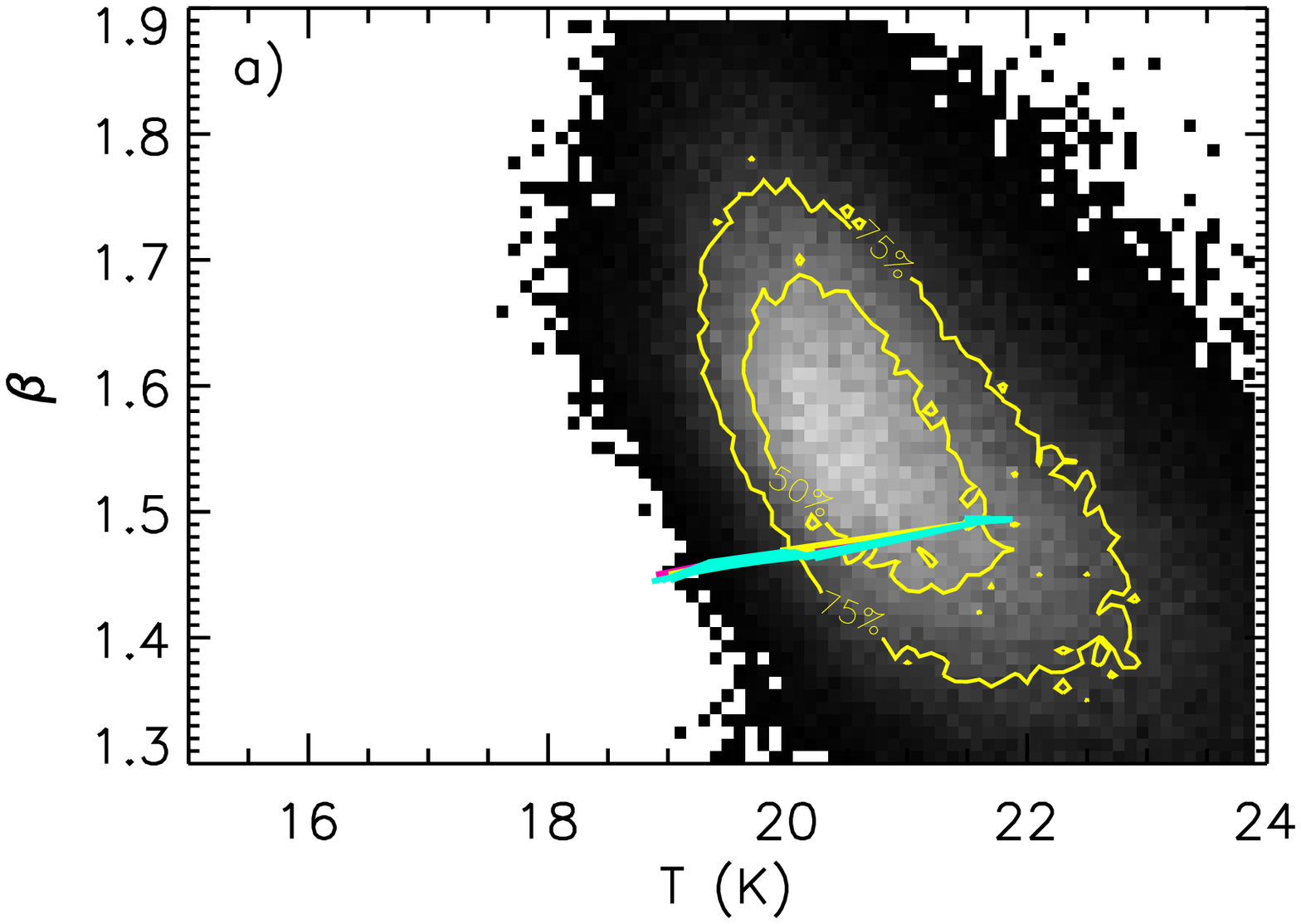} & \includegraphics[width=6cm,height=4.8cm]{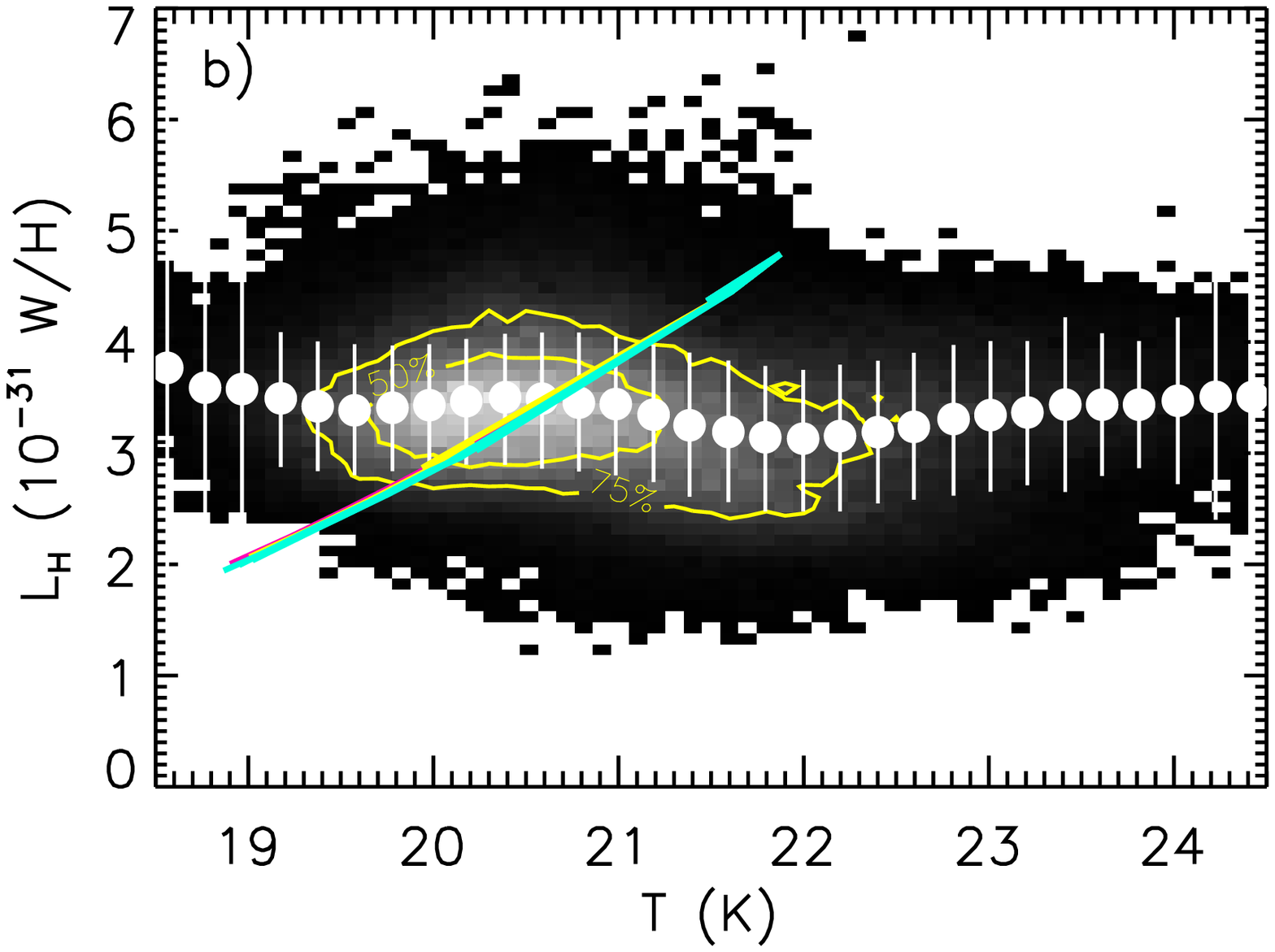}  & \includegraphics[width=6cm,height=4.8cm]{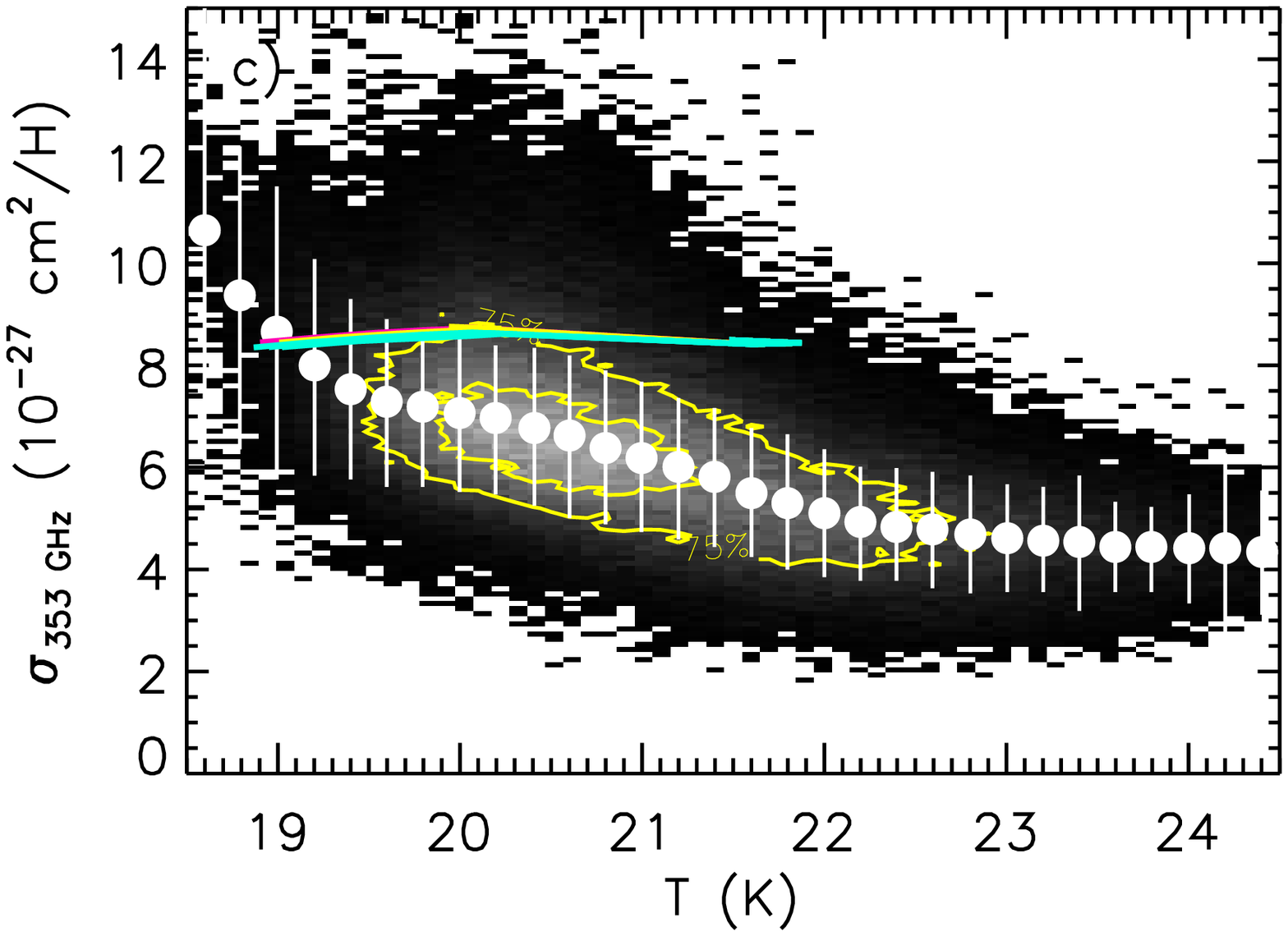} \\
\includegraphics[width=0.34\textwidth]{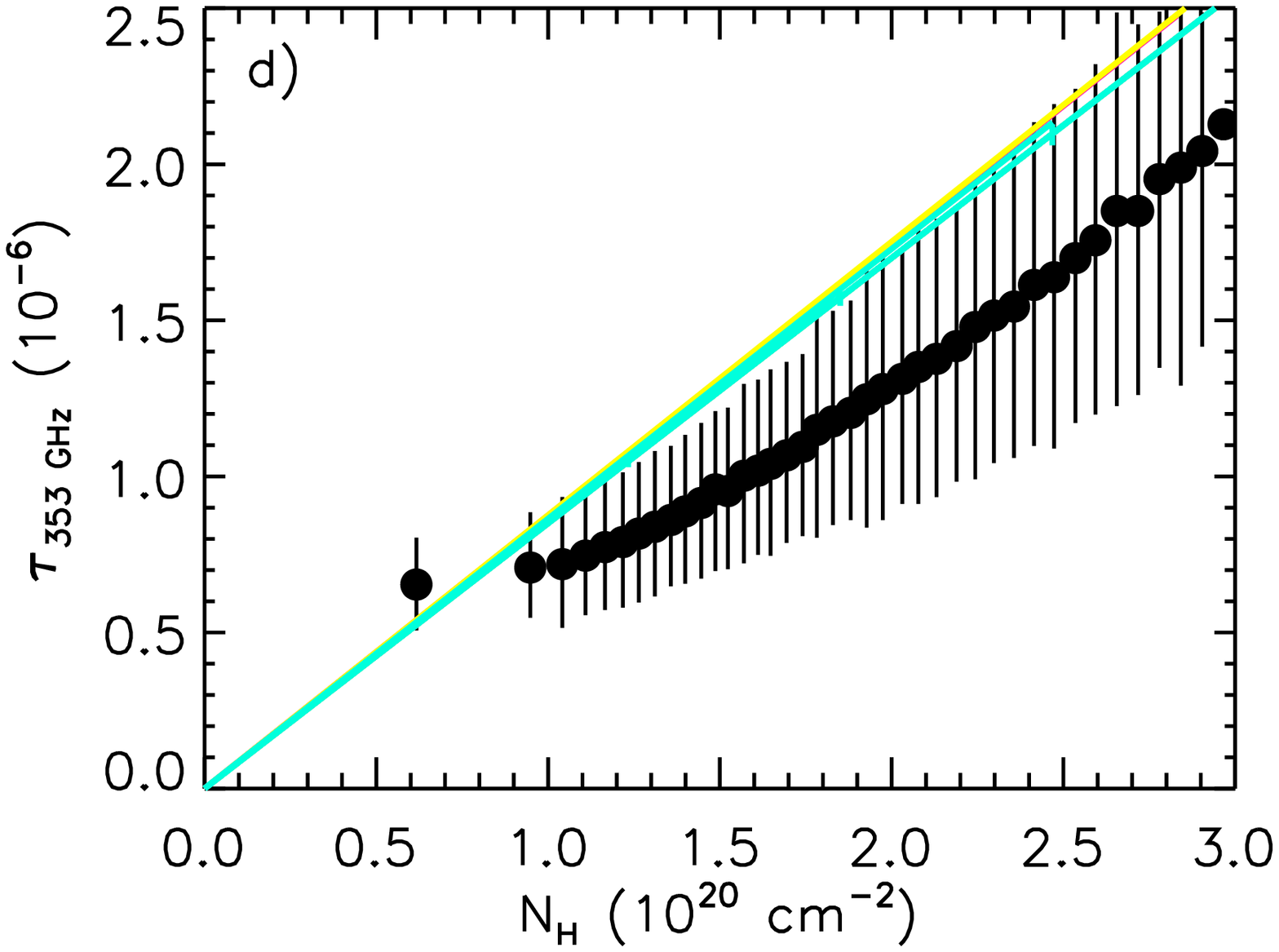} & \includegraphics[width=0.34\textwidth]{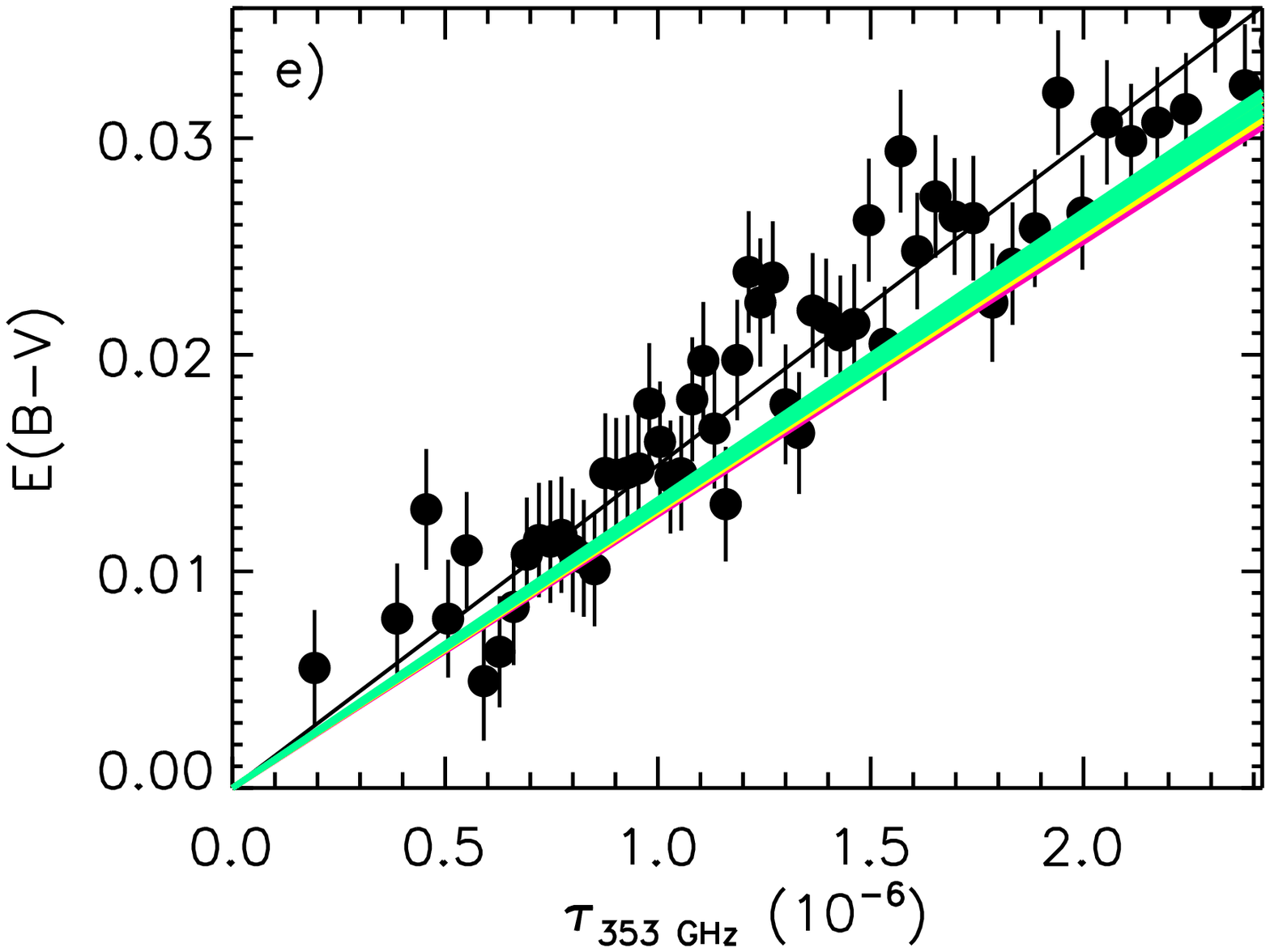} & \includegraphics[width=0.34\textwidth]{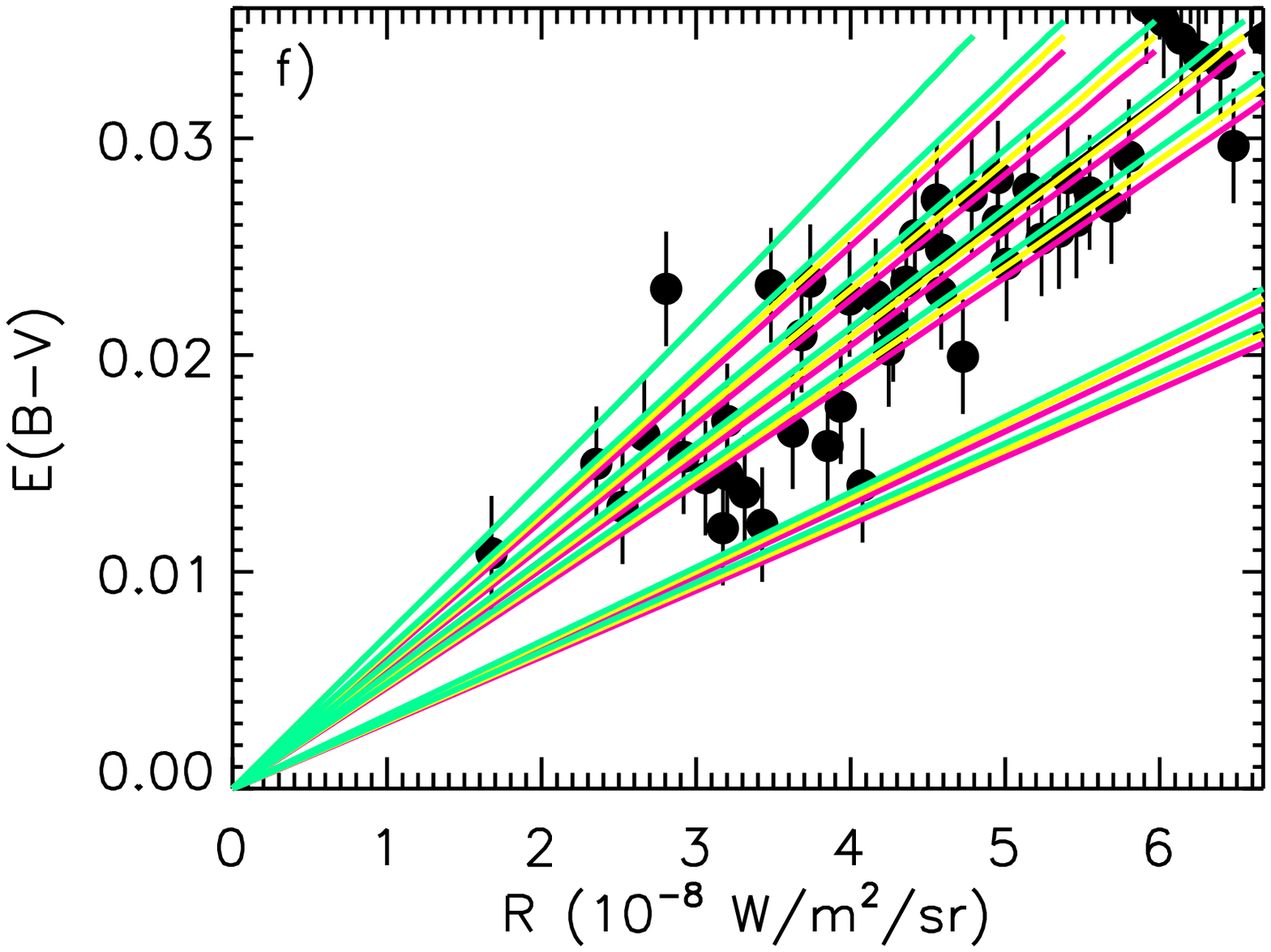}
\end{tabular}}
\caption{Influence of the density distribution in the local ISM (silicates with a 5~nm thick mantle and carbons with a 20~nm thick mantle). Cold Neutral Medium (CNM) with $n_{\rm H} = 25$ to 70~H/cm$^3$ is shown in pink, CNM with $n_{\rm H} = 100$~H/cm$^3$ surrounded by Warm Neutral Medium (WNM) with $n_{\rm H} = 1$~H/cm$^3$ is plotted in yellow, and WNM with $n_{\rm H} = 1$~H/cm$^3$ is displayed in turquoise. The three models overlap in the figures. The radiation field is also varied with $0.8 \leqslant G_0 \leqslant 1.4$. In panels~{\it d}, {\it e}, and {\it f}, seven lines of the same colour are plotted. Each of these lines corresponds to a different value of $G_0$. For instance, in panel {\it f}, the turquoise lines from left to right correspond to $G_0 = 0.8$, 0.9, 1.0, 1.1, 1.2, 1.3, and 1.4, respectively. The models falling below the observed $E(B-V)$ to $\cal{R}$ ratio are those illuminated by an ISRF scaled with $G_0 = 1.3$ and 1.4. See Sect. \ref{methodology} for description of the black and white lines and symbols.}
\label{Fig1} 
\end{figure*}

As suggested by PCXI, one way of explaining the dust observational results measured in the high Galactic latitude diffuse ISM could be to consider variations in the dust environment. Here we investigate variations in the gas density distribution in the medium, in the ISRF intensity, $G_0$\footnote{Scaling factor for the radiation field integrated between 6 and 13.6~eV. The standard radiation field corresponds to $G_0 = 1$ \citep{Draine1985}.}, and in the radiation field hardness.

For the density distribution, we consider three cases. These include: {\it i)} cold neutral medium (CNM): assuming a 2~pc-long box filled with a gas with a density varying linearly from 25 to 70~H/cm$^3$ ; {\it ii)} CNM surrounded by warm neutral medium (WNM): considering a 2~pc-long box filled with a 1~H/cm$^3$ gas with a denser filament at the centre (1~pc wide and $n_{\rm H} = 100$~H/cm$^3$) ; and {\it iii)} WNM: assuming three different boxes filled with a gas with $n_{\rm H} = 1$~H/cm$^3$ and lengths of 20, 30, and 40~pc. The lengths of the boxes were chosen to match the column density range measured along the line of sight, $10^{19} \leqslant N_{\rm H} \leqslant 2.5 \times 10^{20}$~H/cm$^2$, presented in PCXI. These three cloud models are then illuminated by the standard ISRF \citep{Draine1985}, for which we vary the intensity from $G_0 = 0.8$ to 1.4. As seen in Sect. \ref{dust_model}, for all cases, the dust populations are amorphous carbon and silicate grains with aromatic-rich carbon mantles 20 and 5~nm thick, respectively.

\begin{figure*}[!th]
\centerline{
\begin{tabular}{ccc}
\includegraphics[width=6cm,height=4.8cm]{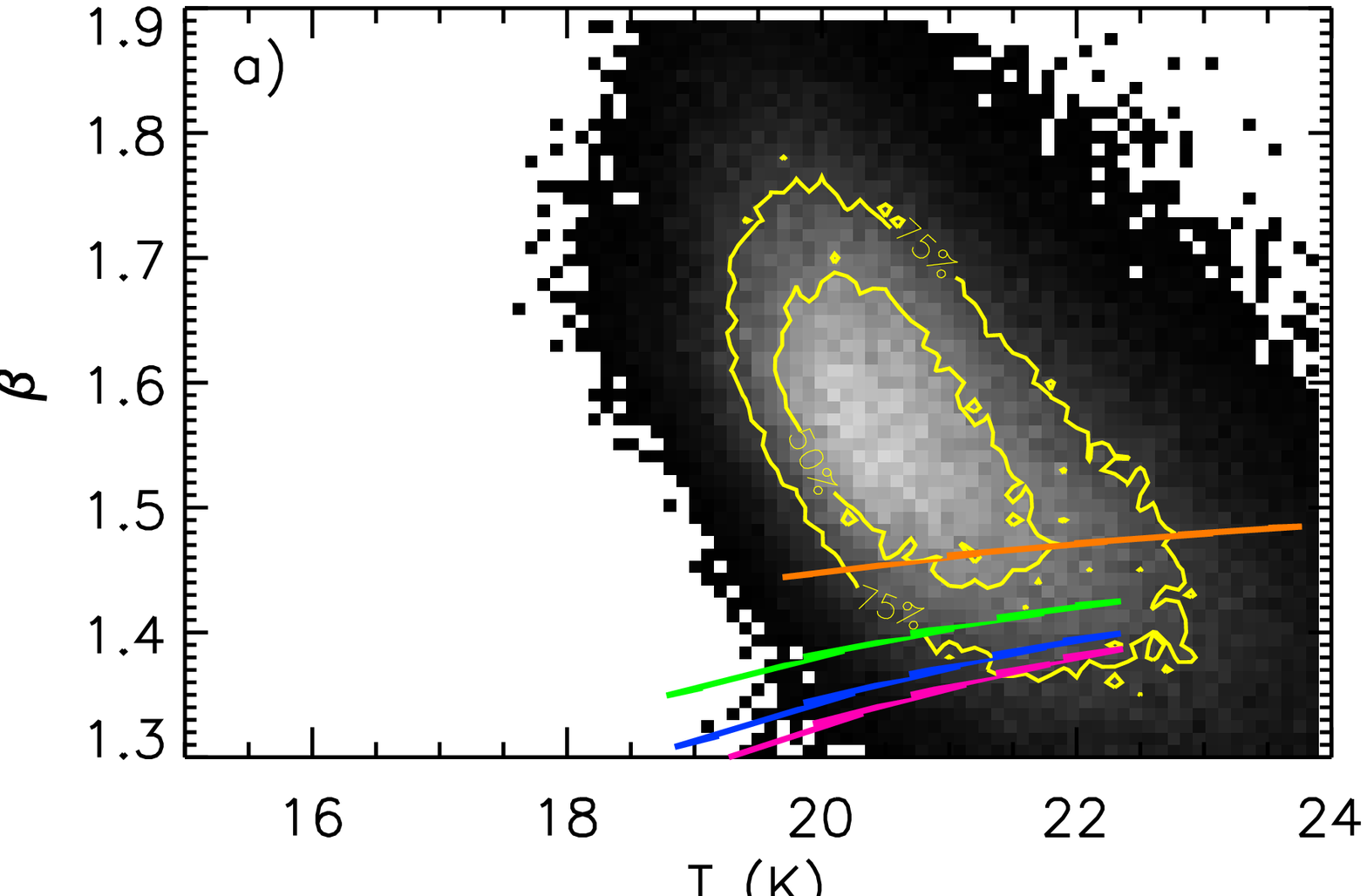} & \includegraphics[width=6cm,height=4.8cm]{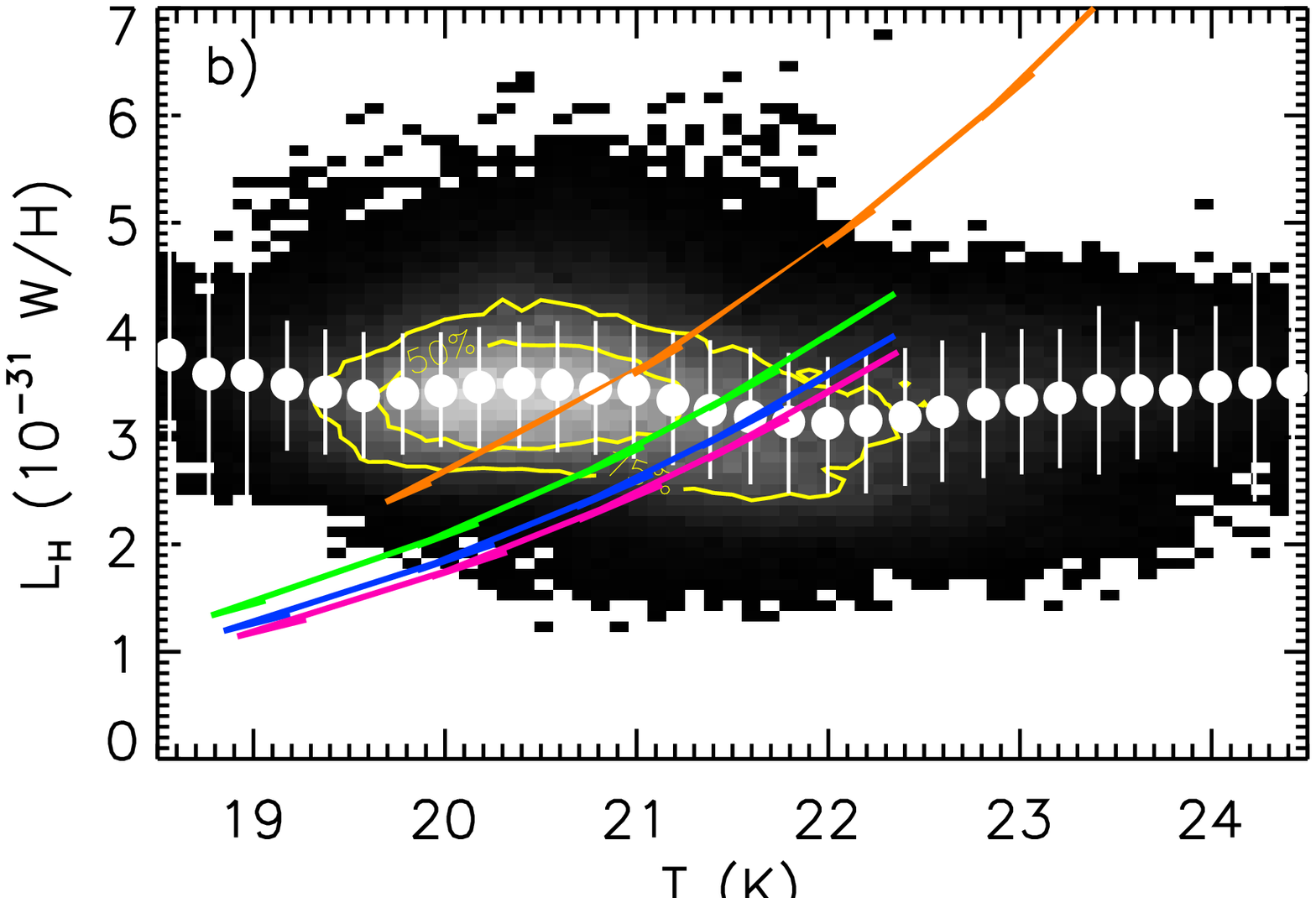}  & \includegraphics[width=6cm,height=4.8cm]{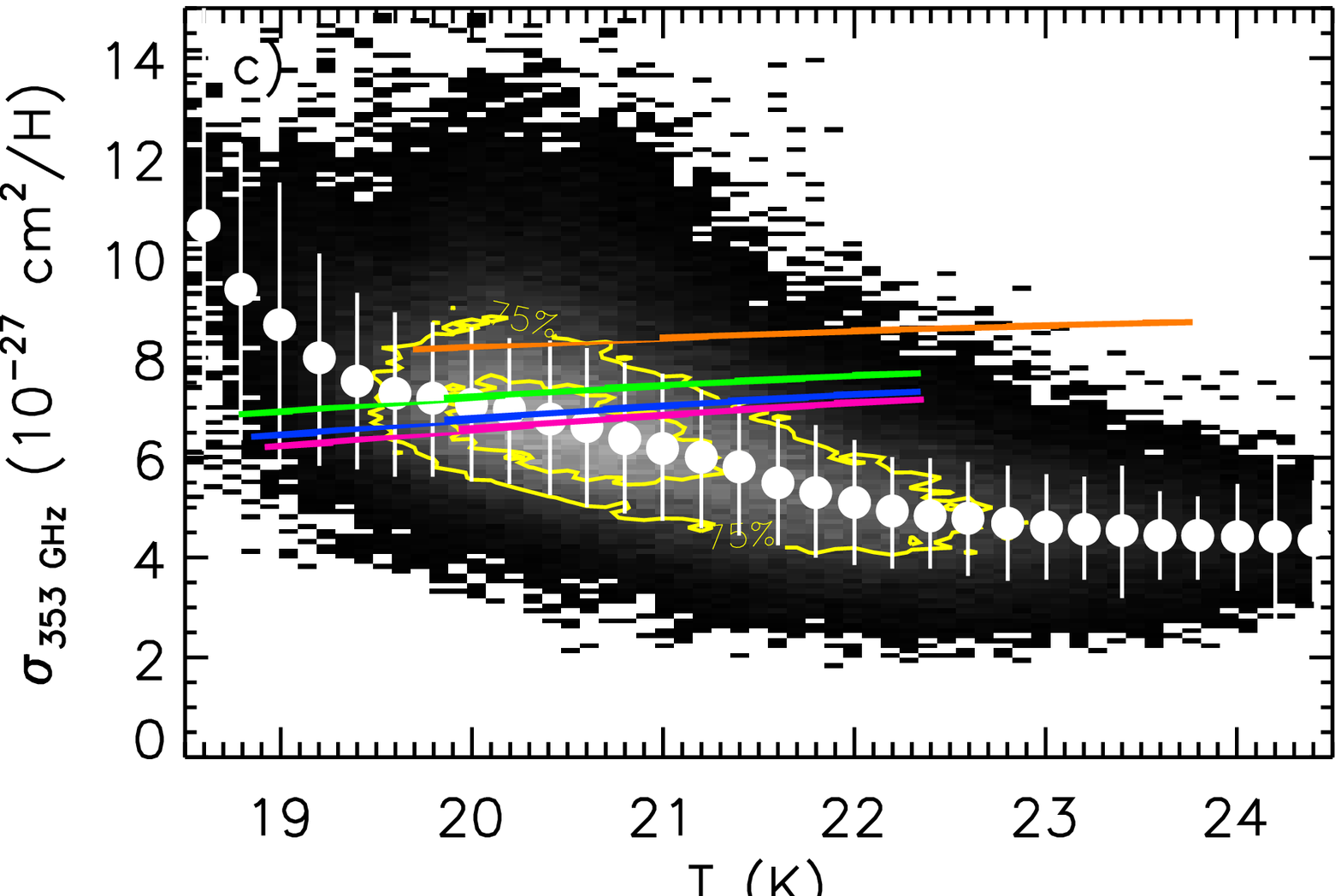} \\
\includegraphics[width=0.34\textwidth]{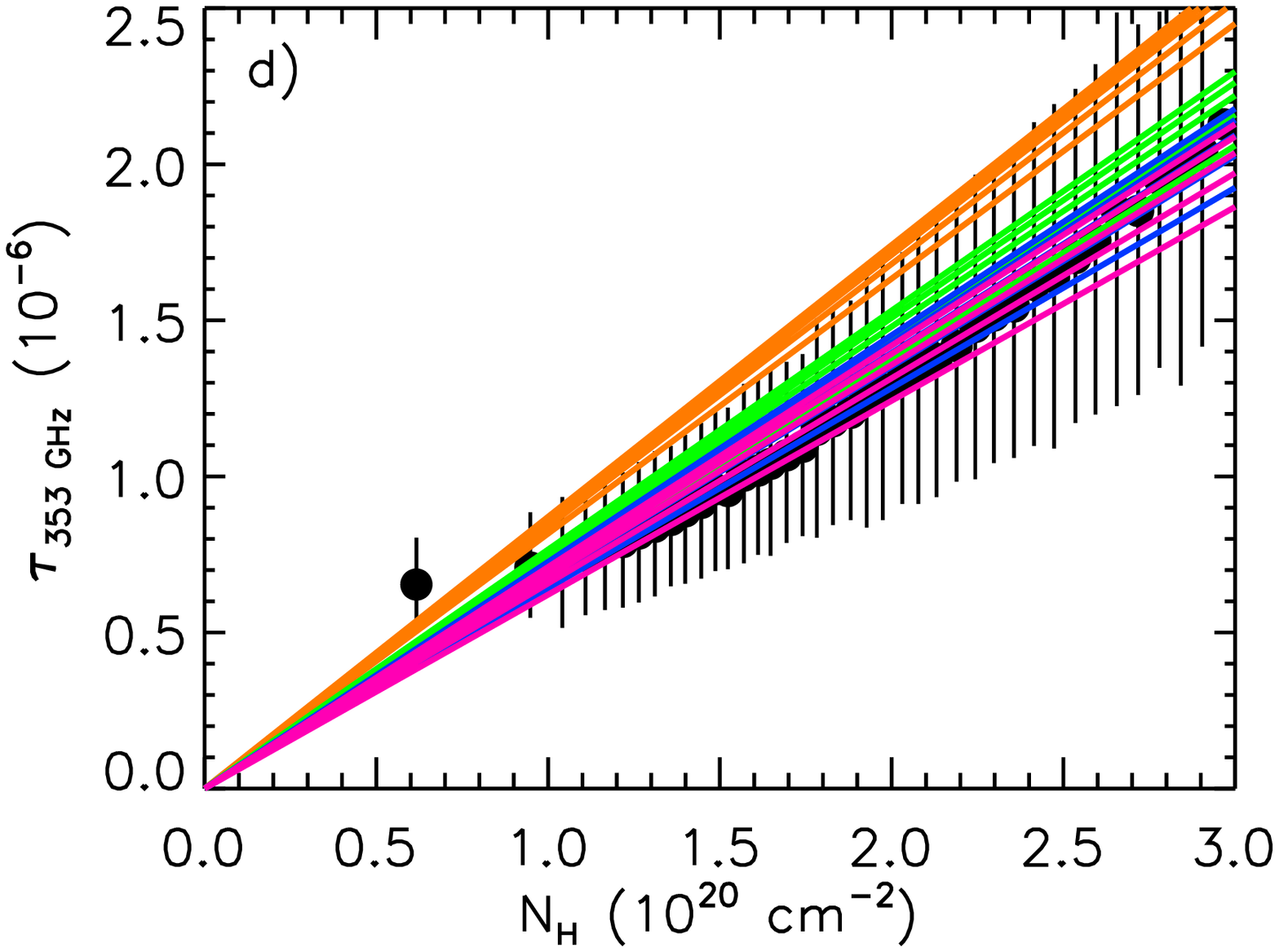} & \includegraphics[width=0.34\textwidth]{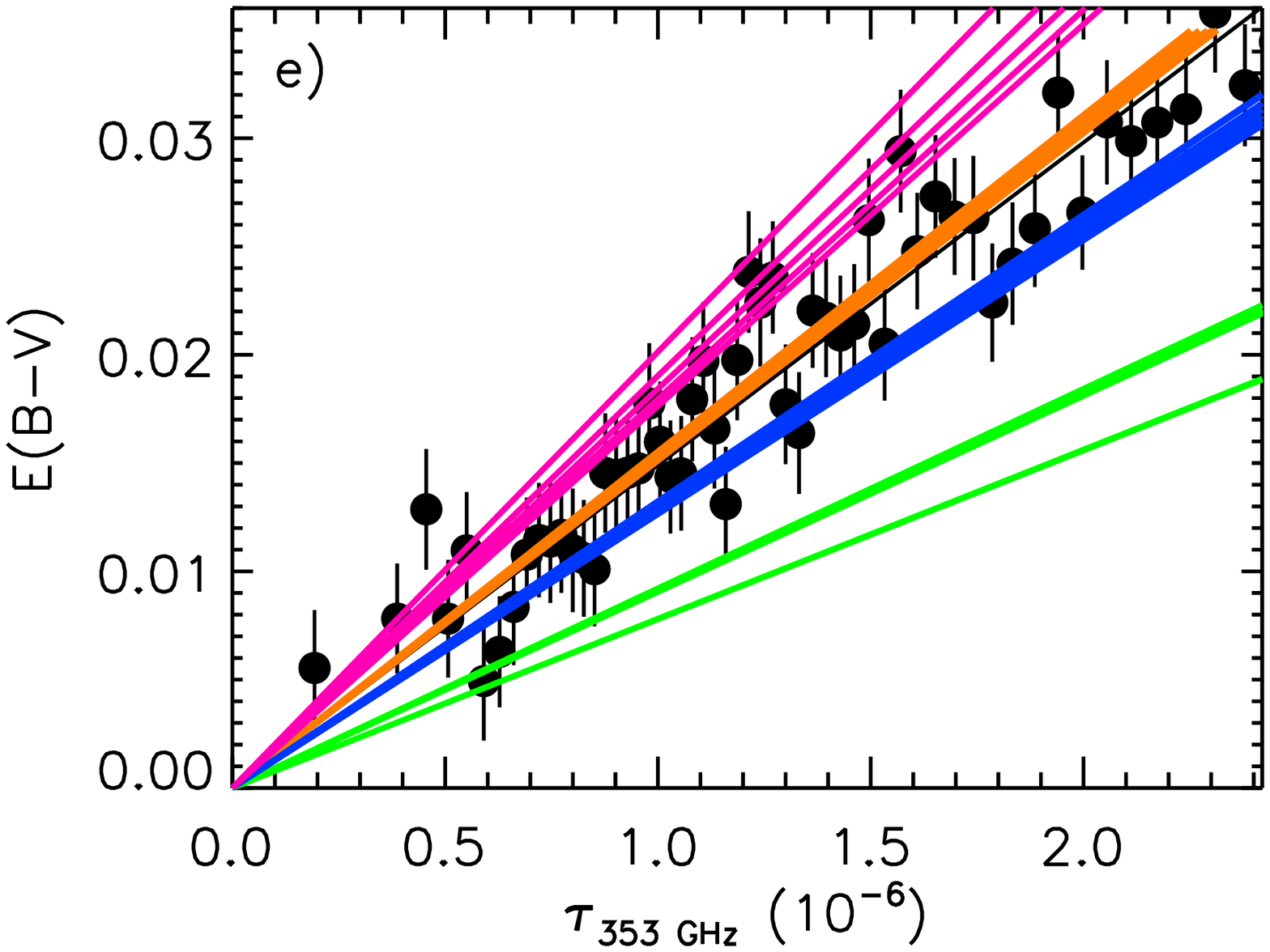} & \includegraphics[width=0.34\textwidth]{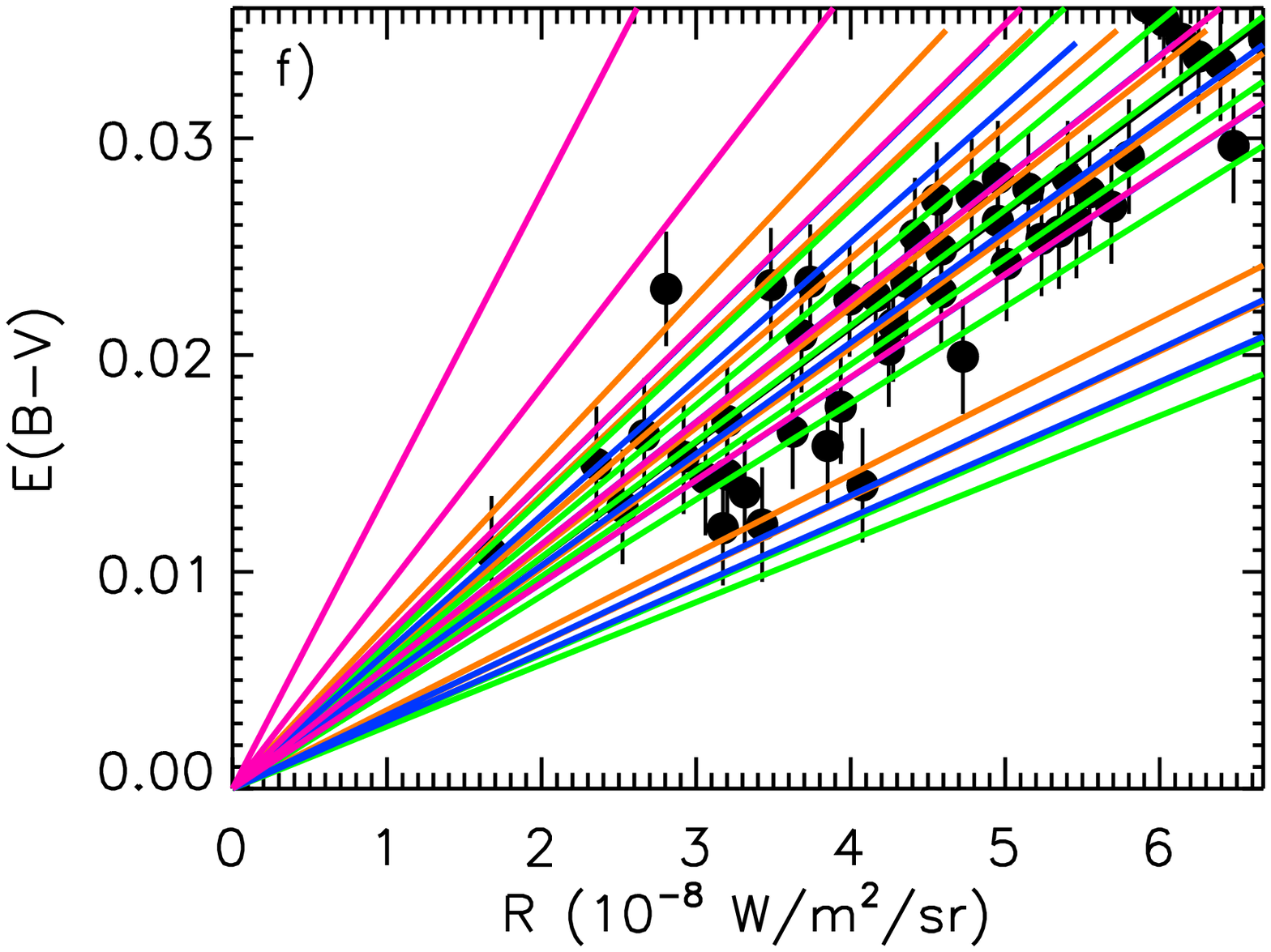}
\end{tabular}}
\caption{Influence of the radiation field hardness (silicates with a 5~nm thick mantle and carbons with a 20~nm thick mantle). The black-body with an effective temperature of 10\,000~K is plotted in orange, 20\,000~K in green, 30\,000~K in blue, and 40\,000~K in pink. The intensity of the radiation field is also varied with $1 \leqslant G_0 \leqslant 3$. In panels~{\it d}, {\it e}, and {\it f}, as in Fig.~\ref{Fig1}, each of the seven lines with the same colour corresponds to a different value of $G_0$. In panels {\it d} and {\it e}, the lines are well separated because with the set of model parameters tested here, the recovered opacity values are not constant with $G_0$ as can be seen in panel {\it c}. See Sect. \ref{methodology} for description of the black and white lines and symbols.}
\label{Fig1bis} 
\end{figure*}

The results are presented in Fig.~\ref{Fig1}.  As expected for optically thin media, it appears that variations in the gas density distribution of the medium do not produce any significant changes in the dust observables (Figs. \ref{Fig1}a, b, c, d, e, f, where all the models more or less overlap). For instance, the variation in dust temperature does not exceed 0.15~K, when comparing the three cases of density distribution. Then, for variations in the ISRF intensity from $G_0 = 0.8$ to 1.4, we see that the colour temperature varies by $\sim 3$~K (from $\sim 19$ to 22~K) but that the spectral index and the opacity are almost constant ($\beta \sim 1.45$ to 1.5 and $\sigma_{{\rm 353\;GHz}} \sim 8.5 \times 10^{-27}$~cm$^2$/H). The two highest values of $G_0$ (1.3 and 1.4) lead to inconsistent $E(B-V)/\cal{R}$ ratios. This means that even if ISRF intensity variations contribute to changing the dust temperature, they cannot explain the observed variations. We also investigate the influence of the radiation field hardness on the dust observables. Indeed, hot stars located at lower Galactic latitudes and in the Galactic plane may participate in the dust heating at high latitudes. These stars have harder radiation fields than the standard ISRF for variable intensities: one can expect that when reaching high latitudes, the radiation field of these stars would be decreased by a significant factor after going through many insterstellar clouds. To test this, we consider black-bodies with effective temperatures of 10\,000, 20\,000, 30\,000, and 40\,000~K. The black-body intensities are normalised to correspond to $G_0 = 1$, 1.5, 2, 2.5, and 3. For a given $G_0$, the absorbed power in the visible/UV range is thus constant, whereas the colour changes (i.e. the energy of the average absorbed photon increases with increasing effective temperature). The results are presented in Fig.~\ref{Fig1bis}. The model illuminated by a black-body at 10\,000~K cannot reproduce the observed extinction to radiance ratio (orange line in Fig.~\ref{Fig1bis}f) but the three other cases fit the data with $T \sim 19-23$~K and $\beta \sim 1.3-1.45$. Variations in the radiation field hardness may thus take part in creating the observed variations in the dust observables.

As can be seen from Fig.~\ref{Fig1}d, the standard model of \citet{Jones2013} as updated by \citet{Koehler2014} does not fit the median results presented in PCXI (black and white circles in Fig. \ref{Fig1}). This was to be expected for two reasons. First, their model was designed to fit pre-Planck data. Second, these authors fitted their model to explain observations of a single line of sight in the solar neighbourhood, for which the conditions are different from those found at higher Galactic latitude, especially regarding the column density and the ISRF. The intensity of the radiation is indeed fundamental to set the thickness of the aromatic-rich carbon mantles on the grains.

\section{Dust properties}
\label{dust_properties}

Changes in the intrinsic dust properties are likely to produce variations in the dust observables. In this section, we investigate the extent of their influence by varying four parameters describing the grain properties: the thickness of the aromatic-rich carbon mantles, the composition of the metallic nano-inclusions in the silicate cores, the total abundance of carbon in dust, and the grain size distribution.

The variations in the mantle thickness lead to 42 possible combinations between grain populations and the composition of the metallic inclusions to 14 combinations (see Sect.~\ref{dust_model}). All these combinations are considered when we vary the carbon abundance and when we change the size distribution. Thus, we explore 280 grain population combinations in total and for each, we calculate the emission for seven different ISRF intensities (i.e. 1\,960 emission+extinction models). To keep the figures and results as easy to understand as possible, we do not show all possible combinations in this section but vary only one property at each step.

\subsection{Aromatic-rich carbon mantle thickness}
\label{mantle_thickness}

Grains in the core-mantle \citet{Jones2013} dust model, and as updated by \citet{Koehler2014}, are covered by an aromatic-rich carbon mantle, which is 20~nm thick for the carbon grains and 5~nm thick for the silicate grains. The aromatisation of the outer layer of the carbonaceous grains is assumed to result from photoprocessing by stellar UV-EUV photons \citep{Jones2012a, Jones2012b}. For the silicate grains, this mantle originates from the accretion of small carbonaceous particles onto the silicate cores. It was shown that these particles are likely to be aromatic-rich \citep{Brooke1999, Jones2013}, either from the start or by the same aromatisation process as for the carbonaceous particles. The thickness of this aromatic-rich mantle may thus vary depending on the grain history, age, and environment. Hence we consider a mantle thickness varying from 0 to 30~nm for the carbonaceous grains and from 0 to 20~nm for the silicates. These thicknesses correspond to abundances of 197 to 273~ppm of carbon and 45 to 25~ppm of silicon in the grains (the variations in the silicon abudance come from the fixed size distribution as explained in Sect. \ref{dust_model}).

\subsubsection{Mantle thickness on amorphous carbon grains}
\label{carbon_mantle}

\begin{figure*}[!th]
\centerline{
\begin{tabular}{ccc}
\includegraphics[width=6cm,height=4.8cm]{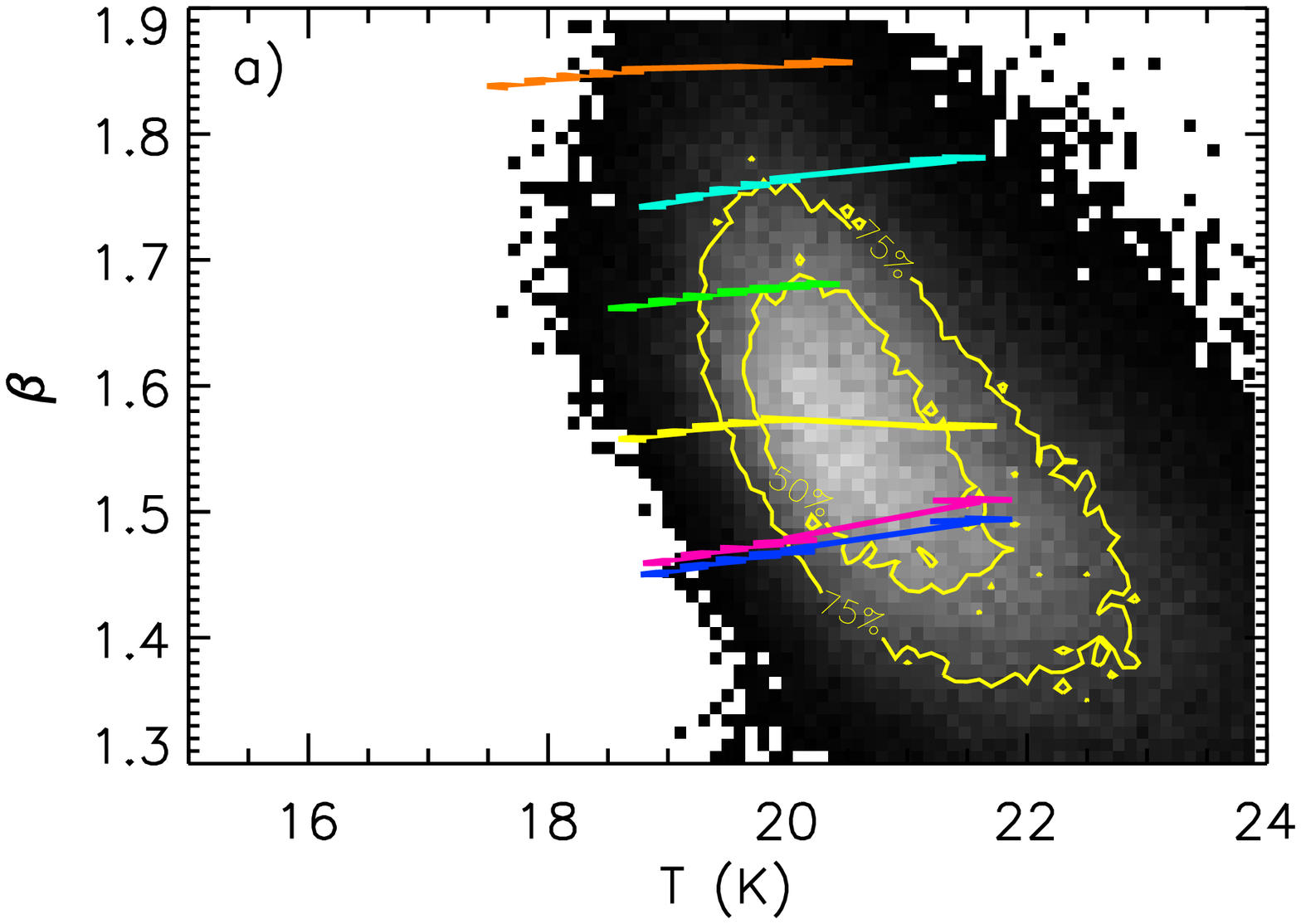} & \includegraphics[width=6cm,height=4.8cm]{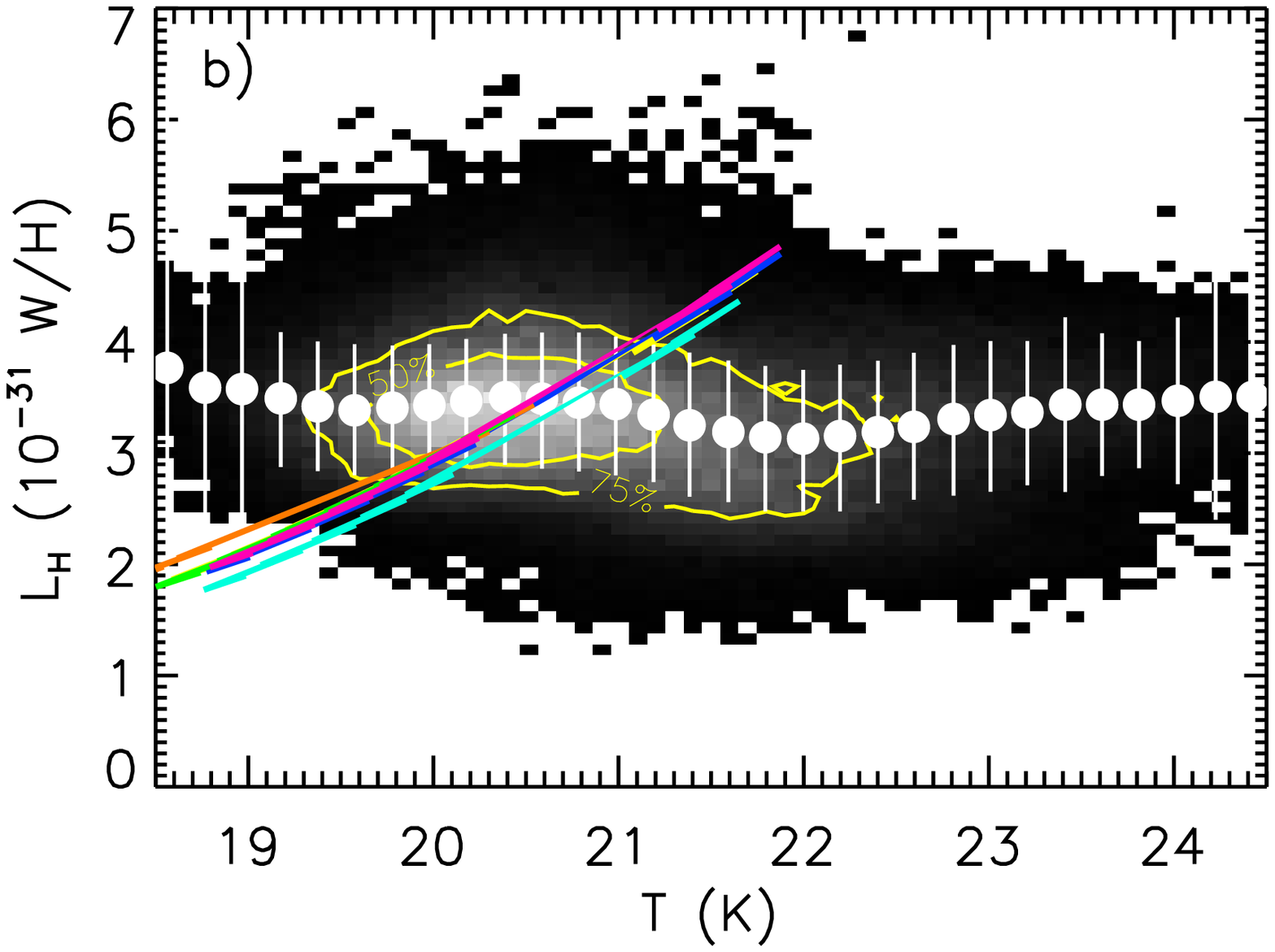}  & \includegraphics[width=6cm,height=4.8cm]{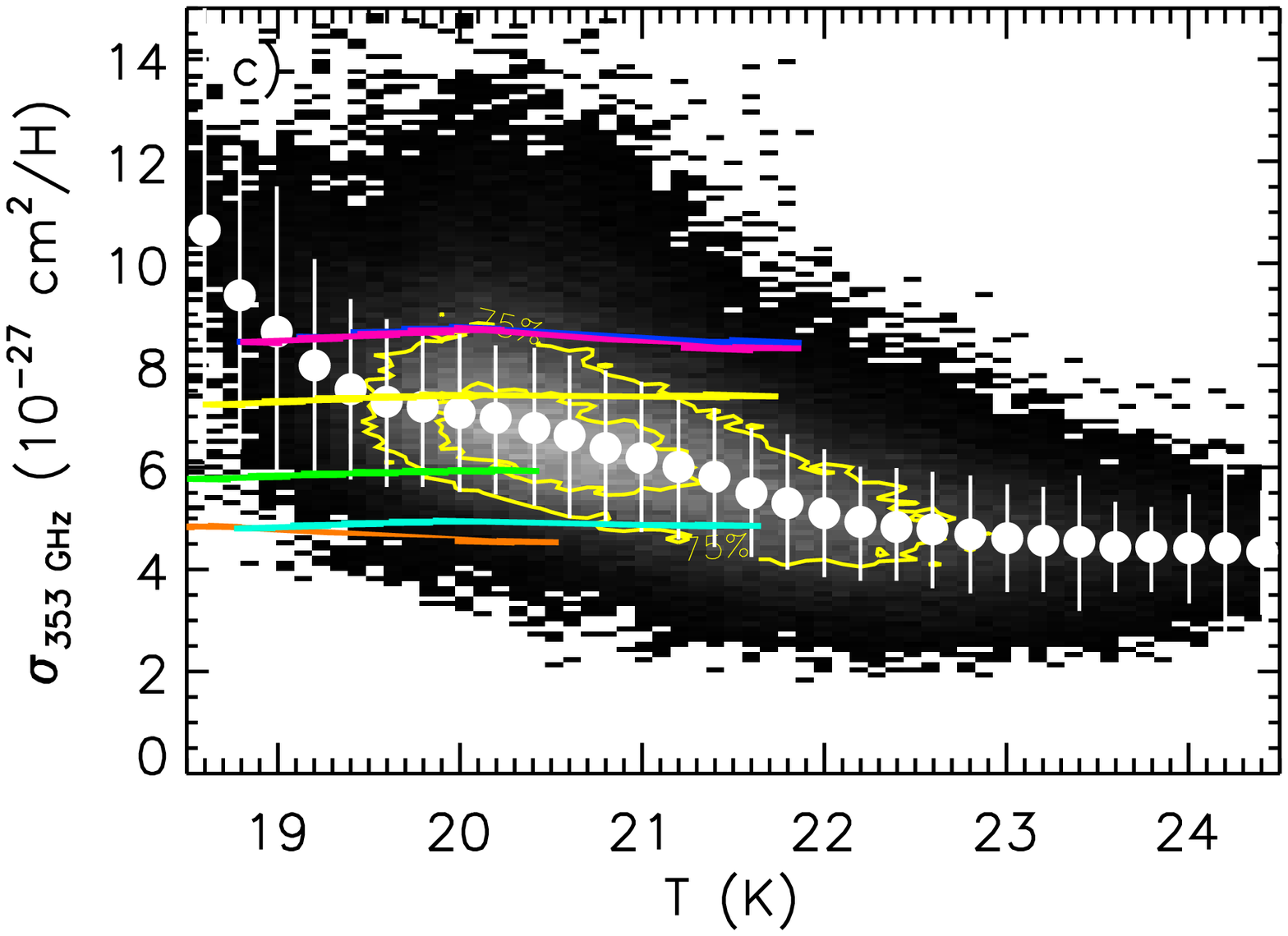} \\
\includegraphics[width=0.34\textwidth]{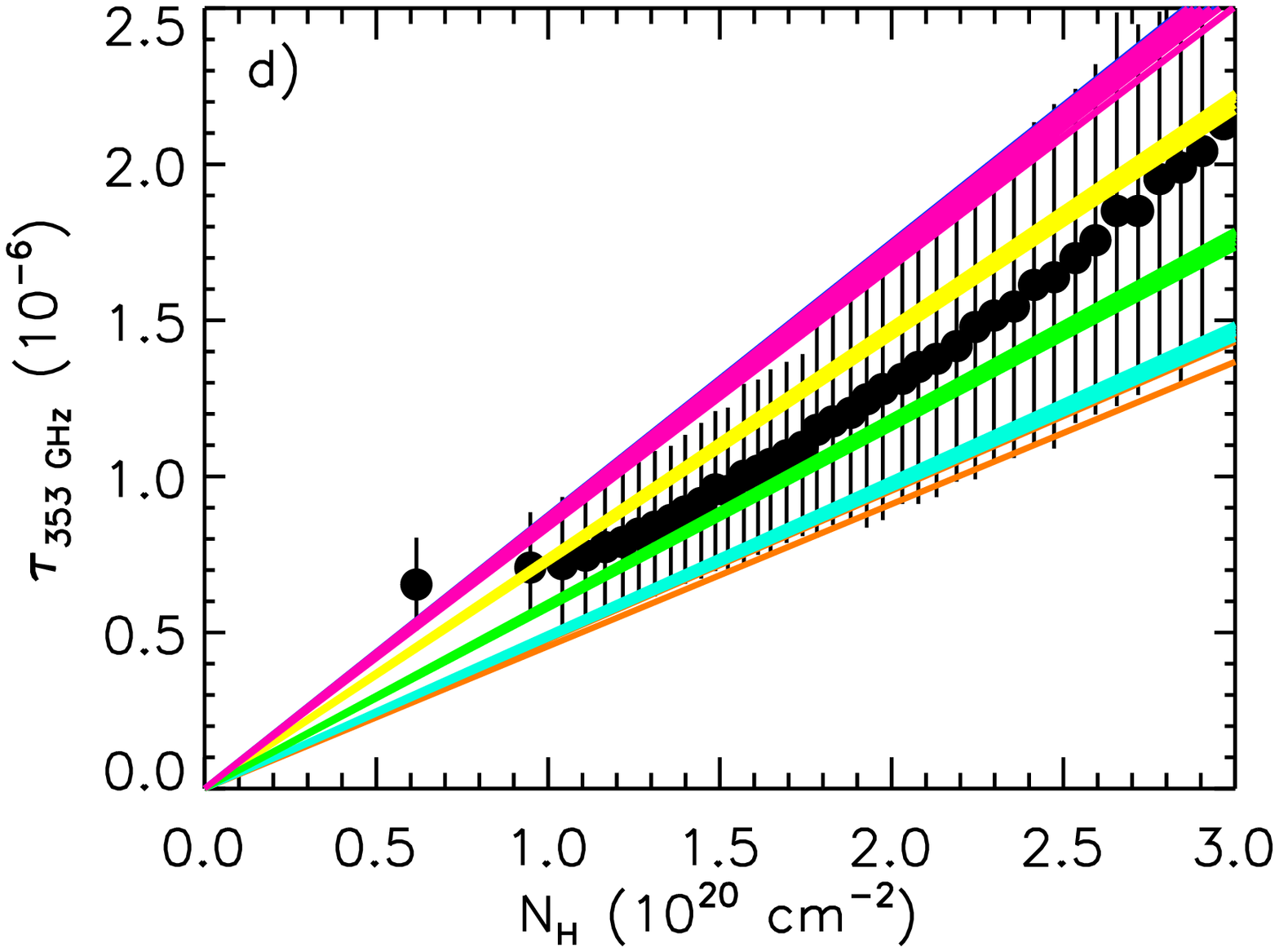} & \includegraphics[width=0.34\textwidth]{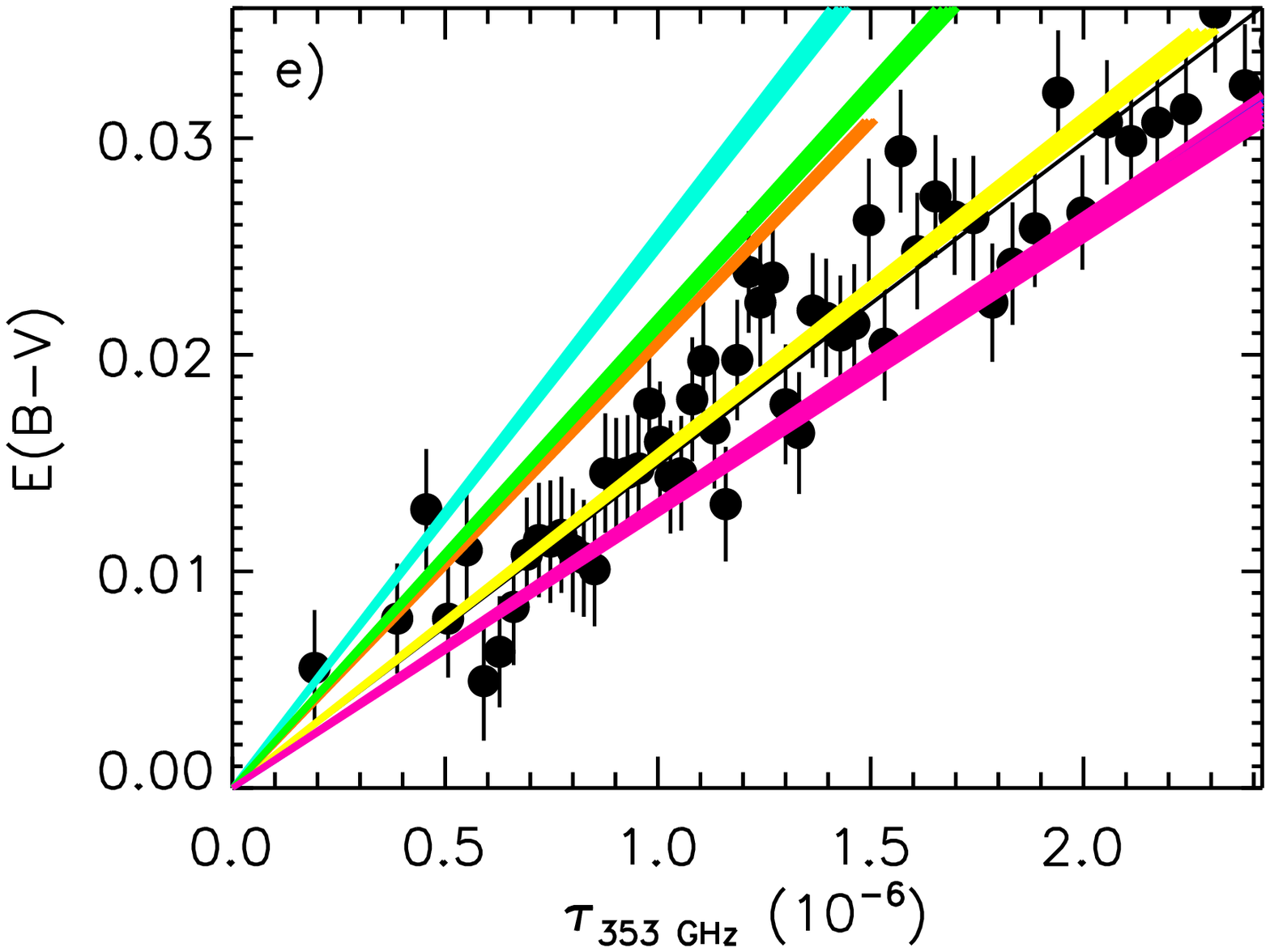} & \includegraphics[width=0.34\textwidth]{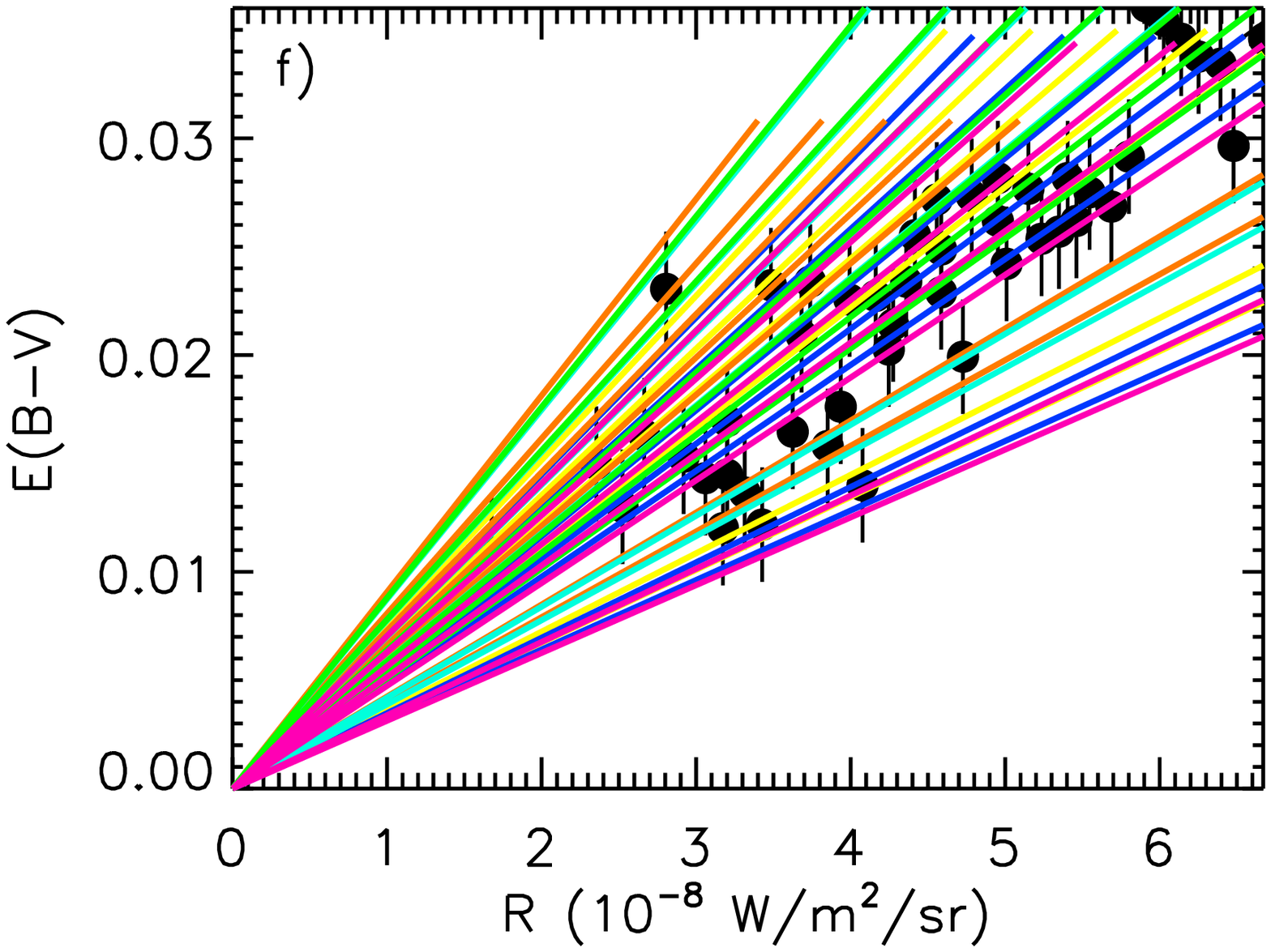}
\end{tabular}}
\caption{Influence of the aromatic-rich carbon mantle thickness on amorphous carbon grains (for silicate grains with a 5~nm thick mantle). Models with no mantle (orange), with mantle thicknesses of 5~nm (turquoise), 7.5~nm (green), 10~nm (yellow), 20~nm (blue), and 30~nm (pink) are displayed. In the panels where the orange lines are not visible, it is because they overlay with the turquoise lines. The radiation field is also varied with $0.8 \leqslant G_0 \leqslant 1.4$. See Sect. \ref{methodology} for description of the black and white lines and symbols.}
\label{Fig2} 
\end{figure*}

The results are presented in Fig.~\ref{Fig2}, where we assume 5~nm thick mantles for the silicate grains. For the carbonaceous grains, we consider six different thicknesses: 0, 5, 7.5, 10, 20, and 30~nm. For 7.5 to 30~nm thick mantles, the models match the observations within the $1\sigma$-dispersion around the median values. The 20 and 30~nm cases yield similar results for all the observables (blue and pink lines in Fig.~\ref{Fig2}), whereas the 7.5 and 10~nm cases exhibit significant differences (green and yellow lines in Fig.~\ref{Fig2}). For instance in the case of a 10~nm thick mantle, the spectral index is higher by $\sim 0.1$ (Fig.~\ref{Fig2}a, $\beta \sim 1.55$ or $\beta \sim 1.65$ for a 7.5~nm thick mantle), while the opacity at 353~GHz, $\tau_{353~{\rm GHz}}/N_{\rm H}$, is lower by $\sim 20$\% and almost perfectly matches the median measurements of PCXI for the optical depth and radiance to extinction ratios (Figs.~\ref{Fig2}c, d, e). As a consequence, the results presented in the following sections are obtained assuming an aromatic-rich carbon mantle thickness of 10~nm for the carbonaceous grains.

We also consider carbonaceous grains without mantles (orange lines in Fig.~\ref{Fig2}). These grains, which are made only of the amorphous aliphatic-rich carbon cores, are much warmer and their emission peaks at wavelengths shorter than 100~$\mu$m. When performing the modified black-body fits in the IRIS and Planck-HFI bands, the temperature and spectral index of the silicate population are are almost exclusively measured. This leads to higher values of $\beta \sim 1.85$ and lower values of $T \sim 17.5-20.5$~K and $\sigma_{\rm 353\; GHz} \sim 4.5 \times 10^{-27}$~cm$^2$/H, which still match the observed dispersion but are far from the most often measured values, which are the areas surrounded by yellow contours in the figures. For an intermediate mantle thickness of 5~nm on top of the carbon grains, we find similar model results but closer to the main bulk of the observational results (turquoise lines in Fig.~\ref{Fig2}, $\beta \sim 1.75$, $\sigma_{\rm 353\; GHz} \sim 5 \times 10^{-27}$~cm$^2$/H). This means that relatively thick aromatic-rich carbon mantles are definitely needed for the carbonaceous grain population to reproduce most of the observations.

\subsubsection{Mantle thickness on amorphous silicate grains}

\begin{figure*}[!th]
\centerline{
\begin{tabular}{ccc}
\includegraphics[width=6cm,height=4.8cm]{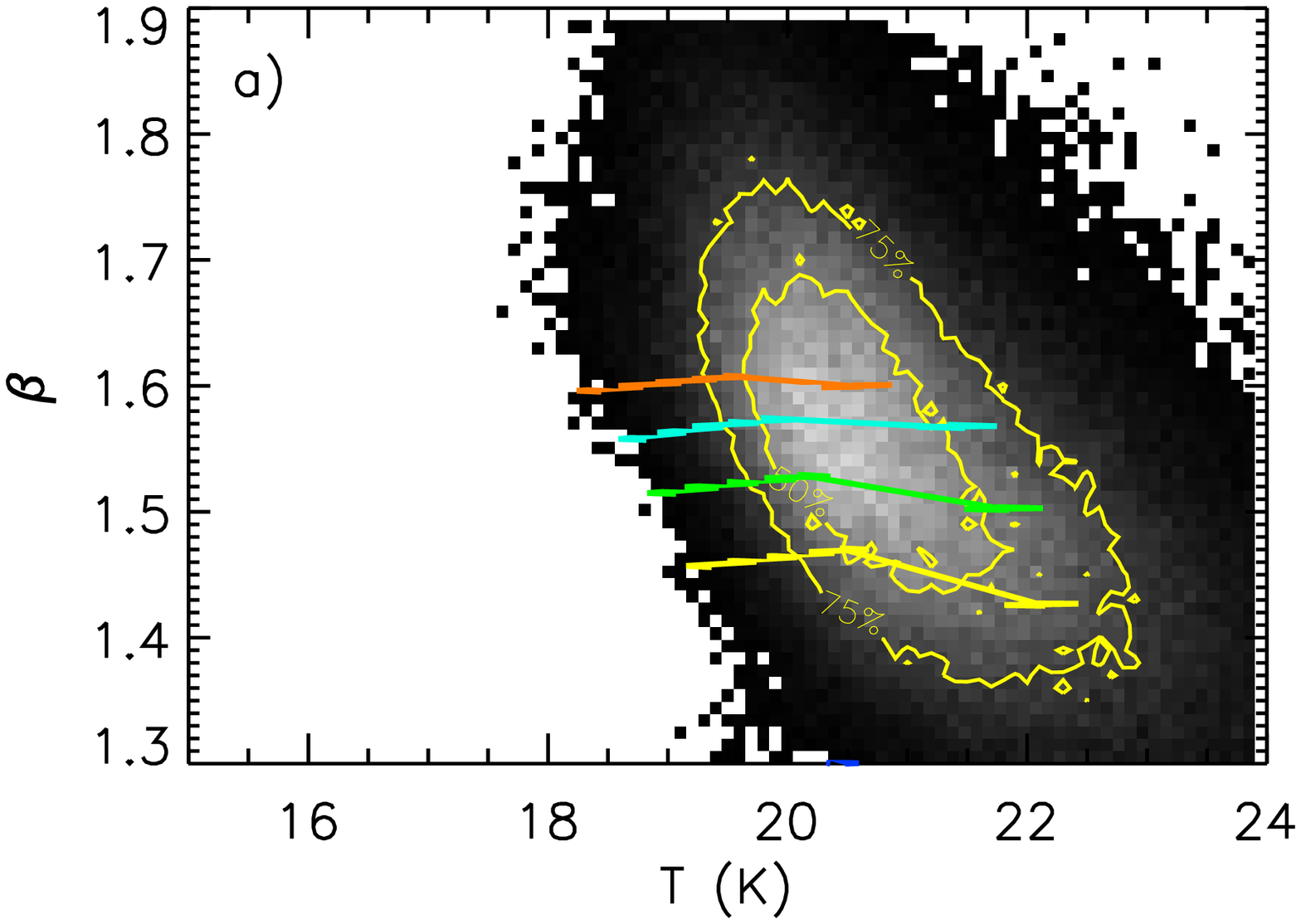} & \includegraphics[width=6cm,height=4.8cm]{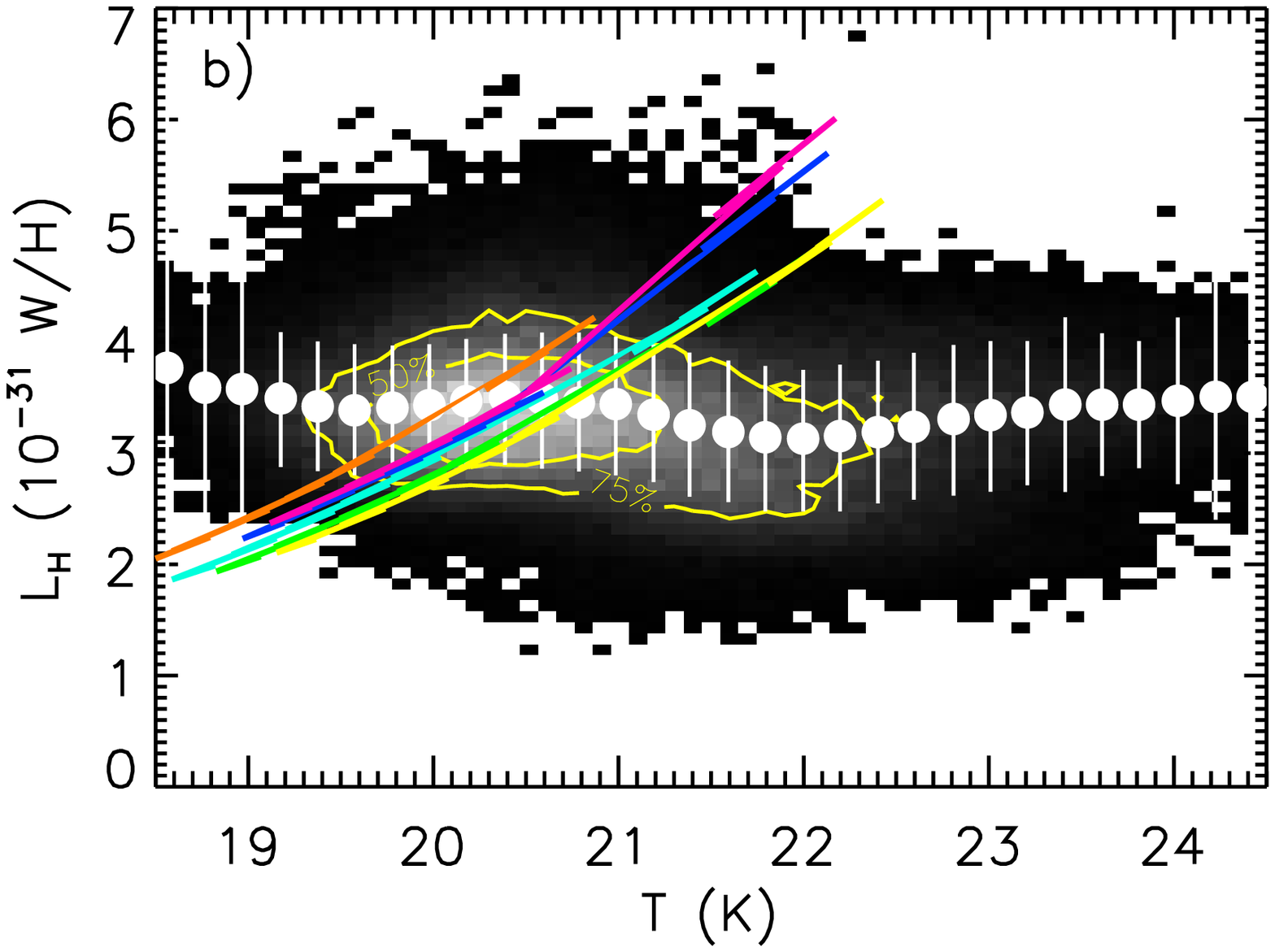}  & \includegraphics[width=6cm,height=4.8cm]{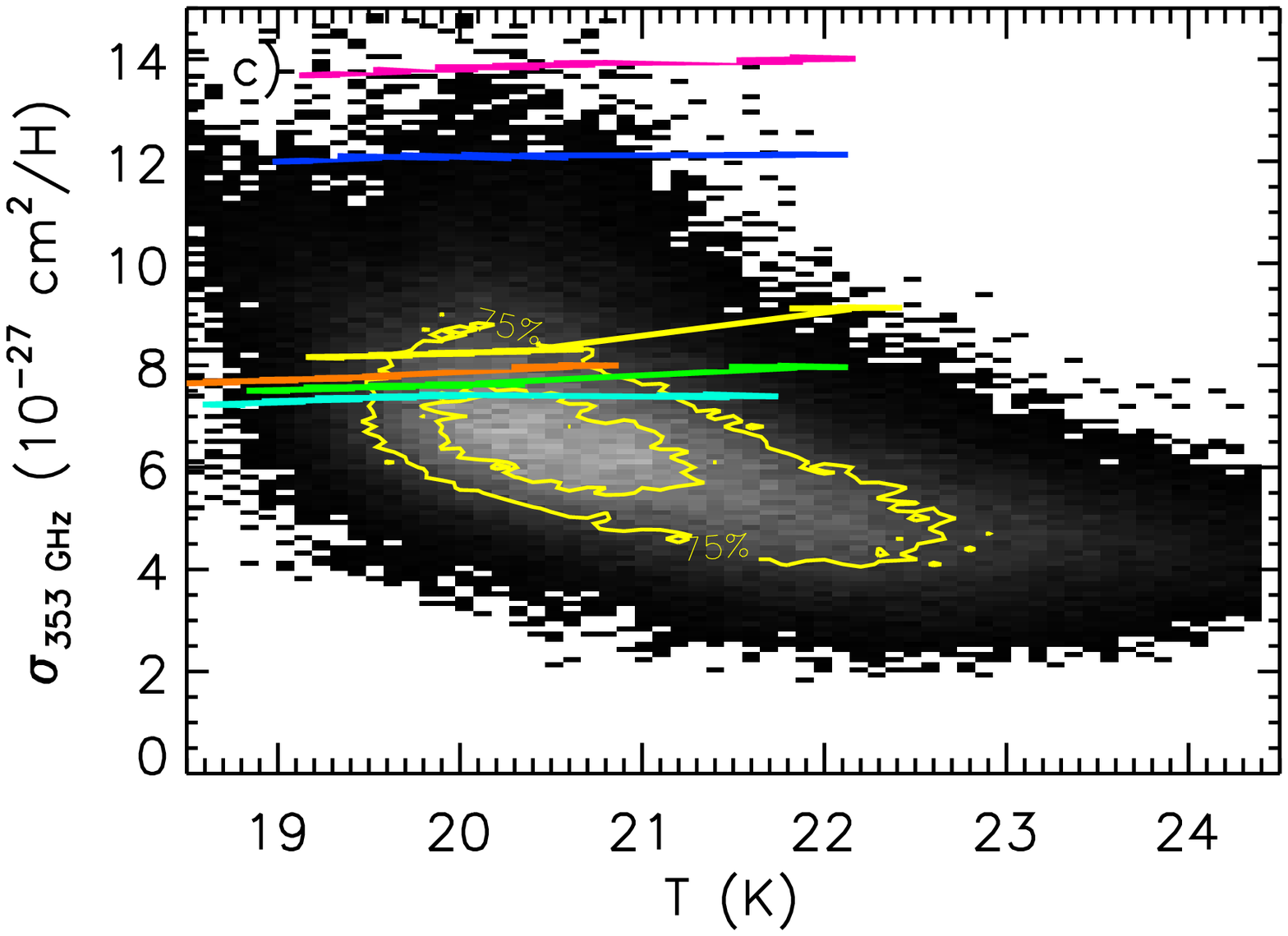} \\
\includegraphics[width=0.34\textwidth]{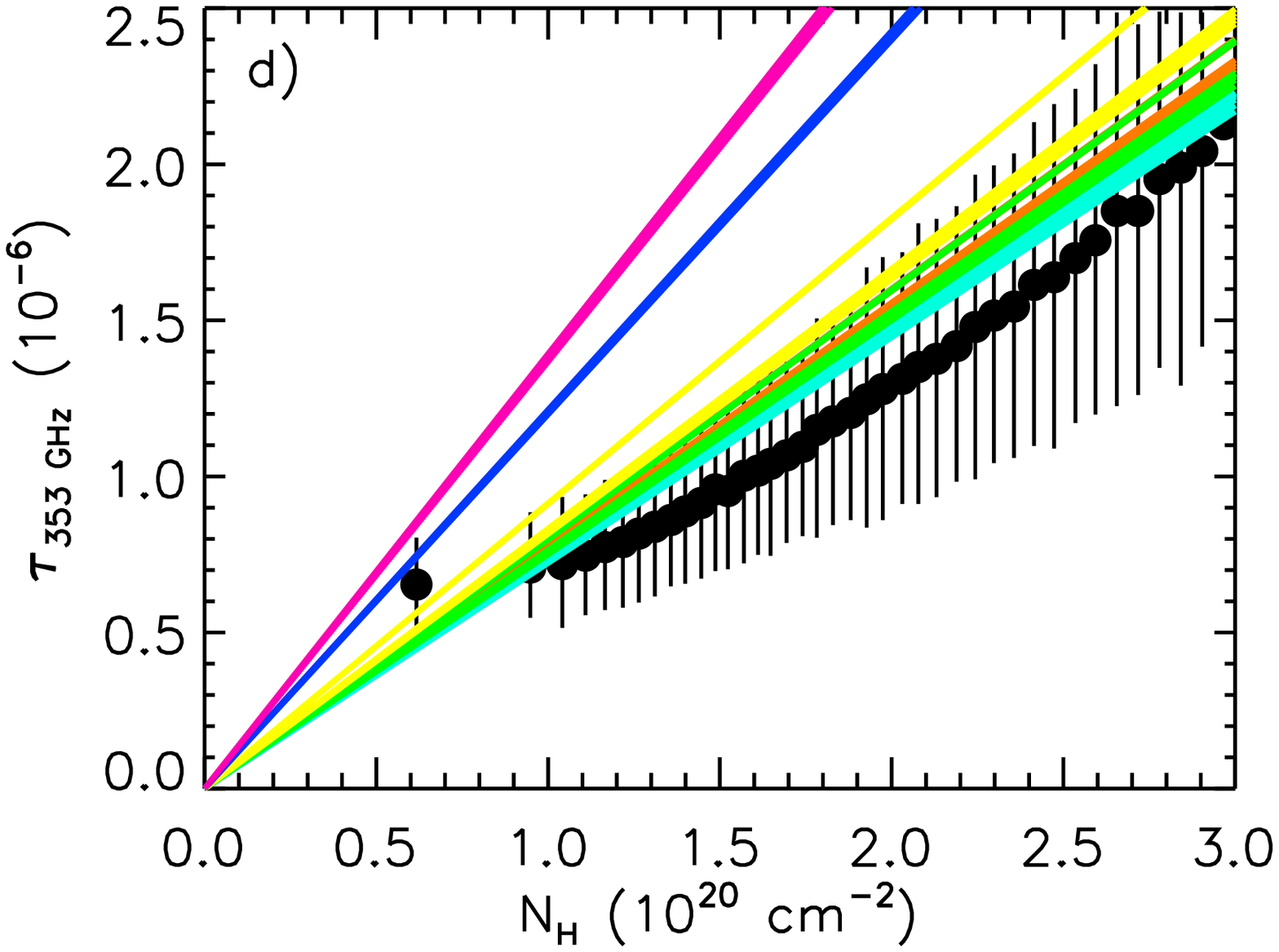} & \includegraphics[width=0.34\textwidth]{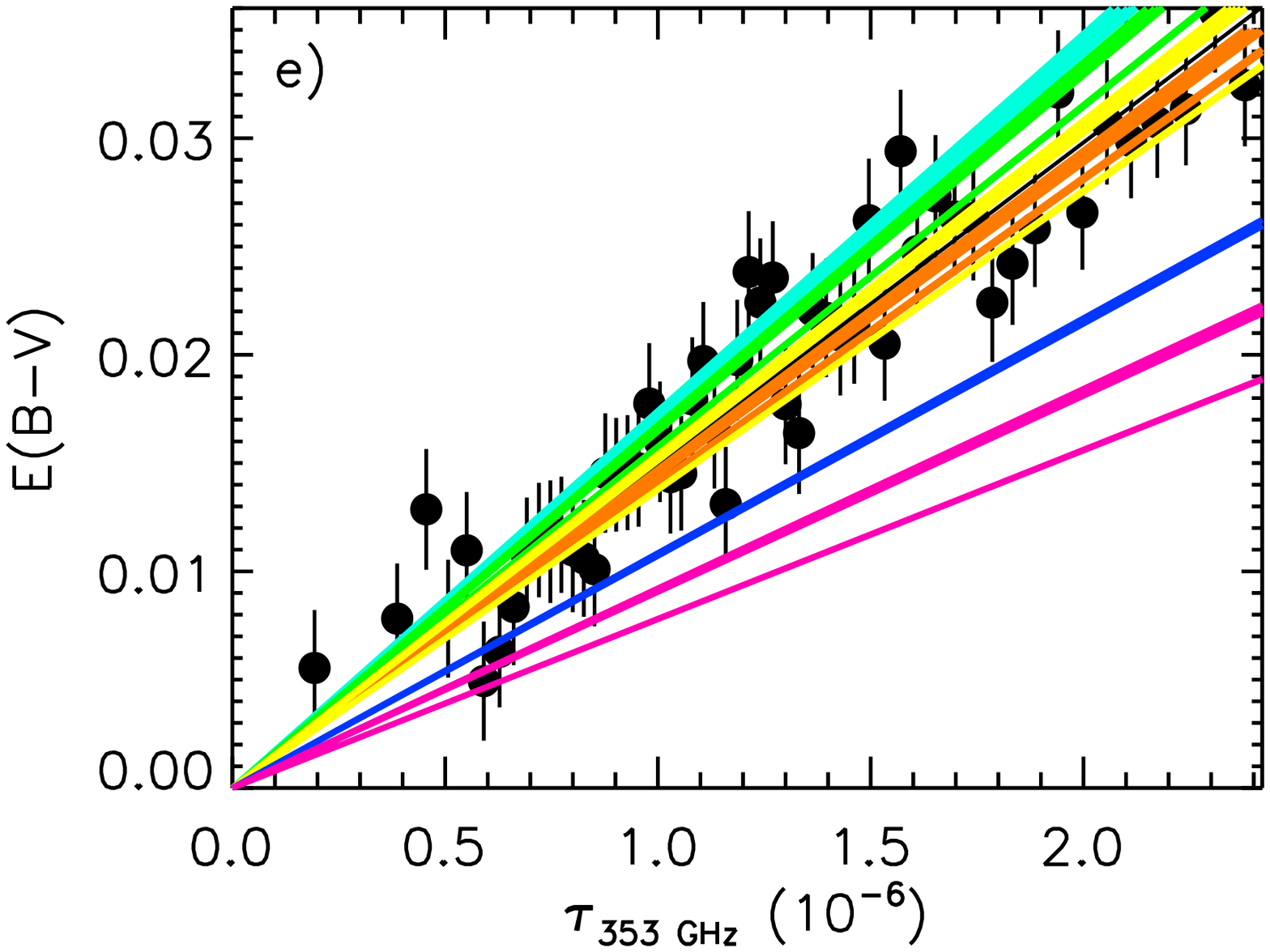} & \includegraphics[width=0.34\textwidth]{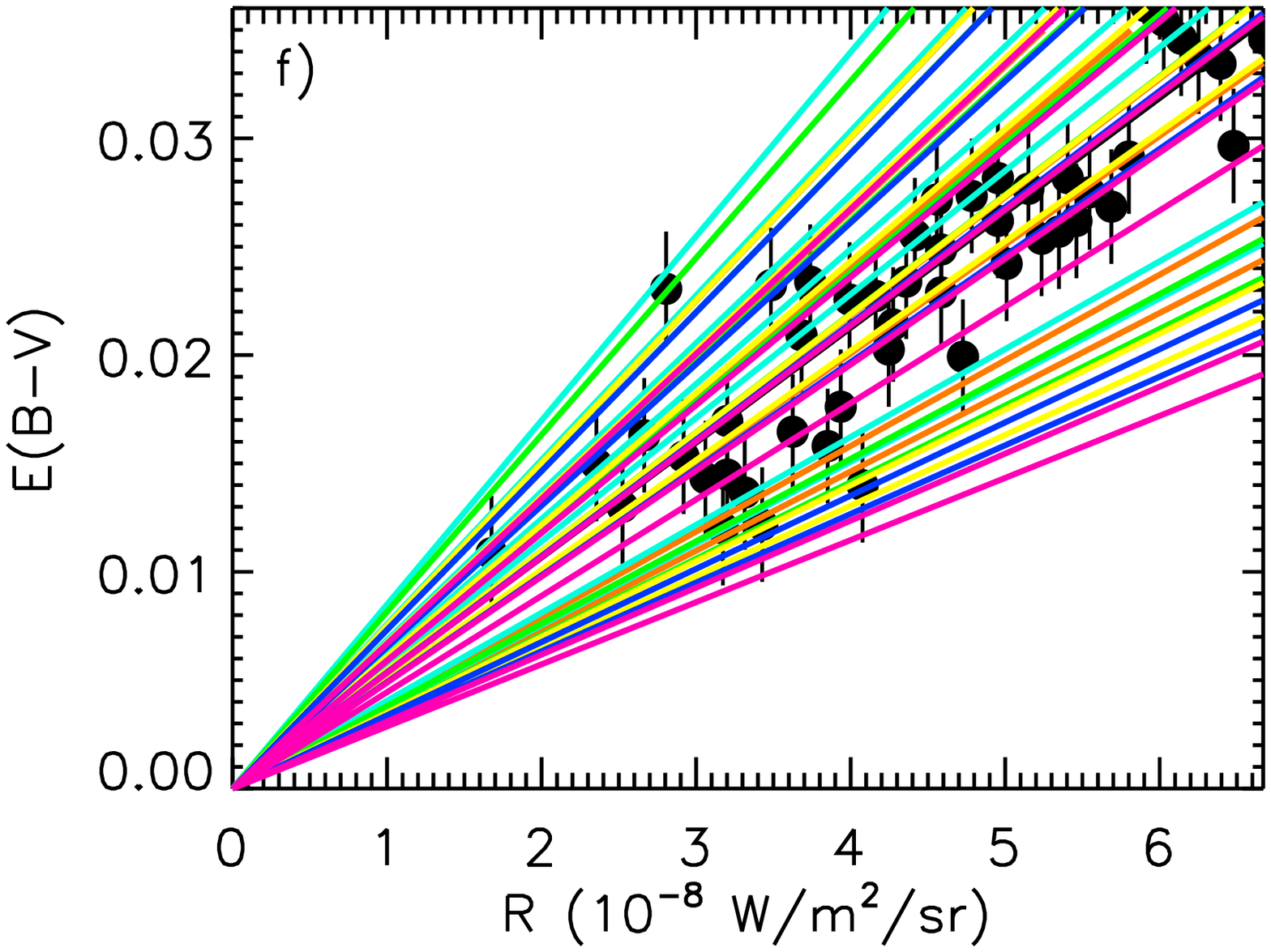}
\end{tabular}}
\caption{Influence of aromatic-rich carbon mantle thickness on amorphous silicate grains (for carbon grains with a 10~nm thick mantle). Models with no mantle (orange) and mantle thicknesses of 5~nm (turquoise), 7.5~nm (green), 10~nm (yellow), 15~nm (blue), and 20~nm (pink) are displayed. In Fig.~{\it a}, the blue and pink lines are out of the plot with $1.2 \leqslant \beta \leqslant 1.3$. The radiation field is also varied with $0.8 \leqslant G_0 \leqslant 1.4$. See Sect. \ref{methodology} for description of the black and white lines and symbols.}
\label{Fig3} 
\end{figure*}

The results are shown in Fig.~\ref{Fig3}, where we assume a 10~nm thick aromatic-rich carbon mantle for the carbonaceous grains. Six thicknesses are considered for the mantles of the silicate grains: 0, 5, 7.5, 10, 15, and 20~nm. A 20~nm thick mantle would mean that 111~ppm of the cosmic carbon is locked in the silicate grains, which means that our models never use more than a total of 280~ppm of carbon, in agreement with the measurements of cosmic elemental abundances \citep{Parvathi2012, Asplund2009, Sofia2001, Dwek1997}.

For thicknesses up to 10~nm, the models match the observations. Thicker mantles (15 and 20~nm) must be rejected as their optical depth at 353~GHz is too high by $\sim 60$ to 90\% (blue and pink lines in Figs.~\ref{Fig3}c, d) and their $\beta$-values are far from the main bulk of the observed dispersion with $1.2 \leqslant \beta \leqslant 1.3$. The extinction to submm optical depth ratios are also too low by factors $\sim 1.4$ and 1.6, for 15 and 20~nm thick mantles, respectively. These results hold when considering combinations with carbonaceous grains having 20 and 30~nm thick mantles. For the carbonaceous grains without mantle or with a thin 5~nm mantle, we can no longer reject the silicates with 15~nm thick mantles. This can be understood from Figs.~\ref{Fig2}c, d. Indeed, the effect of dramatically decreasing the thickness of the aromatic-rich layer on carbonaceous grains is to decrease the opacity at 353~GHz. This is enough to couterbalance the effect of the 15~nm thick mantles on top of the silicate grains. However, it seems physically unrealistic to assume that the mantles are removed from the carbonaceous grains, while, at the same time, increasing their thickness on the silicate grains.

\subsection{Composition of the metallic Fe/FeS inclusions}
\label{metallic_inclusions}

\begin{figure*}[!th]
\centerline{
\begin{tabular}{ccc}
\includegraphics[width=6cm,height=4.8cm]{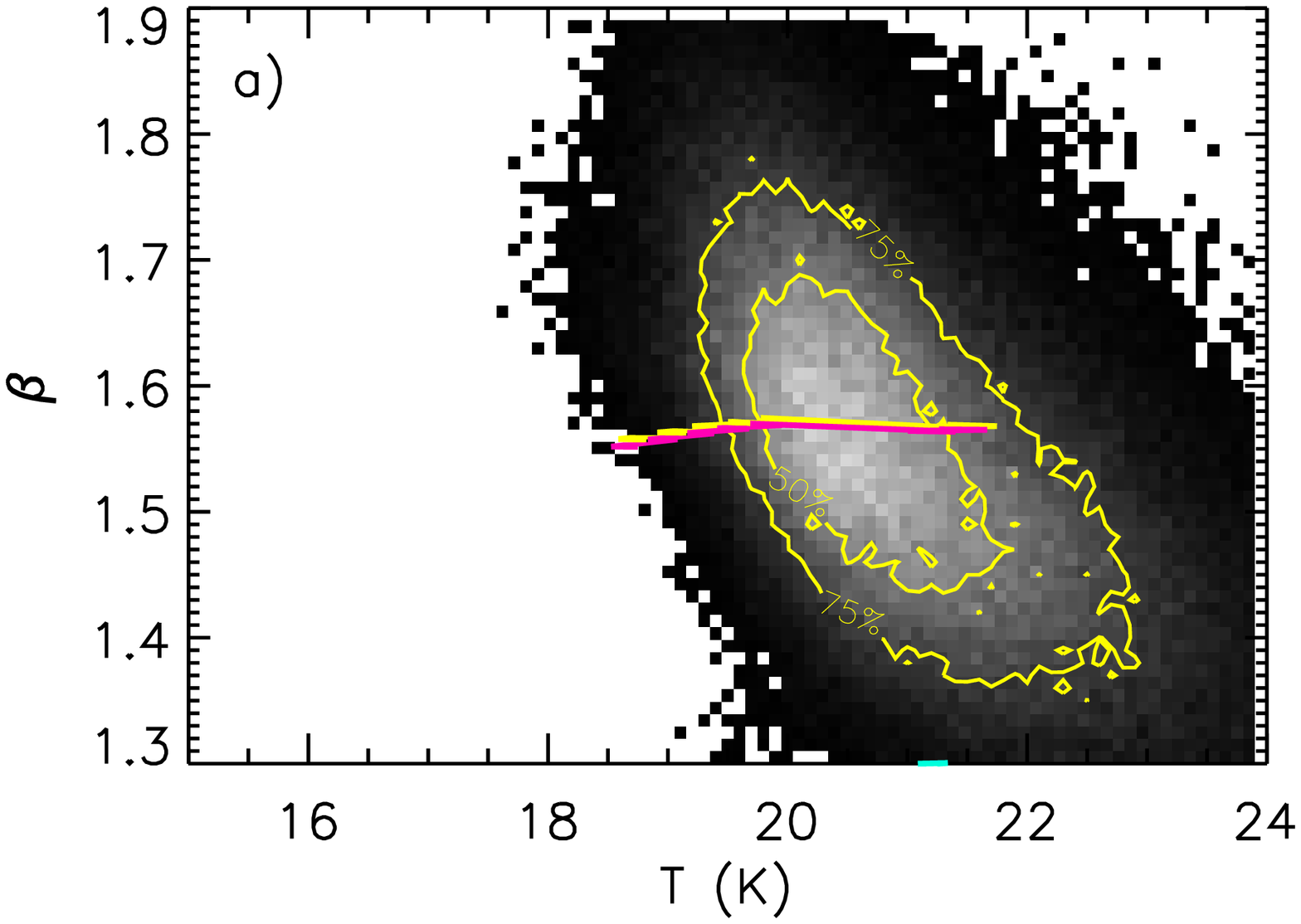} & \includegraphics[width=6cm,height=4.8cm]{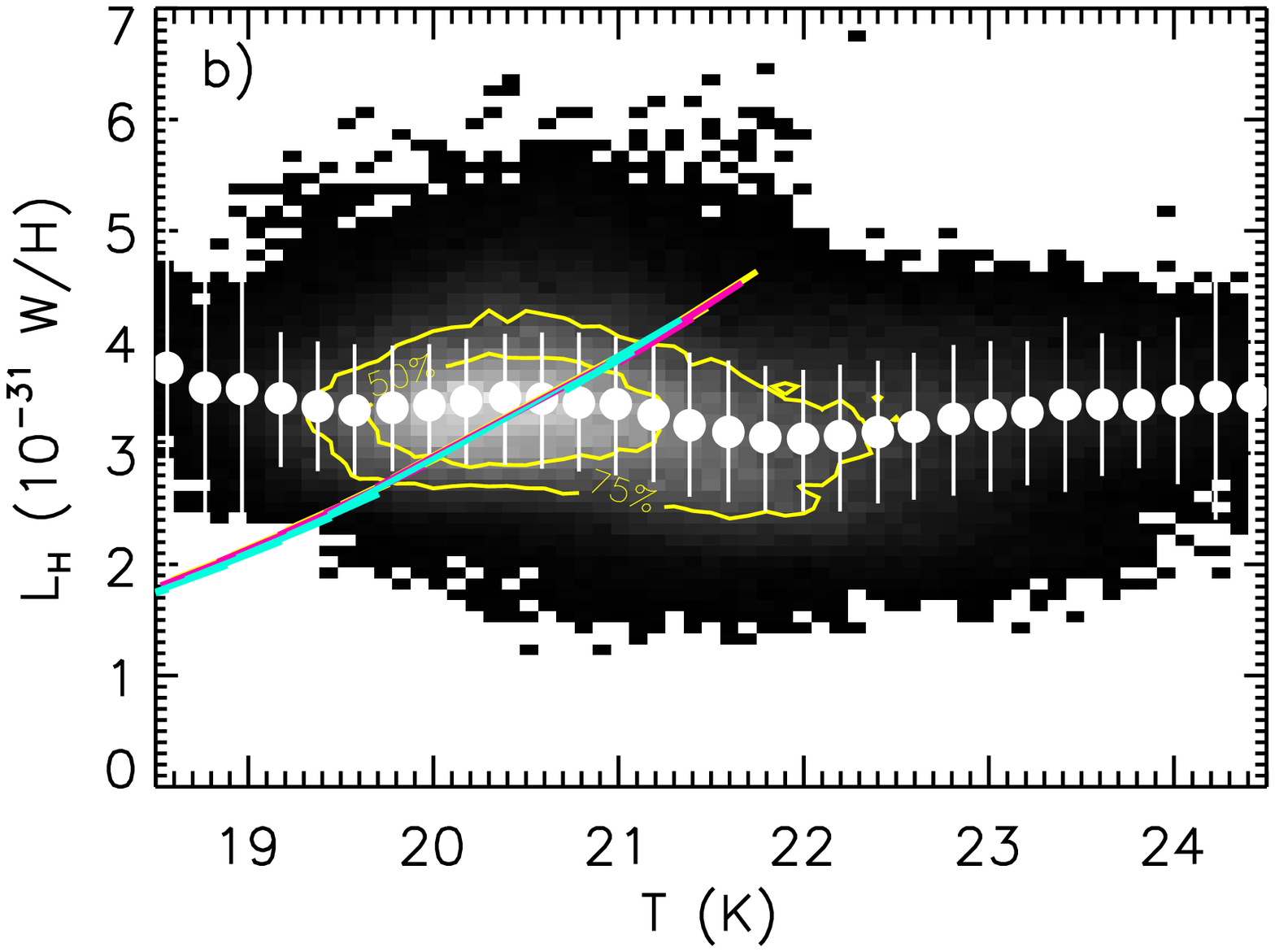}  & \includegraphics[width=6cm,height=4.8cm]{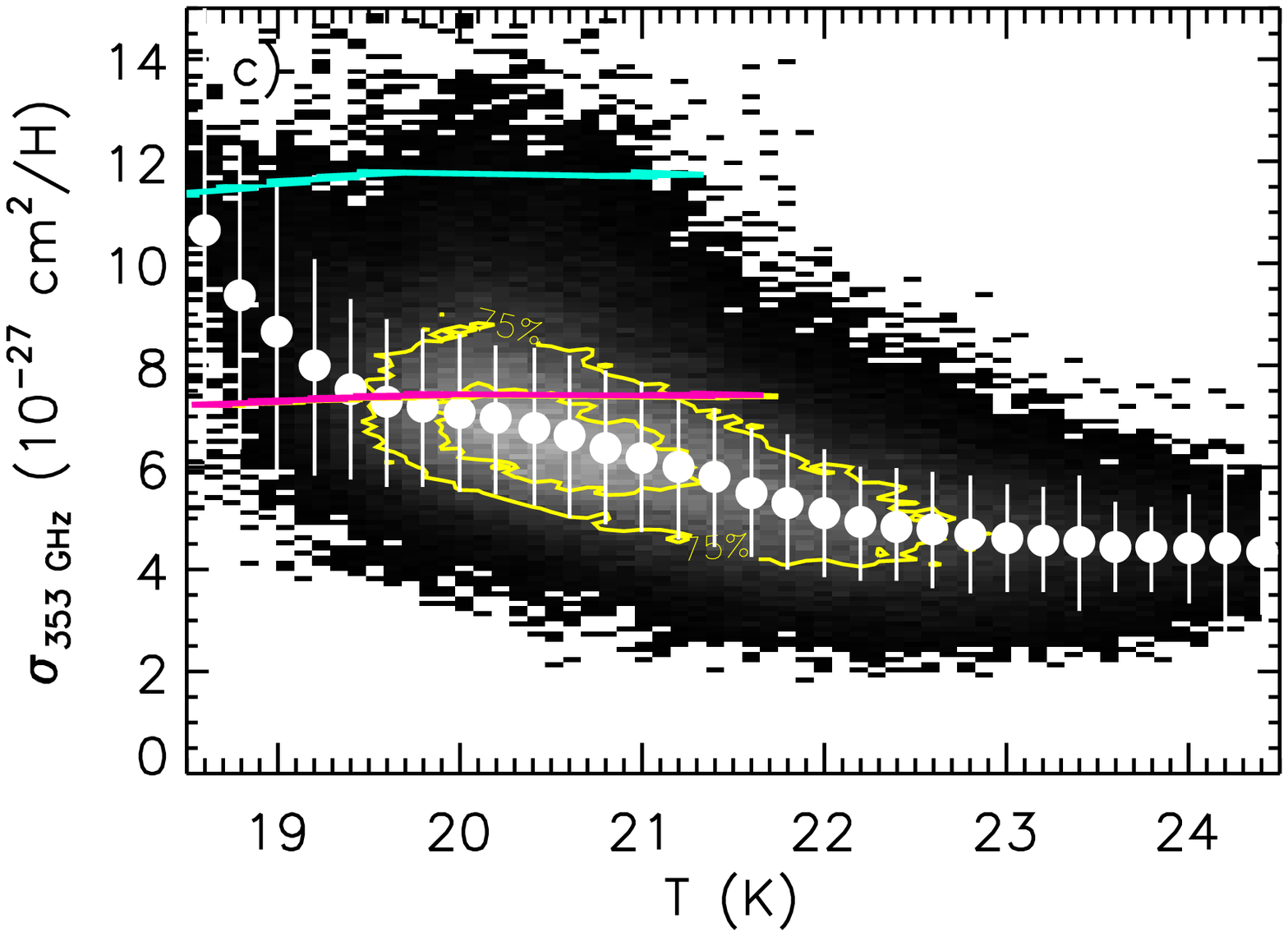} \\
\includegraphics[width=0.34\textwidth]{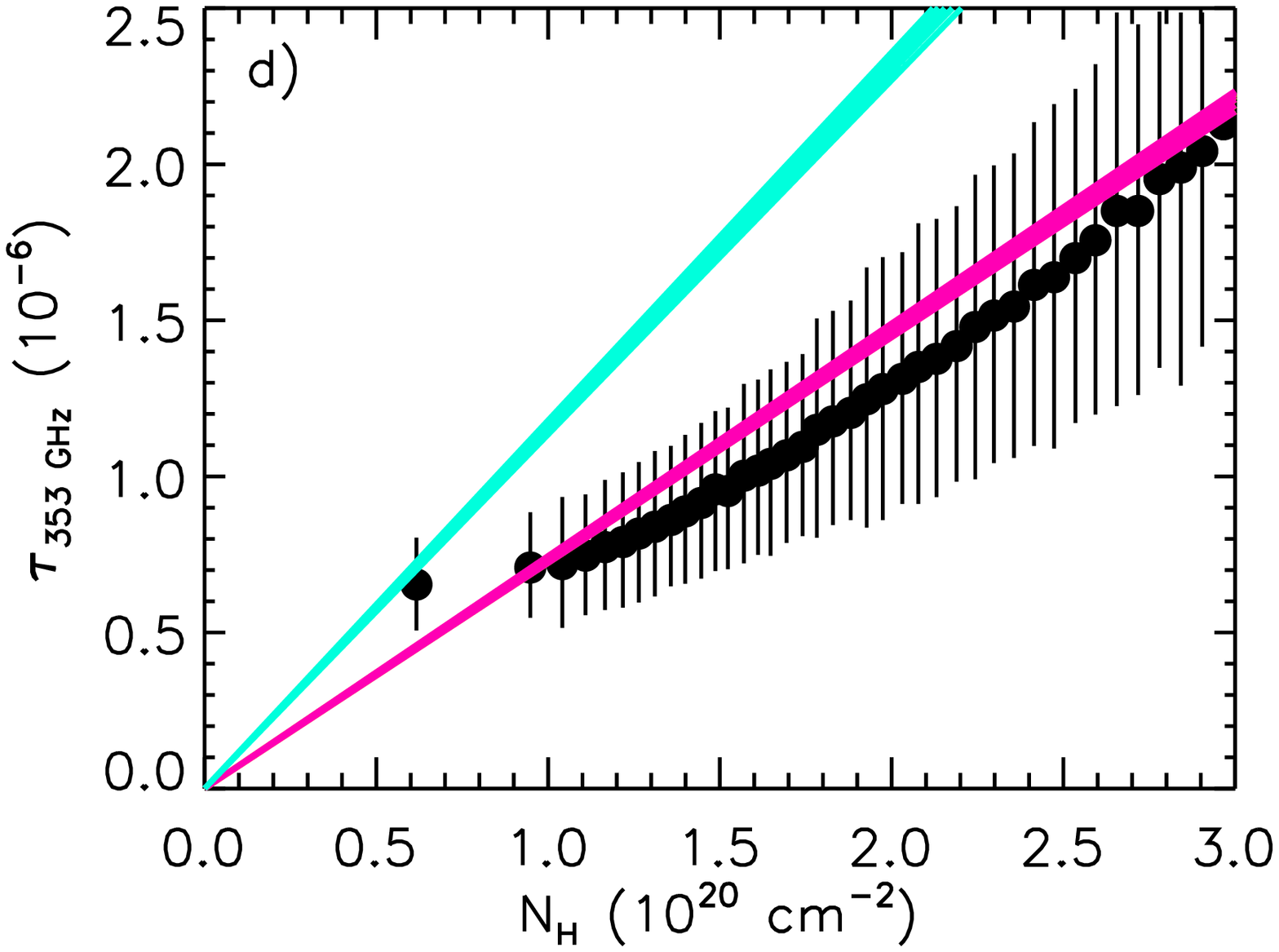} & \includegraphics[width=0.34\textwidth]{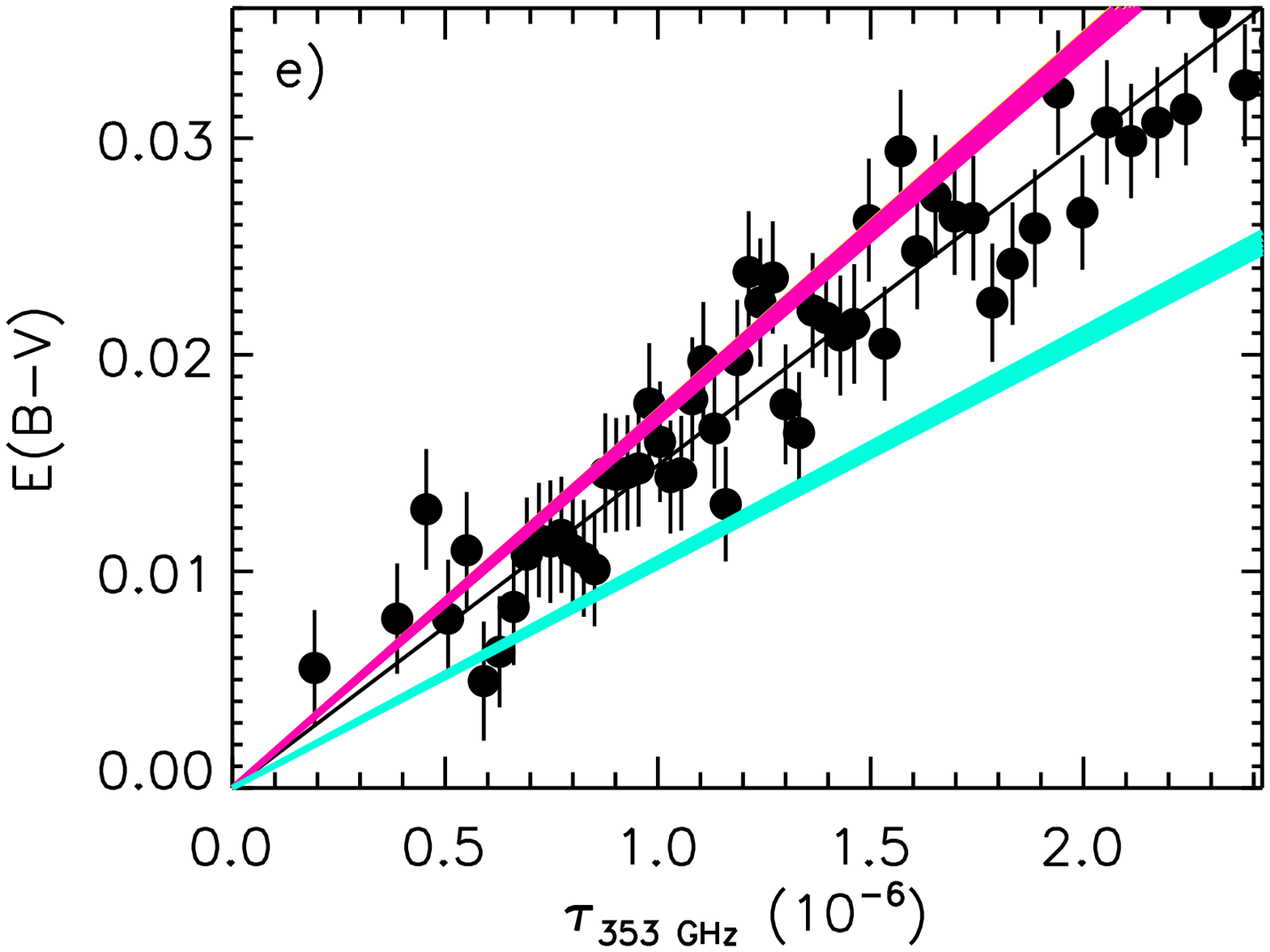} & \includegraphics[width=0.34\textwidth]{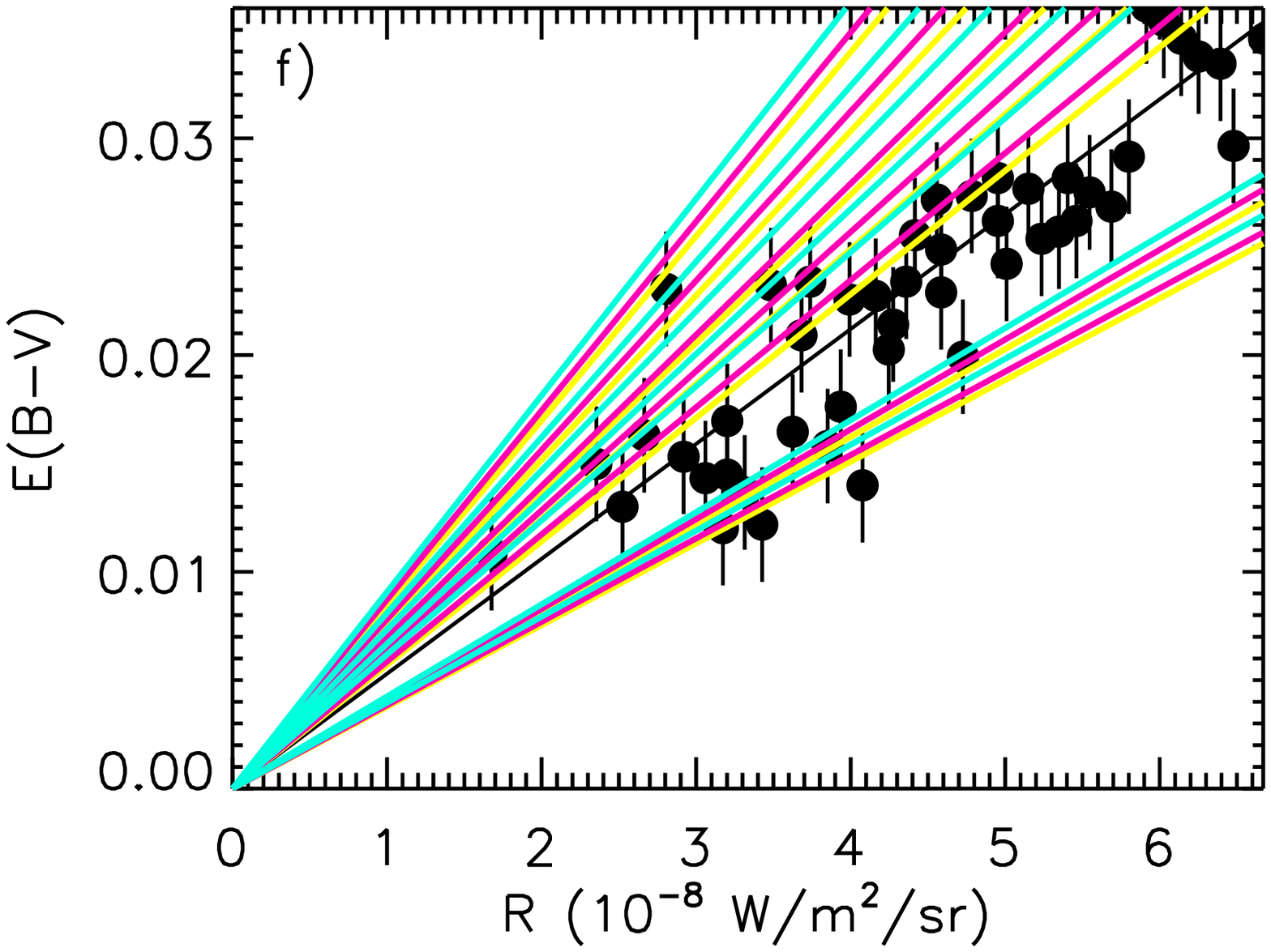}
\end{tabular}}
\caption{Influence of the composition of metallic iron inclusions (for carbonaceous grains with a 10~nm thick aromatic-rich carbon mantle and silicates with a 5~nm mantle). We show models, where 30\% (yellow), 50\% (pink), and 100\% (turquoise) of the iron inclusions are made of FeS. The first two models almost perfectly overlap in the figures (yellow and pink lines). In Fig.~{\it a}, the turquoise line is out of the plot with $\beta \sim 1.25$. The radiation field is also varied with $0.8 \leqslant G_0 \leqslant 1.4$. See Sect. \ref{methodology} for description of the black and white lines and symbols.}
\label{Fig4} 
\end{figure*}

The \citet{Jones2013} dust model includes metallic iron nano-inclusions inside the silicate grains. They are included in the form of metal making up 10\% of the grain volume and using $70\% \cong 25$~ppm of the cosmic iron. Subsequently, \citet{Koehler2014} showed that silicate grains can be a reservoir for interstellar sulphur in the form of Fe/FeS inclusions and that as much as 5~ppm of sulphur can be included without major changes in the dust emission and extinction (i.e. 30\% of the volume of the metallic inclusions is FeS). However, the amount of sulphur included in the grains is still an open question and, for that reason, we investigate the influence of FeS inclusions on the dust observables. We consider three different volume fractions of FeS in Fe inclusions: 30\%, 50\%, or 100\%. This corresponds to 25~ppm of Fe and 5~ppm of S, 23~ppm of Fe and 9~ppm of S, and 18~ppm of Fe and 18~ppm of S, respectively.

The results are shown in Fig.~\ref{Fig4}. The two first cases with 30 and 50\% of the inclusion volume filled with FeS do not show any difference (pink and yellow lines, which overlap in Fig.~\ref{Fig4}). However, the case, where 100\% of the inclusions are FeS, must be excluded as its spectral index is off by $\sim 0.1$ to 0.2 with $\beta \sim 1.25$, its optical depth at 353~GHz is too high by $\sim 60$\% for $N_{\rm H} > 10^{20}$~H/cm$^2$ (turquoise line in Fig.~\ref{Fig4}d), and its extinction to optical depth ratio is also too low by a factor $\sim 1.5$ (Fig.~\ref{Fig4}e). We conclude that a variable composition of the metallic nano-inclusions cannot account for the variations in the dust observables in the diffuse ISM. Conversely, the far-IR/submm observations do not provide a strong constraint on the FeS content (in nano-inclusions) in interstellar dust.

\subsection{Carbon abundance}
\label{carbon_abundance}

\begin{figure*}[!th]
\centerline{
\begin{tabular}{ccc}
\includegraphics[width=6cm,height=4.8cm]{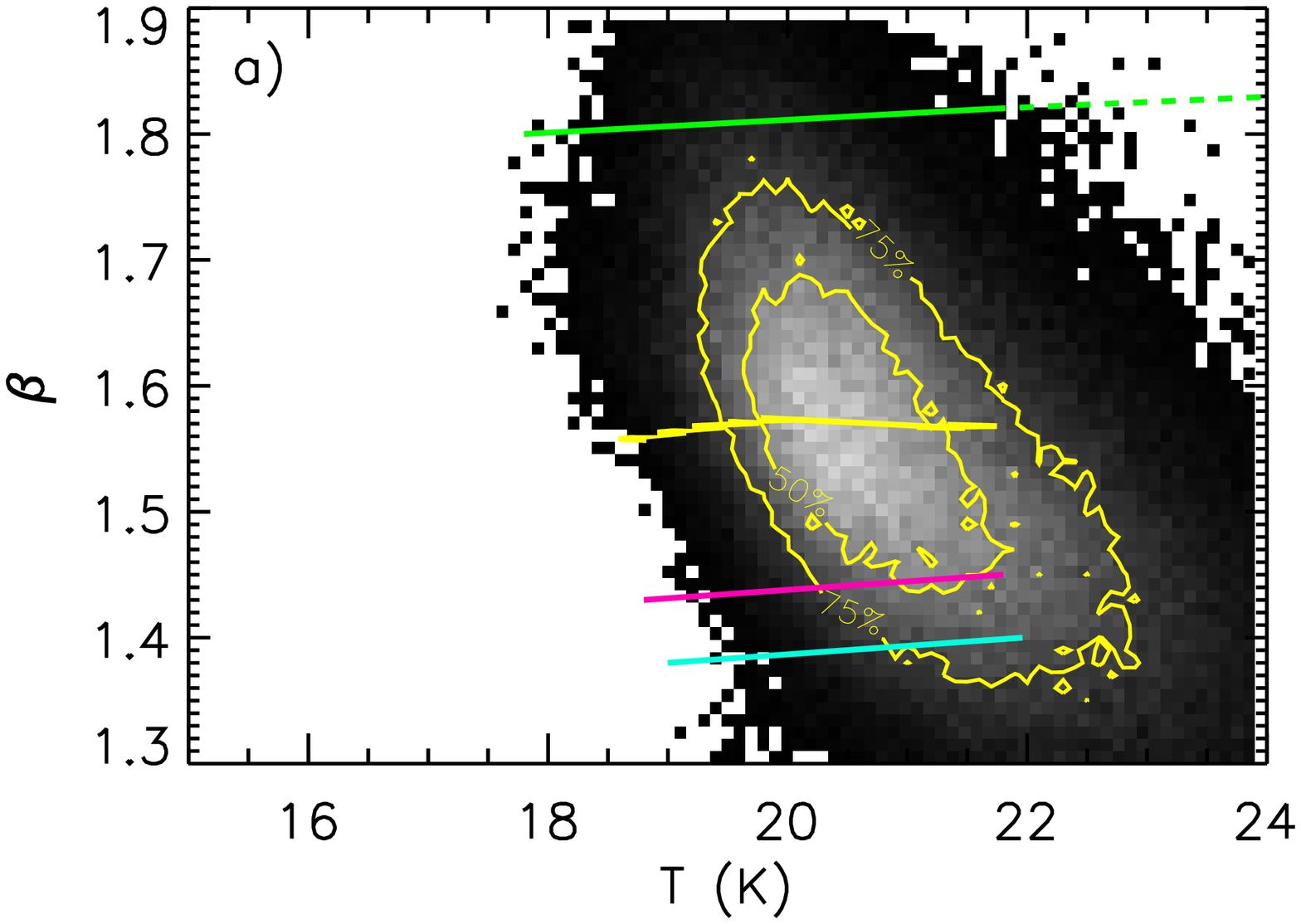} & \includegraphics[width=6cm,height=4.8cm]{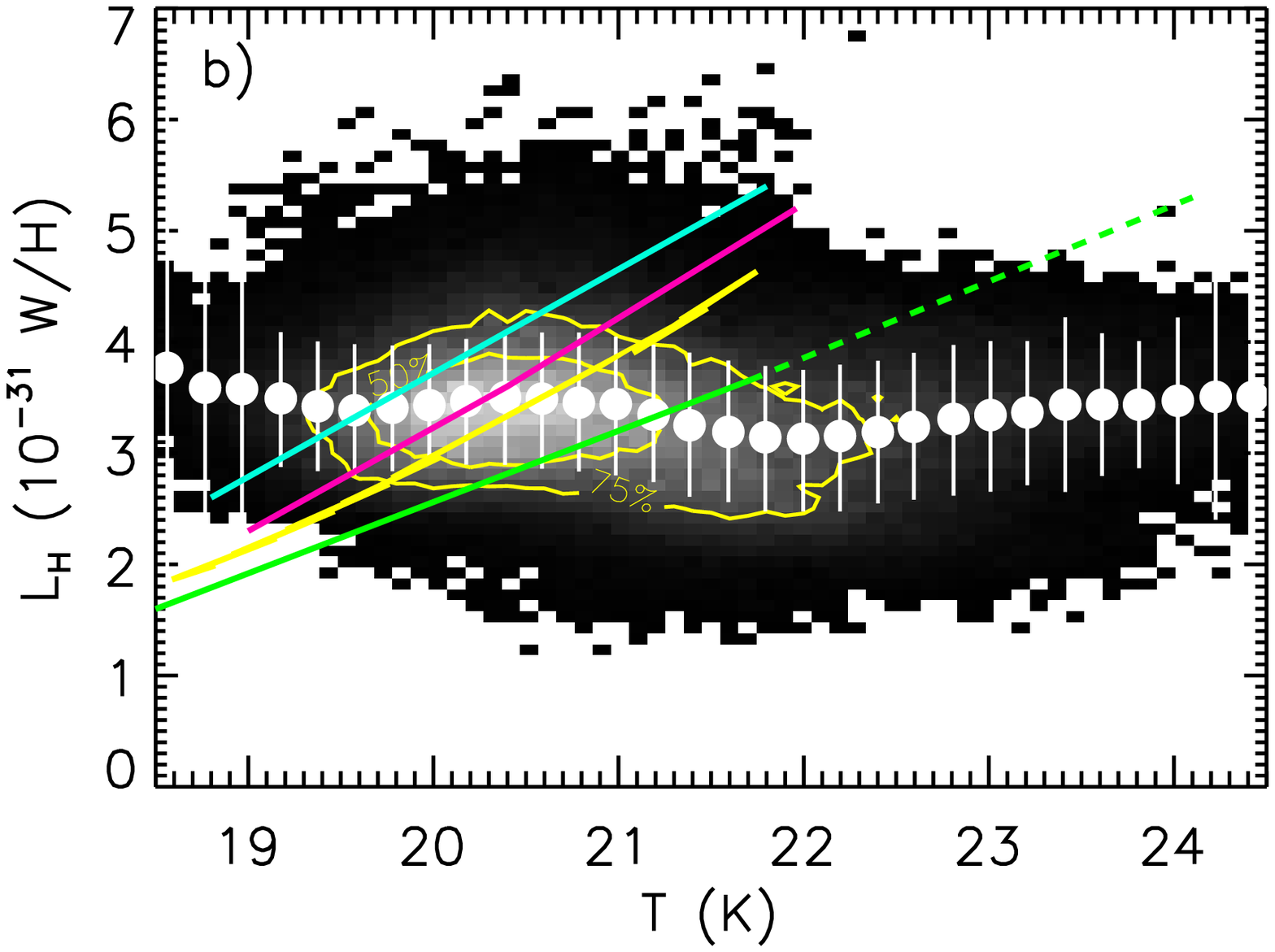}  & \includegraphics[width=6cm,height=4.8cm]{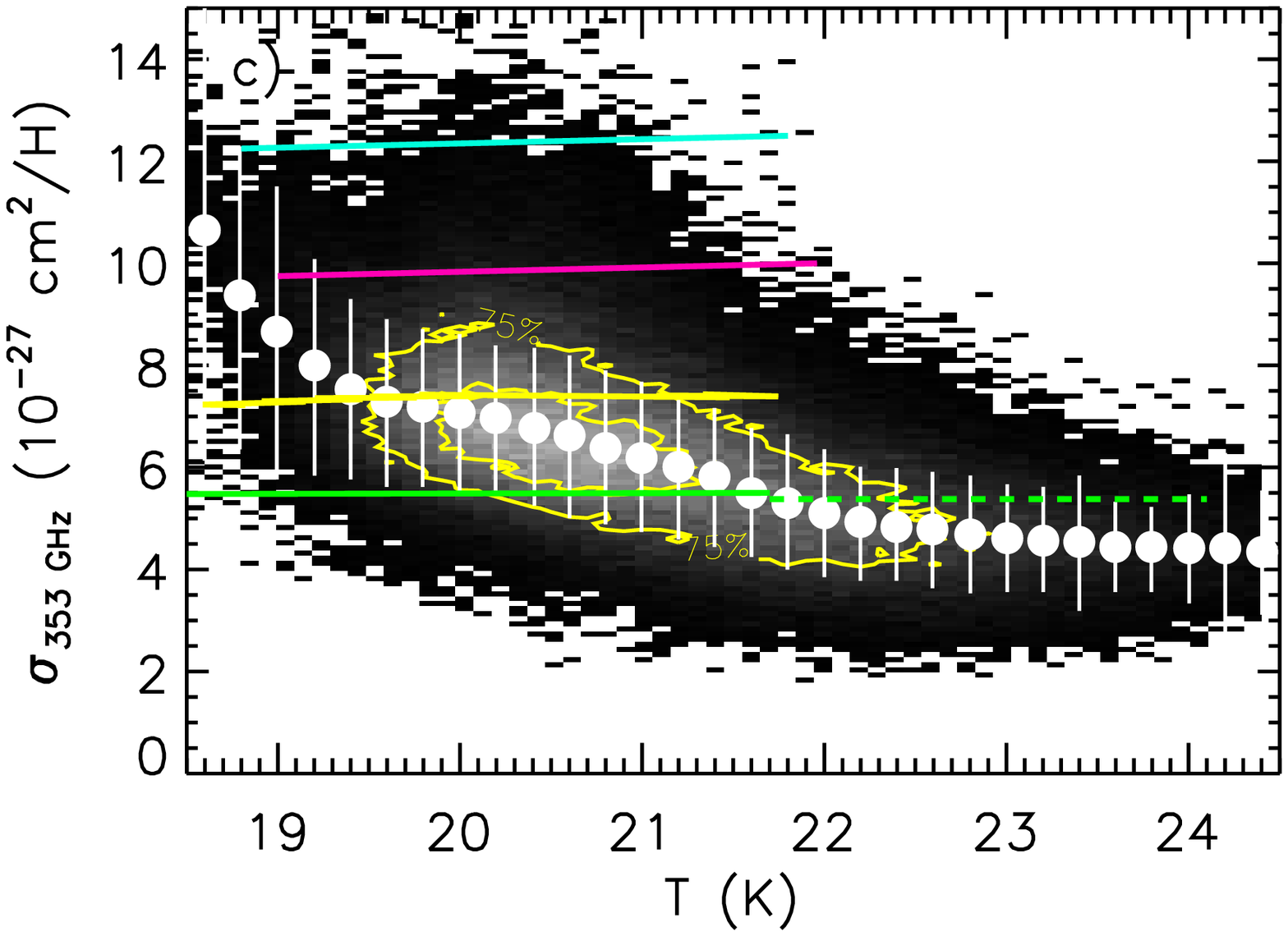} \\
\includegraphics[width=0.34\textwidth]{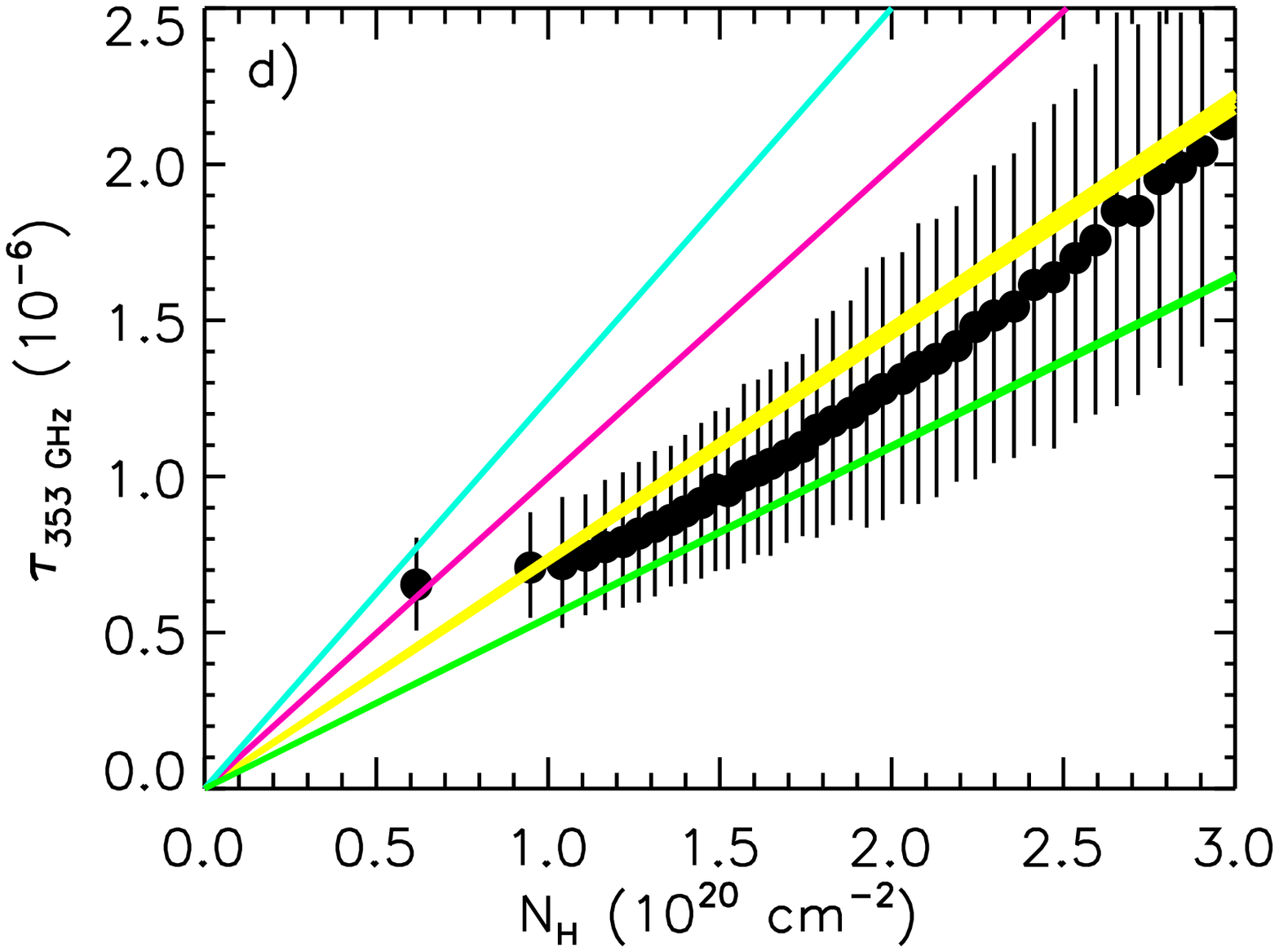} & \includegraphics[width=0.34\textwidth]{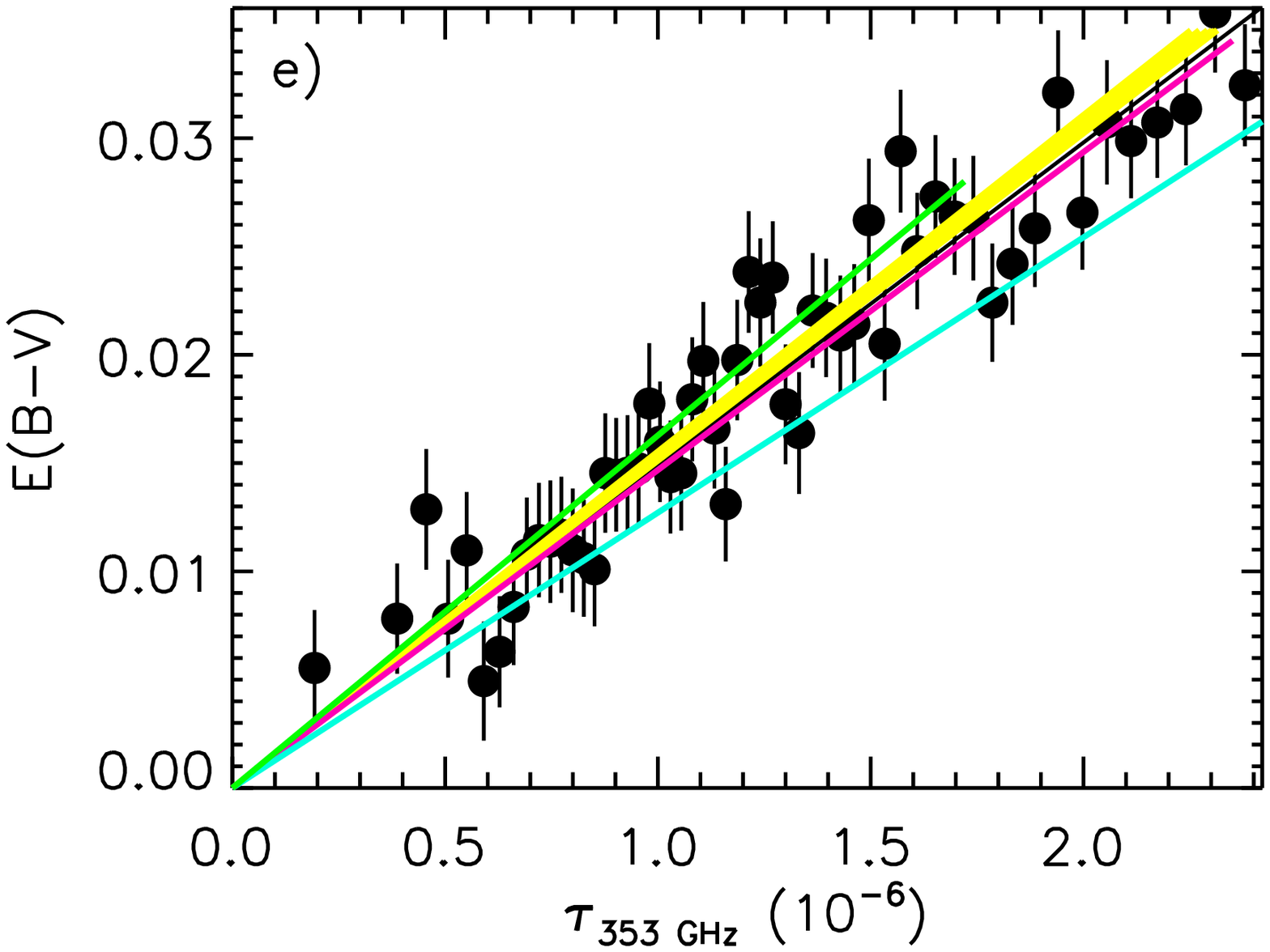} & \includegraphics[width=0.34\textwidth]{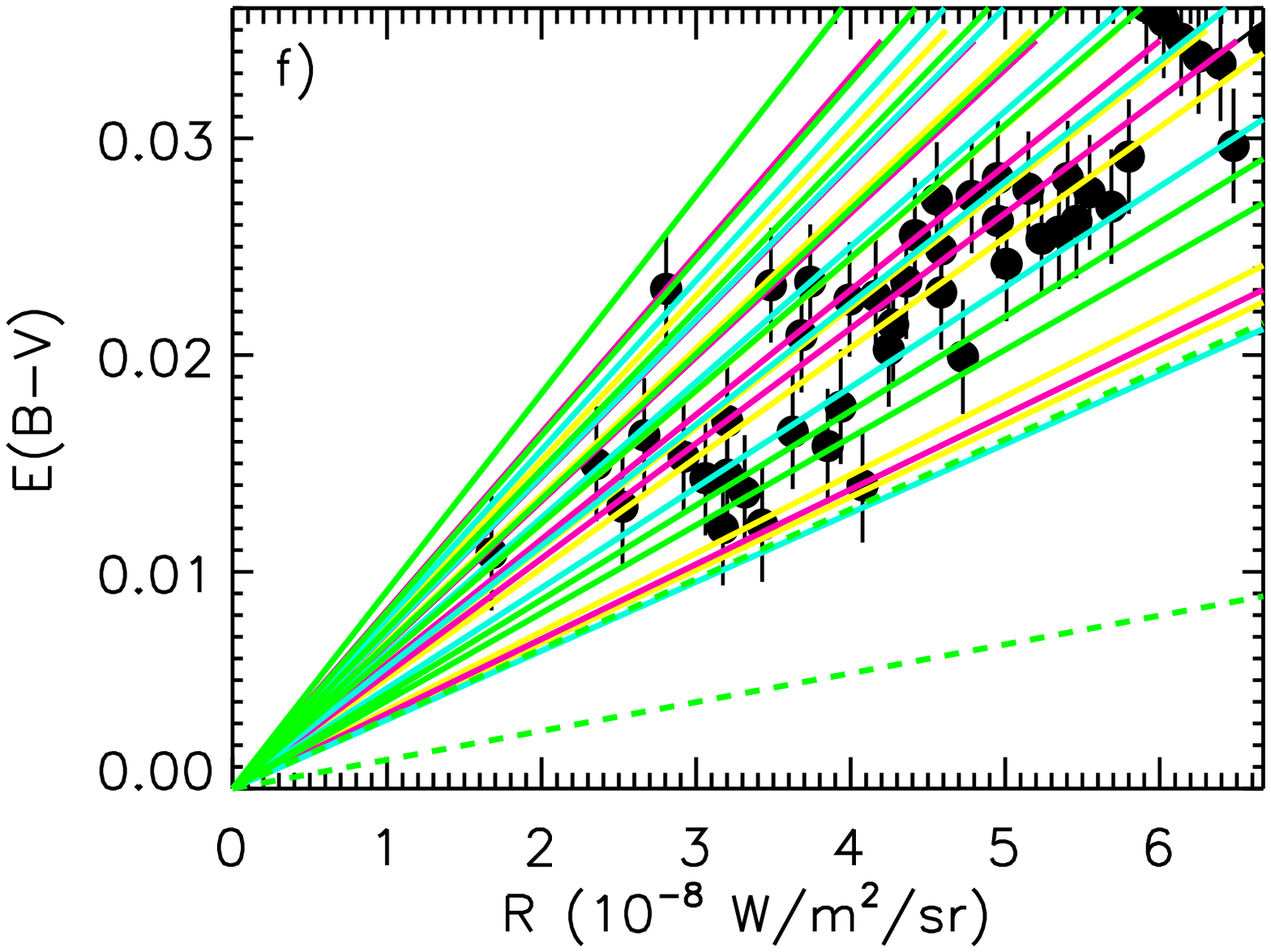}
\end{tabular}}
\caption{Influence of the carbon abundance (for carbonaceous grains with a 10 nm thick aromatic mantle and silicates with a 5 nm aromatic mantle). The solid yellow lines show the model with the standard carbon abundance of \citet{Jones2013}, the solid pink lines with an extra 60 ppm in both the small and big carbonaceous grains, and the solid turquoise lines with an extra 120 ppm in the big carbonaceous grains only. The solid green lines show the results for the case, where only 12~ppm of carbon are included in the big carbonaceous grains (see Sect. \ref{carbon_abundance} for details). The radiation field is varied with $0.8 \leqslant G_0 \leqslant 1.4$. The dashed green lines present the results for the same carbon abundance as the solid green lines but for $1.4 < G_0 \leqslant 5$. See Sect. \ref{methodology} for description of the black and white lines and symbols.}
\label{Fig5} 
\end{figure*}

In the \citet{Jones2013} dust model, about 200~ppm of the cosmic carbon is in the grains. The observational estimate of the carbon abundance inside the grains varies depending on the observed line of sight. For instance, \citet{Parvathi2012} found that as much as $395 \pm 61$~ppm of the cosmic carbon is in dust in the direction of HD~207198. Considering this, we explore the effect of adding more carbon in our dust model by making two different assessments: first, we add 120~ppm of carbon in the big carbonaceous grains ($M_d / M_{\rm H} = 1.98 \times 10^{-3}$) ; and second, we add 60~ppm in both big ($M_d / M_{\rm H} = 1.26 \times 10^{-3}$) and small carbonaceous grains ($M_d / M_{\rm H} = 2.13 \times 10^{-3}$). \citet{Parvathi2012} also found lines of sight, where the carbon abundance in the dust is very low. These observations could be explained by local variations in the cosmic carbon abundance or by the destruction of carbonaceous grains. Indeed, \citet{Bocchio2014} showed that in supernova shock waves carbon dust is easily destroyed, leading to the complete destruction of the smallest grains and to a decrease in the amount of big carbon grains. They further show that silicate grains are more resilient to shocks than carbon grains. Considering this, we explore the effect of including less carbon in the grains. We thus remove all the small carbonaceous grains and assume that only 12~ppm of carbon are inside the big carbonaceous grains ($M_d / M_{\rm H} = 0.14 \times 10^{-3}$).

The results are presented in Fig.~\ref{Fig5} for carbon and silicate grains with 10 and 5~nm thick aromatic-rich carbon mantles, respectively. For an increase in the carbon abundance in the grains, both models match only the optical depth deduced for the lines of sight with the lowest column densities ($N_{\rm H} \sim 6 \times 10^{19}$ to $10^{20}$~H/cm$^2$) but otherwise deviate from the data points (Fig.~\ref{Fig5}d). Increasing the abundance of carbon in the grains may thus explain part of the variations measured towards the most diffuse areas of the ISM but cannot be the main explanation for the entire range of variations. The results for a decrease in the carbon abundance in dust are represented by the green lines in Fig. \ref{Fig5}. For that particular case, we considered radiation field intensities up to $G_0 = 5$. Indeed, if we assume that the carbon grains are destroyed by supernova shock waves \citep{Bocchio2014}, the local radiation field is likely to be higher than in the standard diffuse ISM. We find that decreasing the amount of big carbon grains lead to a higher $\beta \sim 1.8$ and a lower optical depth $\sigma_{353 \; {\rm GHz}}\sim 5.5 \times 10^{-27}$~cm$^2$/H, while the luminosity is decreased by $\sim 10$ to 15\%. This is thus a likely scenario to explain part of the observed variations in the dust properties.

\subsection{Grain size distribution}
\label{size_distribution}

\begin{figure*}[!th]
\centerline{
\begin{tabular}{ccc}
\includegraphics[width=6cm,height=4.8cm]{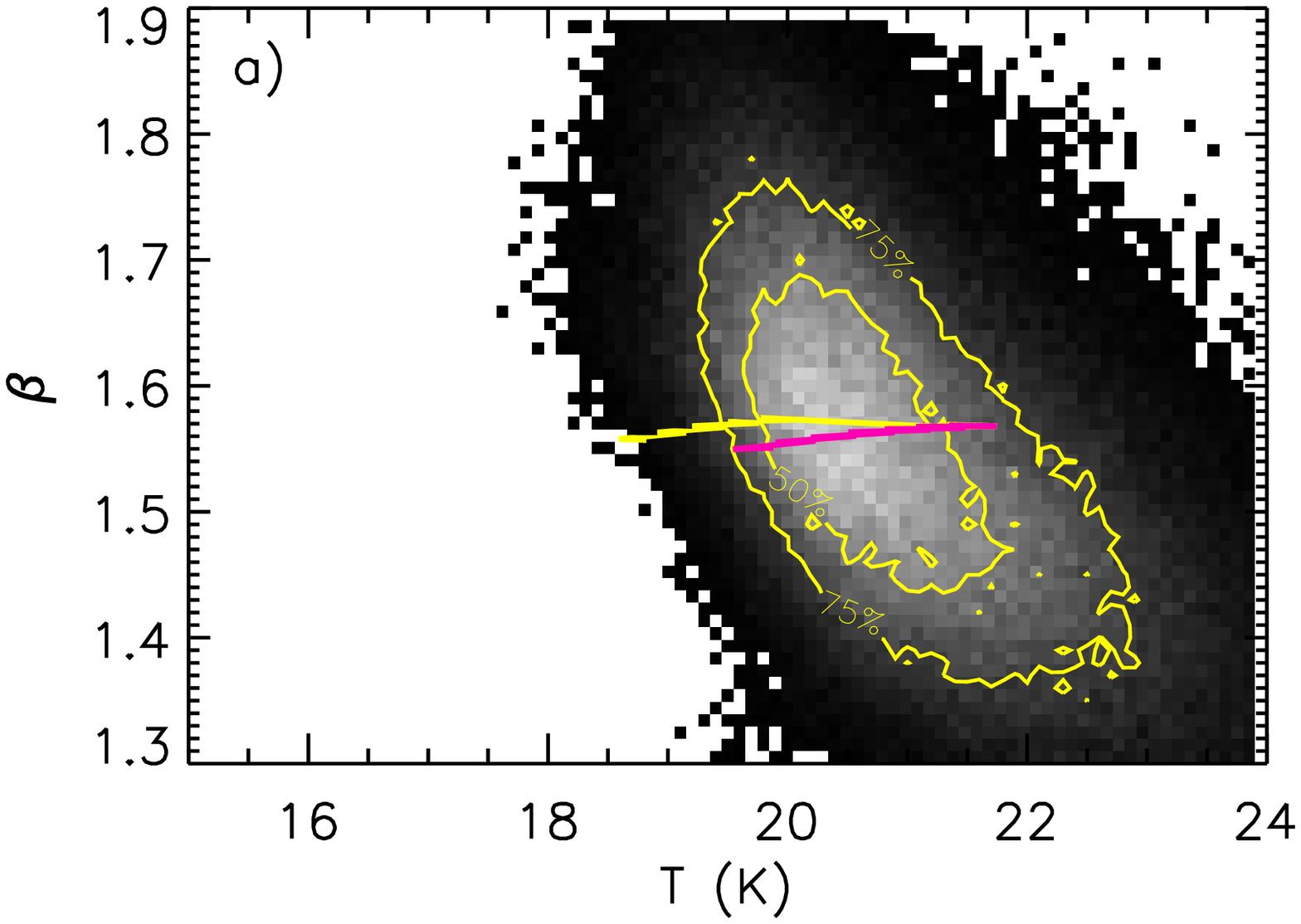} & \includegraphics[width=6cm,height=4.8cm]{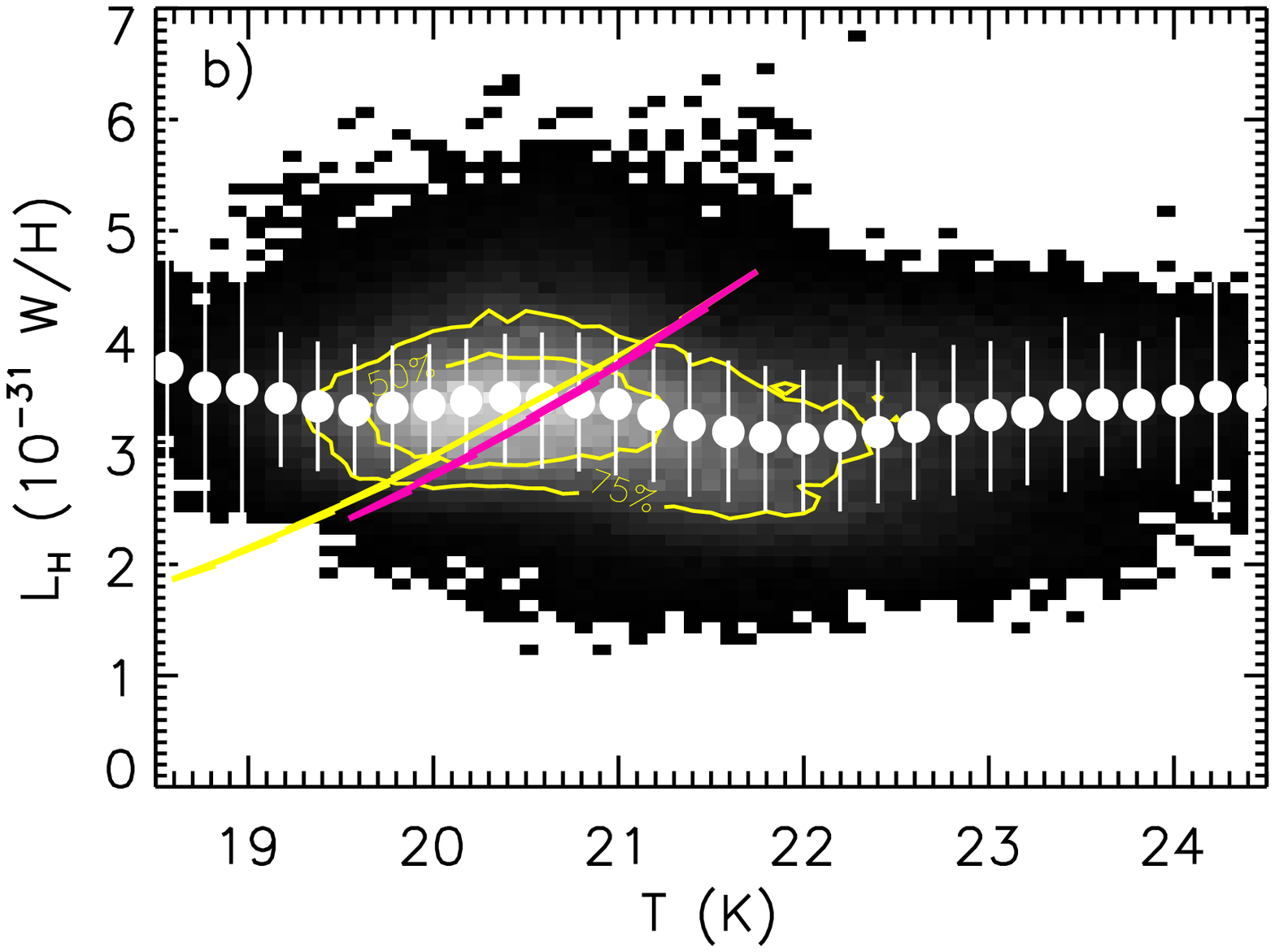}  & \includegraphics[width=6cm,height=4.8cm]{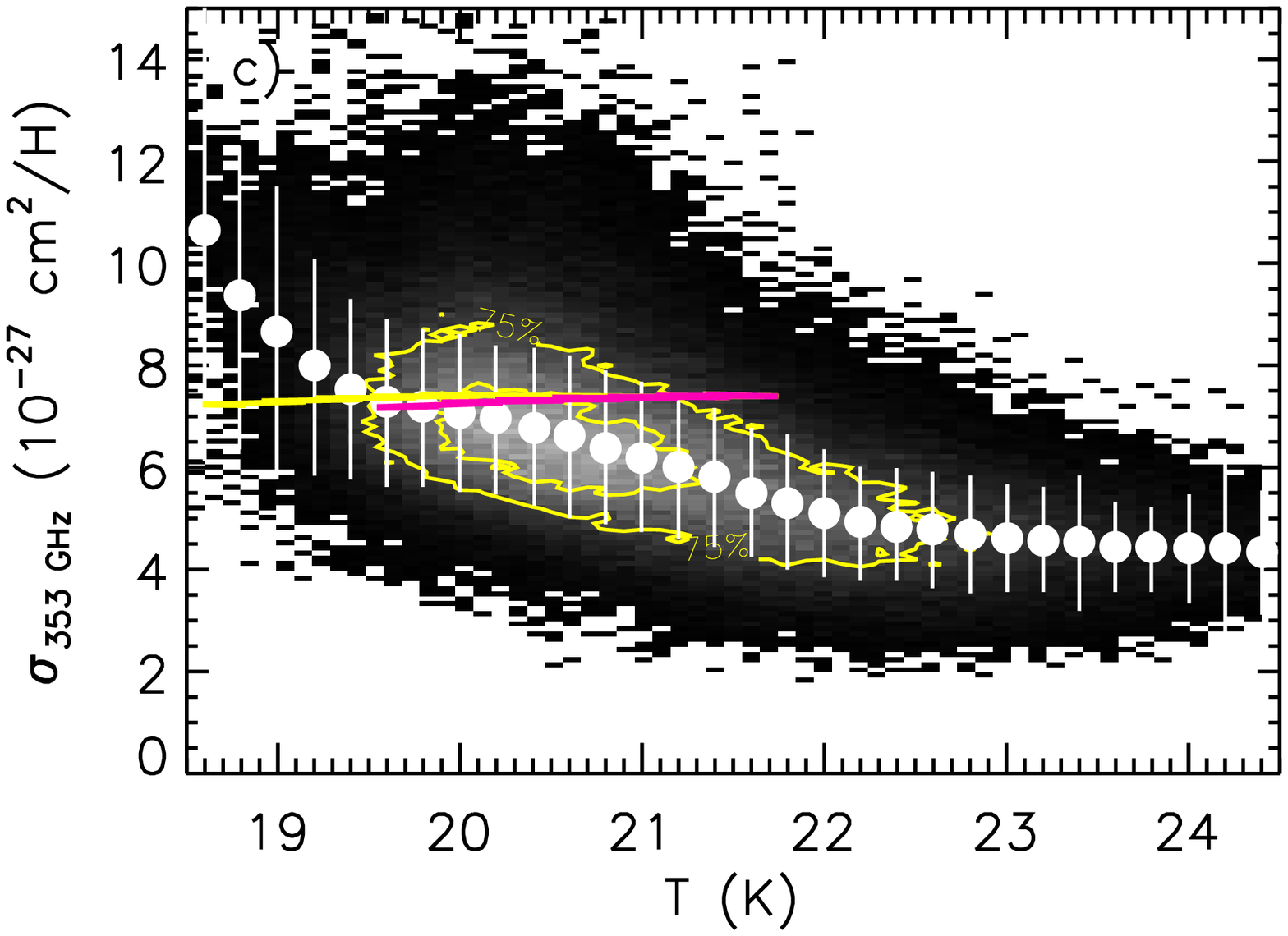} \\
\includegraphics[width=0.34\textwidth]{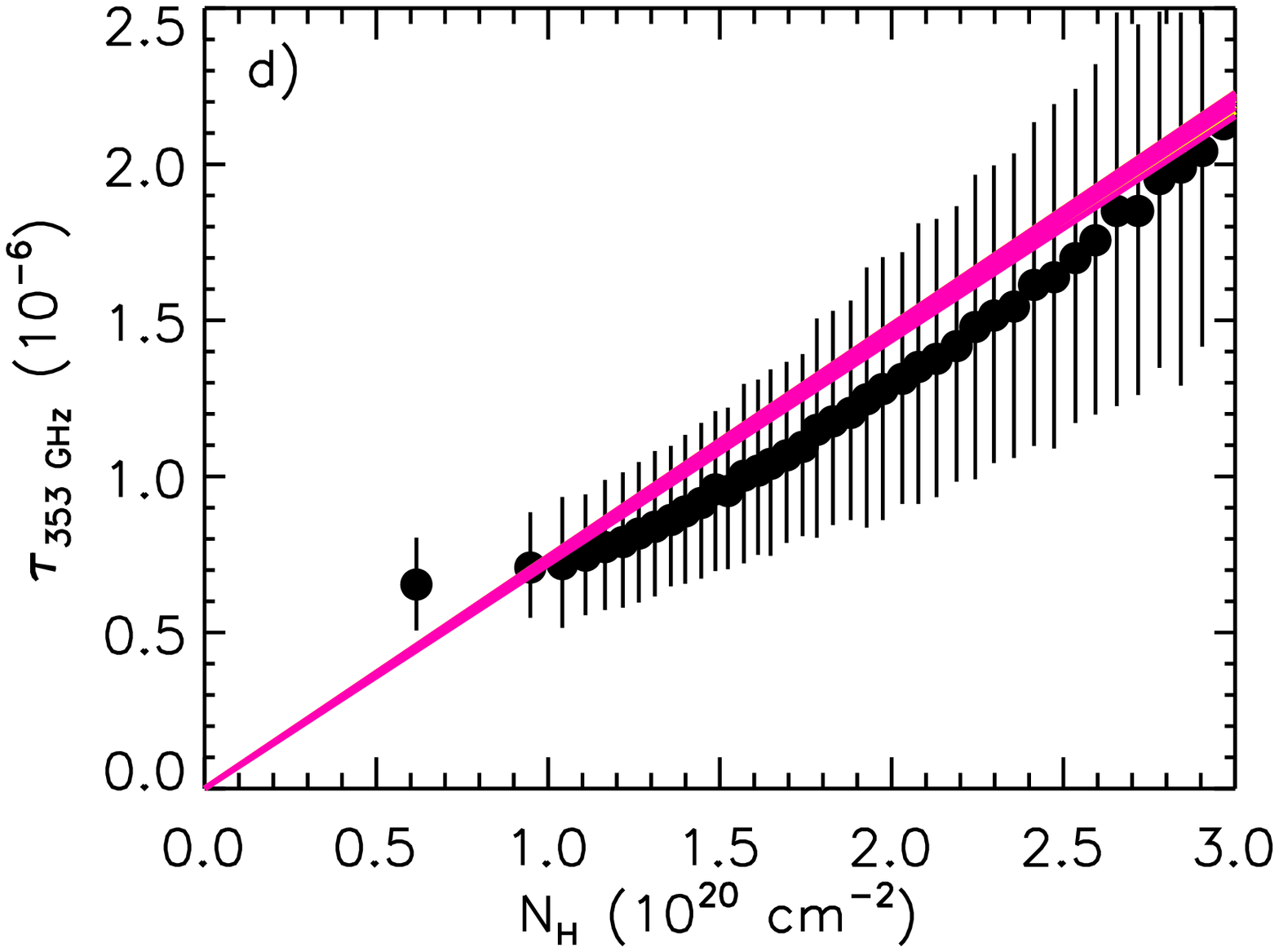} & \includegraphics[width=0.34\textwidth]{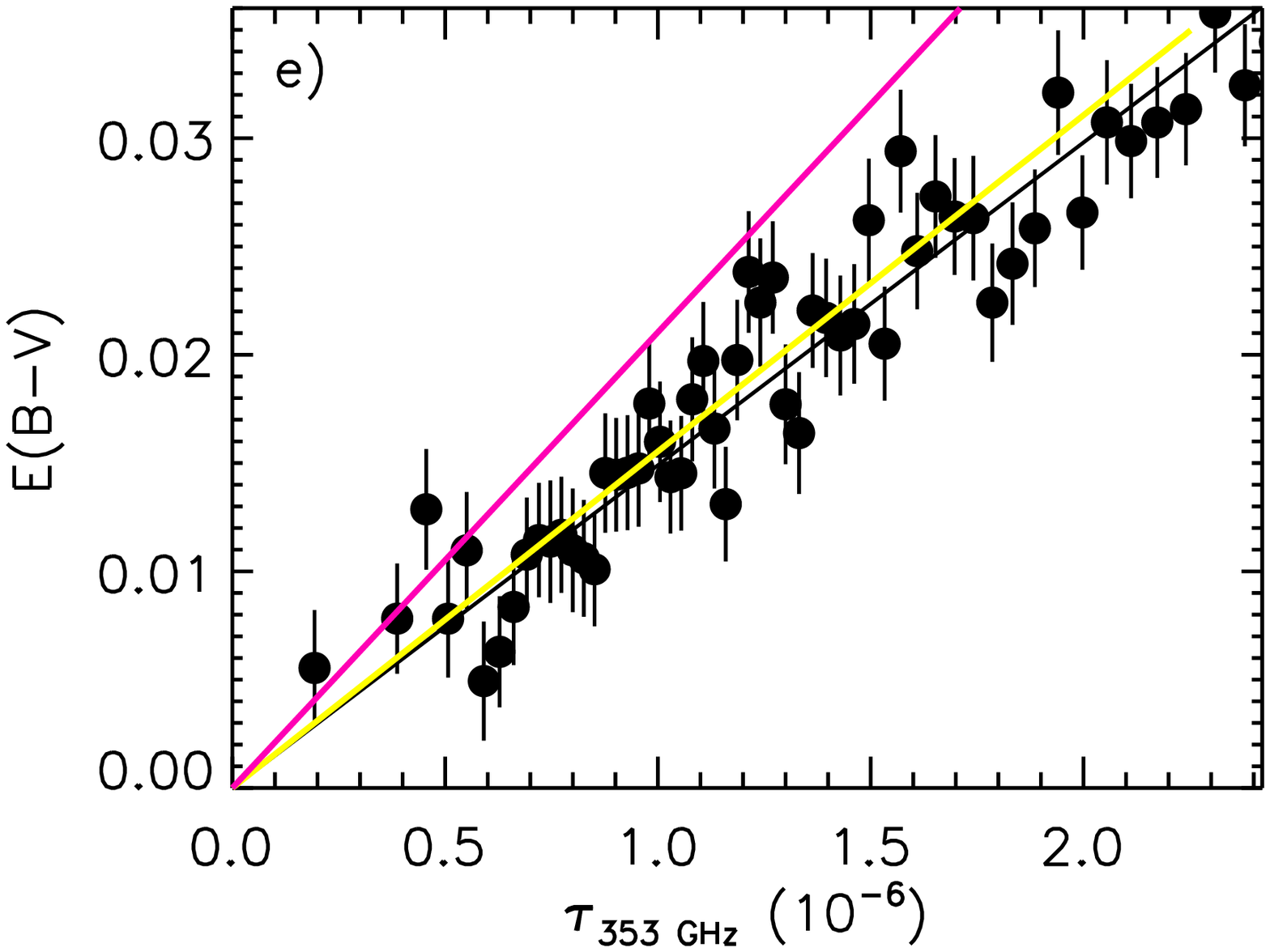} & \includegraphics[width=0.34\textwidth]{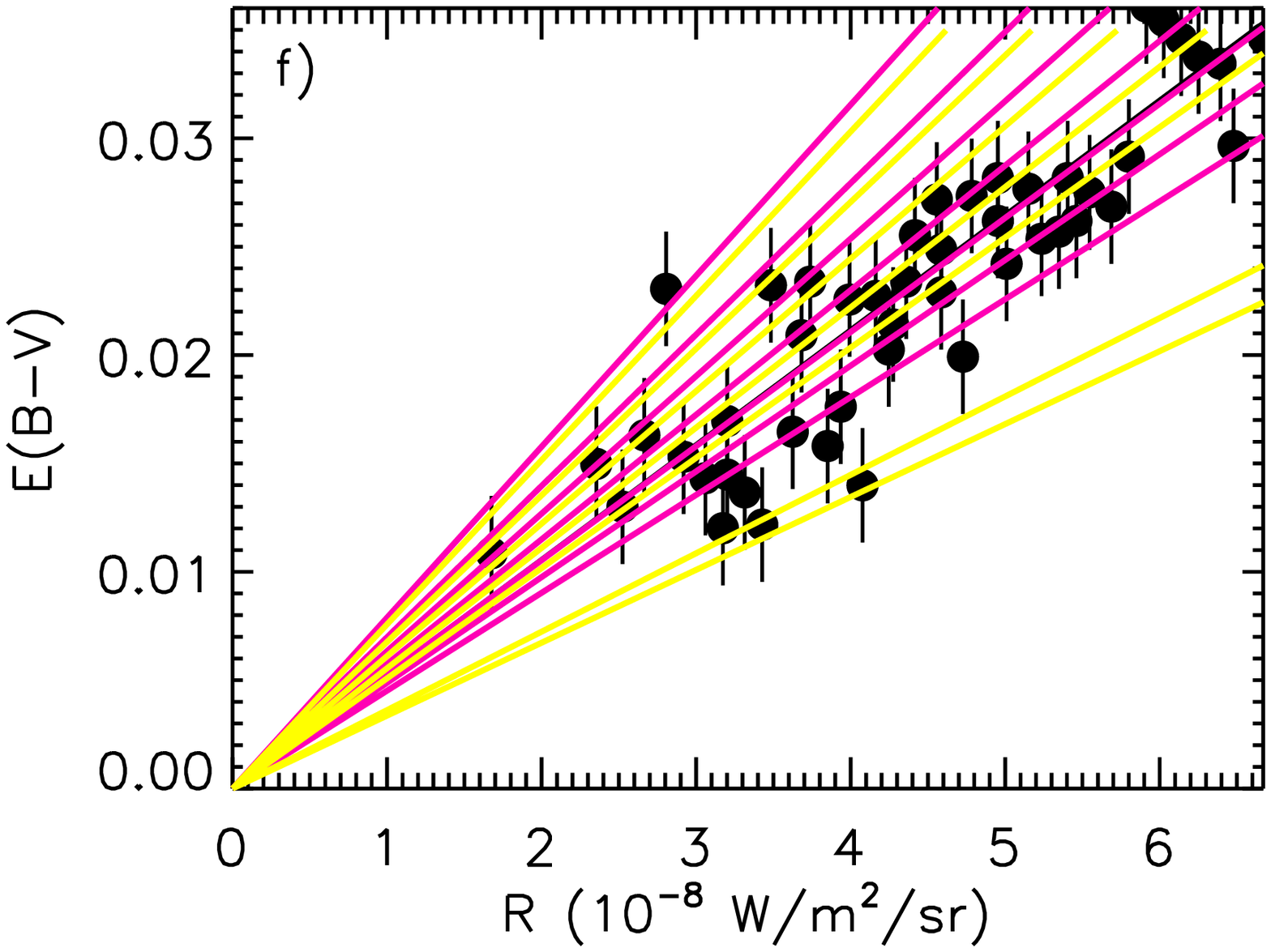}
\end{tabular}}
\caption{Influence of the grain size distribution (carbonaceous grains with a 10~nm thick aromatic-rich carbon mantle and silicates with a 5~nm mantle). The yellow lines show the model with the standard size distribution of \citet{Jones2013} and the pink lines the case where the centre radius of the log-normal size distributions of both carbon and silicate grains have been divided by two. The radiation field is also varied with $0.8 \leqslant G_0 \leqslant 1.4$. See Sect. \ref{methodology} for description of the black and white lines and symbols.}
\label{Fig6} 
\end{figure*}

The grain size distributions can be altered by processes such as shattering and sputtering in the diffuse ISM \citep{Bocchio2014, Bocchio2012, Jones1996, McKee1989, Seab1987}. To investigate the effect of variations in the grain sizes on the dust observables, we test this by dividing the centre radius of the log-normal size distributions by a factor of 2 for both silicate and carbonaceous grains.

The results are shown in Fig.~\ref{Fig6} for carbon and silicate grains with 10 and 5~nm thick aromatic-rich carbon mantles, respectively. The smaller grain populations lead to colour temperatures about 1~K higher than the standard populations of \citet{Jones2013}, while the spectral index and the opacity at 353~GHz remain almost the same (Figs. \ref{Fig6}a and c, $\beta \sim 1.55$, $T \sim 20-23$~K, and $\sigma_{\rm 353 \; GHz} \sim 7 \times 10^{-27}$~cm$^2$/H). This allows us to reproduce some of the observational results for the diffuse ISM and variations in the grain size distribution are thus a likely scenario to explain at least part of the observed variations.

\subsection{Extinction}
\label{extinction}

\begin{figure}[!t]
\centerline{
\includegraphics[width=0.4\textwidth]{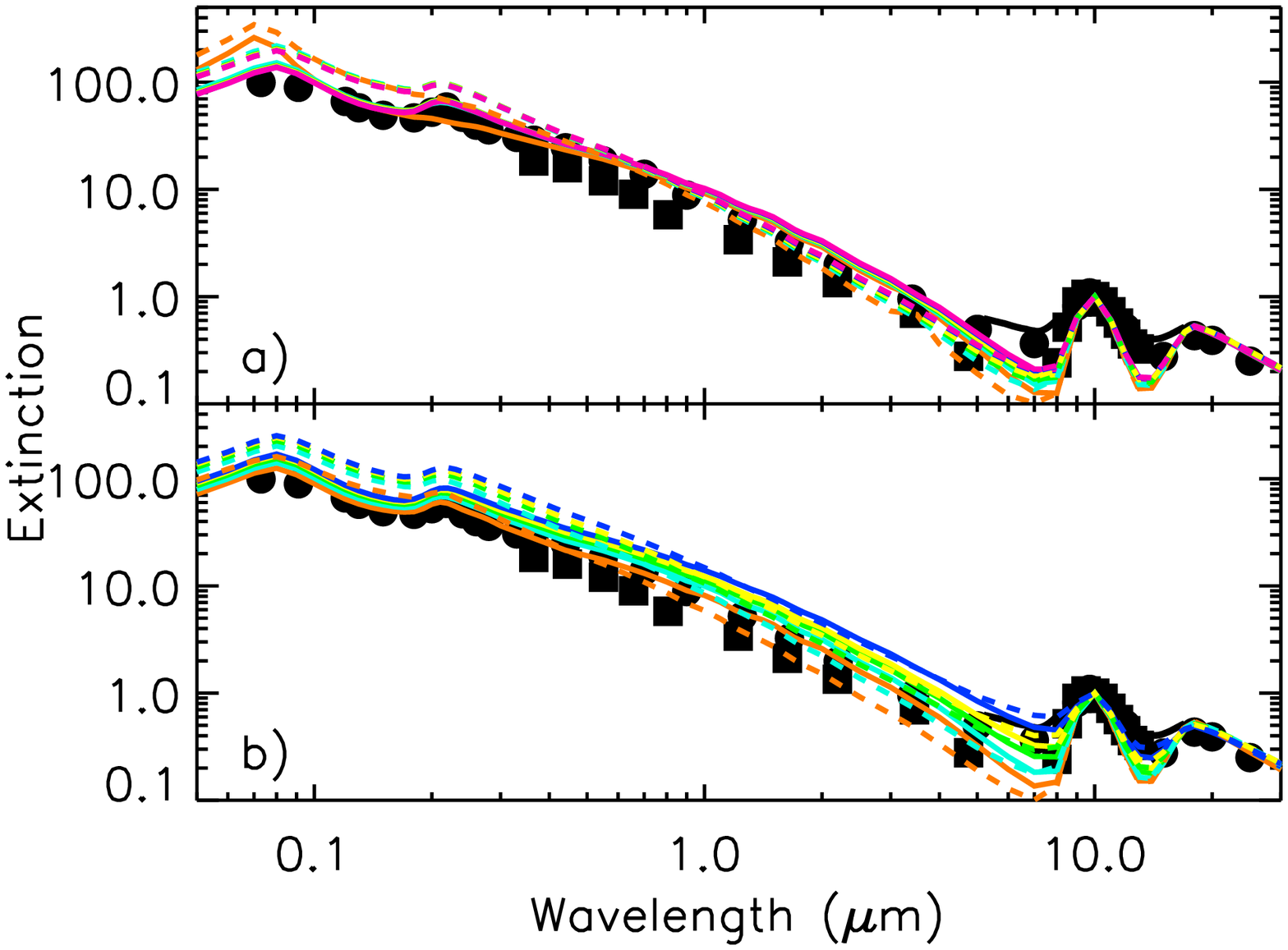}}
\caption{Extinction normalised at 10 $\mu$m. The black circles, boxes, and the thick black line are observational data from \citet{Mathis1990}, \citet{Rieke1985}, and \citet{McClure2009}, respectively. The solid coloured lines are for models with the standard size distributions from \citet{Jones2013}, whereas the dashed coloured lines show models with smaller sizes as in Sect. \ref{size_distribution}. a) The models presented here are for silicate grains with a 5 nm thick aromatic-rich carbon mantle and carbonaceous grains without mantle (orange), with a 5~nm (turquoise), 7.5~nm (green), 10~nm (yellow), 20~nm (blue), and 30~nm (pink) thick aromatic-rich carbon mantle. b) The models are for carbonaceous grains with a 10~nm thick aromatic-rich carbon mantle and silicate grains with no mantle (orange), a 5~nm (turquoise), 7.5 nm (green), 10 nm (yellow), and 15 nm (blue) thick aromatic mantle.}
\label{Fig7} 
\end{figure}

For consistency, we finally check that the extinction curves corresponding to our various dust models are qualitatively in agreement with the extinction measurements towards the diffuse ISM made by \citet{Rieke1985}, \citet{Mathis1990}, and \citet{McClure2009}. We are only looking for qualitative agreement as these measurements were done for lines of sight less diffuse than those presented in PCXI, with $A_{\rm K} \sim 0.1$, 0.4, and 0.3, respectively (to be compared with our models:  $10^{-3} \leqslant A_{\rm K} \leqslant 1.5 \times 10^{-2}$ for carbon grains with 10~nm thick mantles and silicates with 5~nm mantles and $10^{19} \leqslant N_{\rm H} \leqslant 2.5 \times 10^{20}$~H/cm$^2$). The results are presented in Fig.~\ref{Fig7} for extinction curves normalised at 10~$\mu$m. The standard dust model of \citet{Jones2013} reproduces the observations of \citet{Mathis1990}. Variations in the aromatic-rich carbon mantle thickness of the carbonaceous grains have only a small influence on the extinction curve. However, varying the mantle thickness on the silicate grains significantly changes the height and width of the 9.7~$\mu$m silicate feature. While thicknesses up to 5~nm agree with the data points from \citet{Mathis1990}, thicker mantles (7.5, 10, and 15~nm) lead to a broader feature, which moves towards explaining the observational results of \citet{Rieke1985} and \citet{McClure2009}.

Fig.~\ref{Fig7} also shows the extinction curves for the grain size distributions presented in Sect.~\ref{size_distribution}. The main result is that these smaller grains lead to increased extinction at short wavelengths (visible/UV) due to an increase in absorption. This explains why these grains are hotter than those of the standard model of \citet{Jones2013} as shown in Fig.~\ref{Fig6}. We conclude that our dust model can qualitatively explain the extinction measurements as well as most of the variations observed in the dust SED in the Galactic diffuse ISM.

\section{Conclusions}
\label{conclusions}

\begin{figure*}[!th]
\centerline{
\begin{tabular}{ccc}
\includegraphics[width=6cm,height=4.8cm]{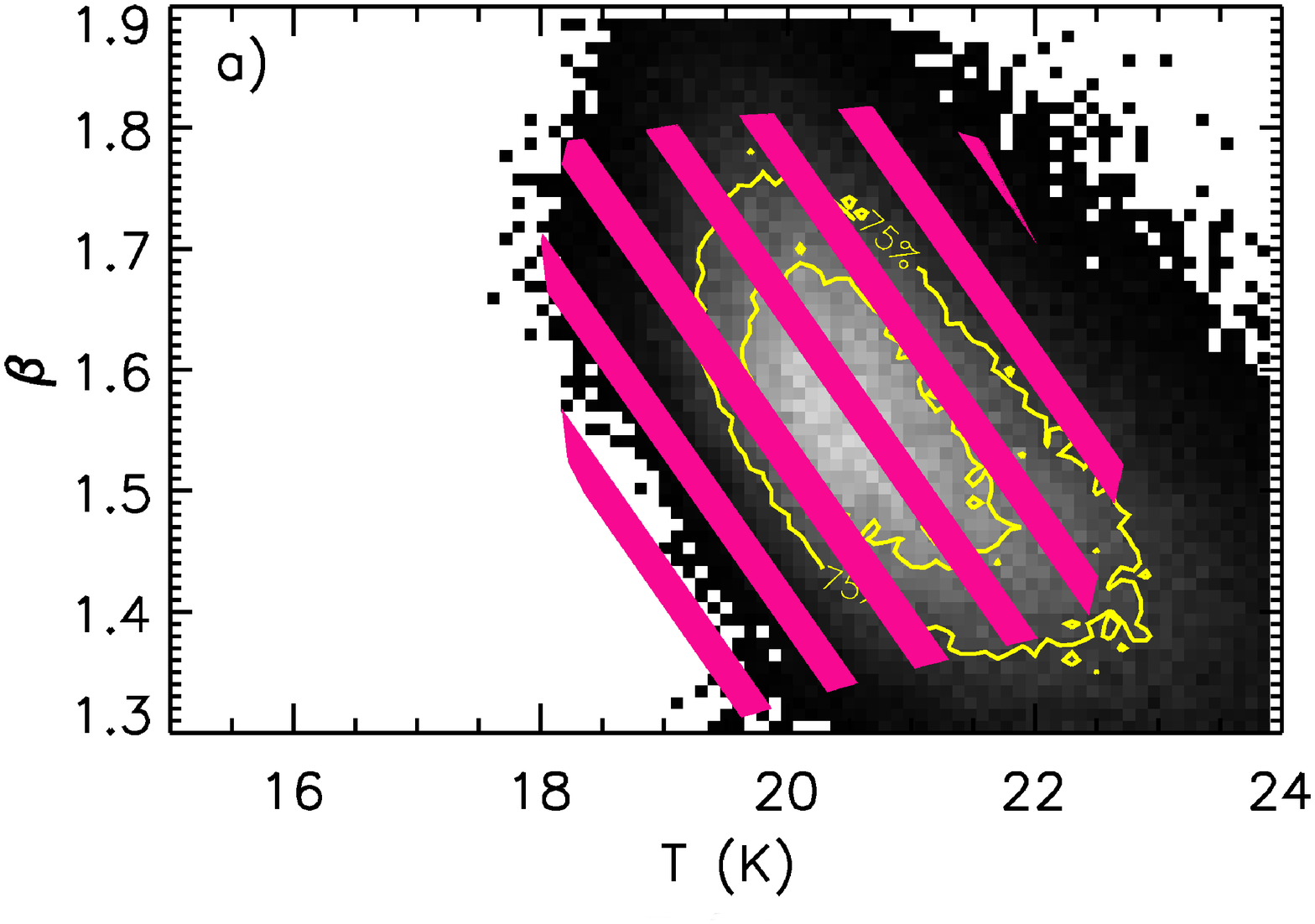} & \includegraphics[width=6cm,height=4.8cm]{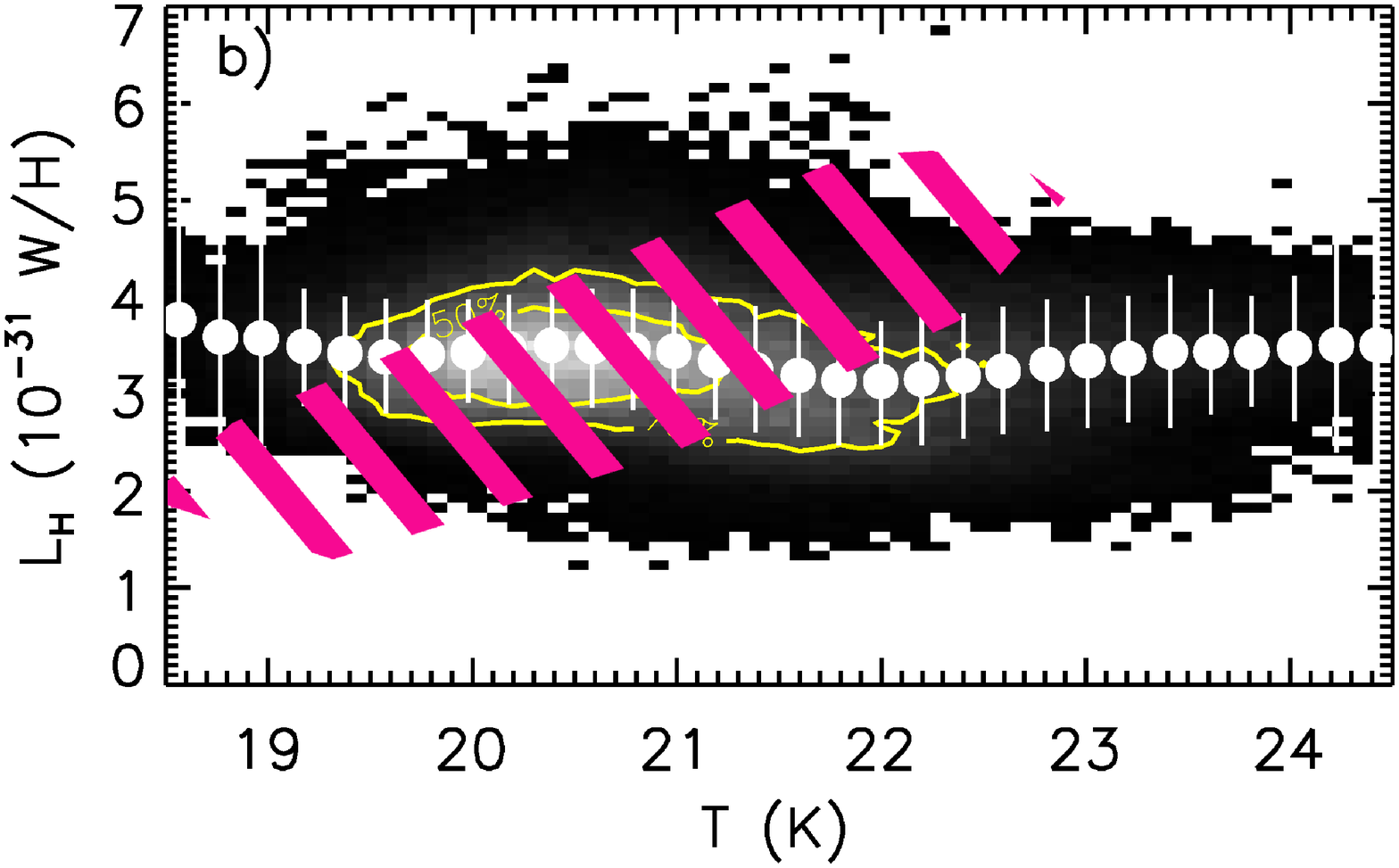}  & \includegraphics[width=6cm,height=4.8cm]{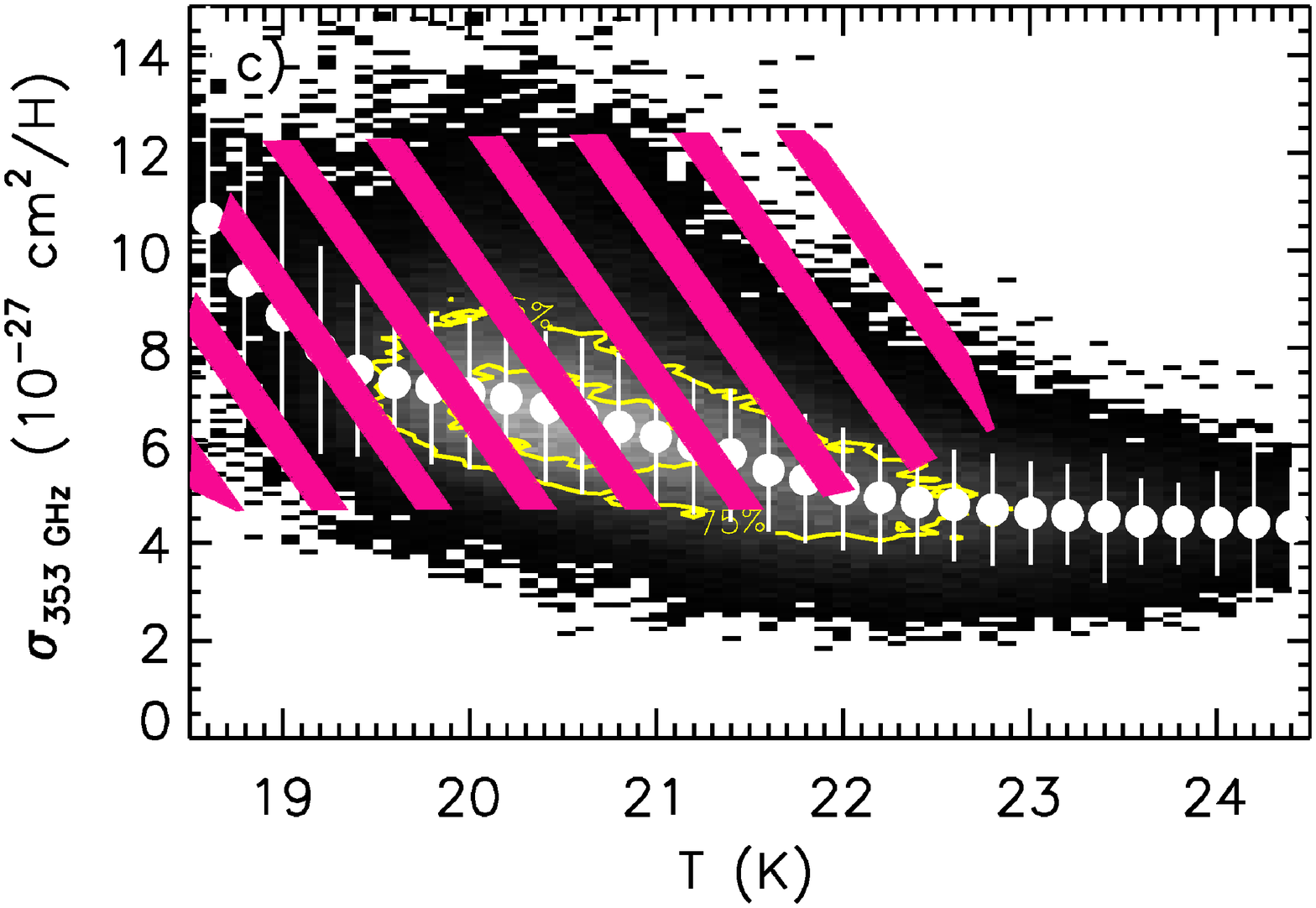} \\
\includegraphics[width=0.34\textwidth]{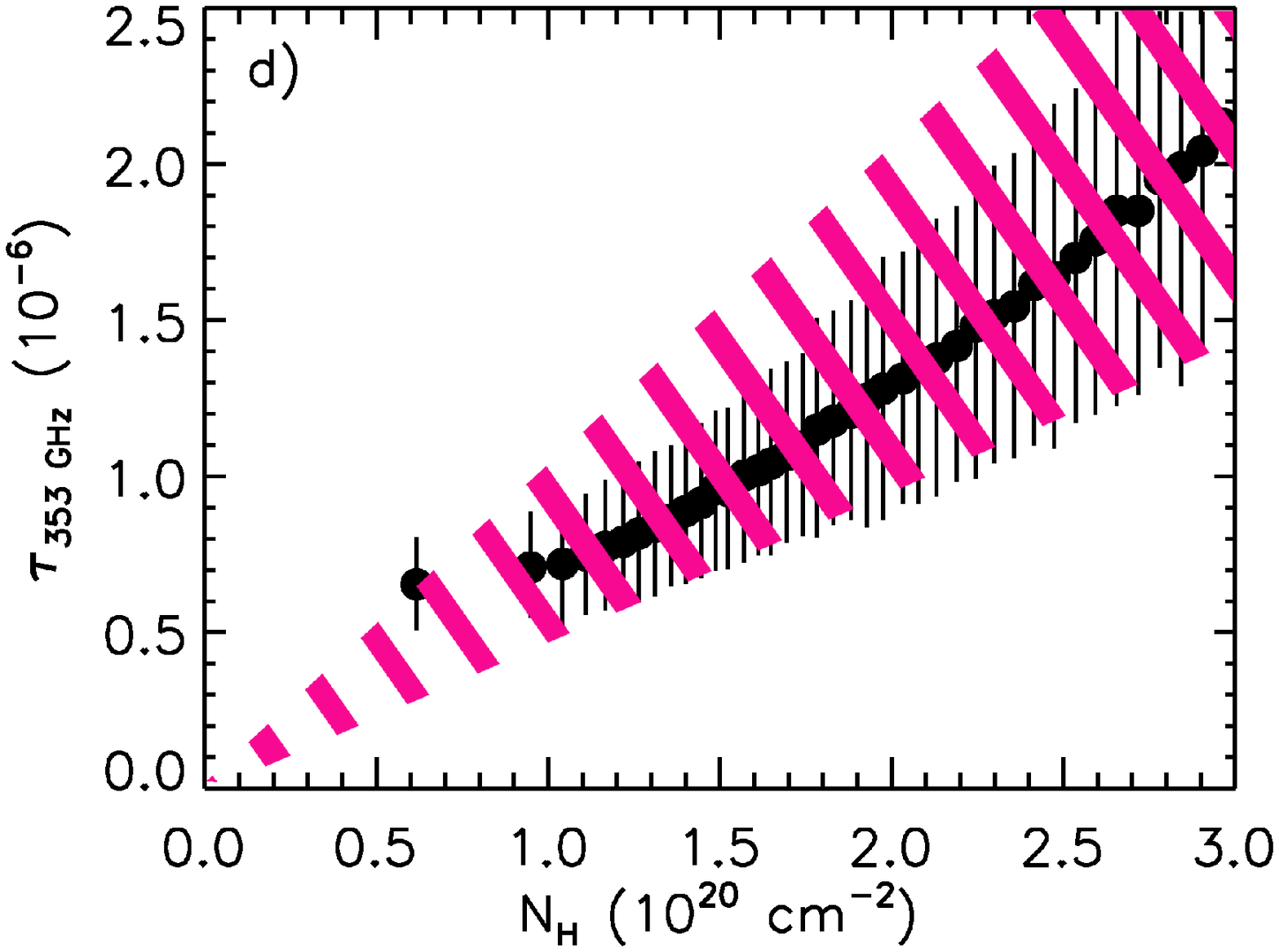} & \includegraphics[width=0.34\textwidth]{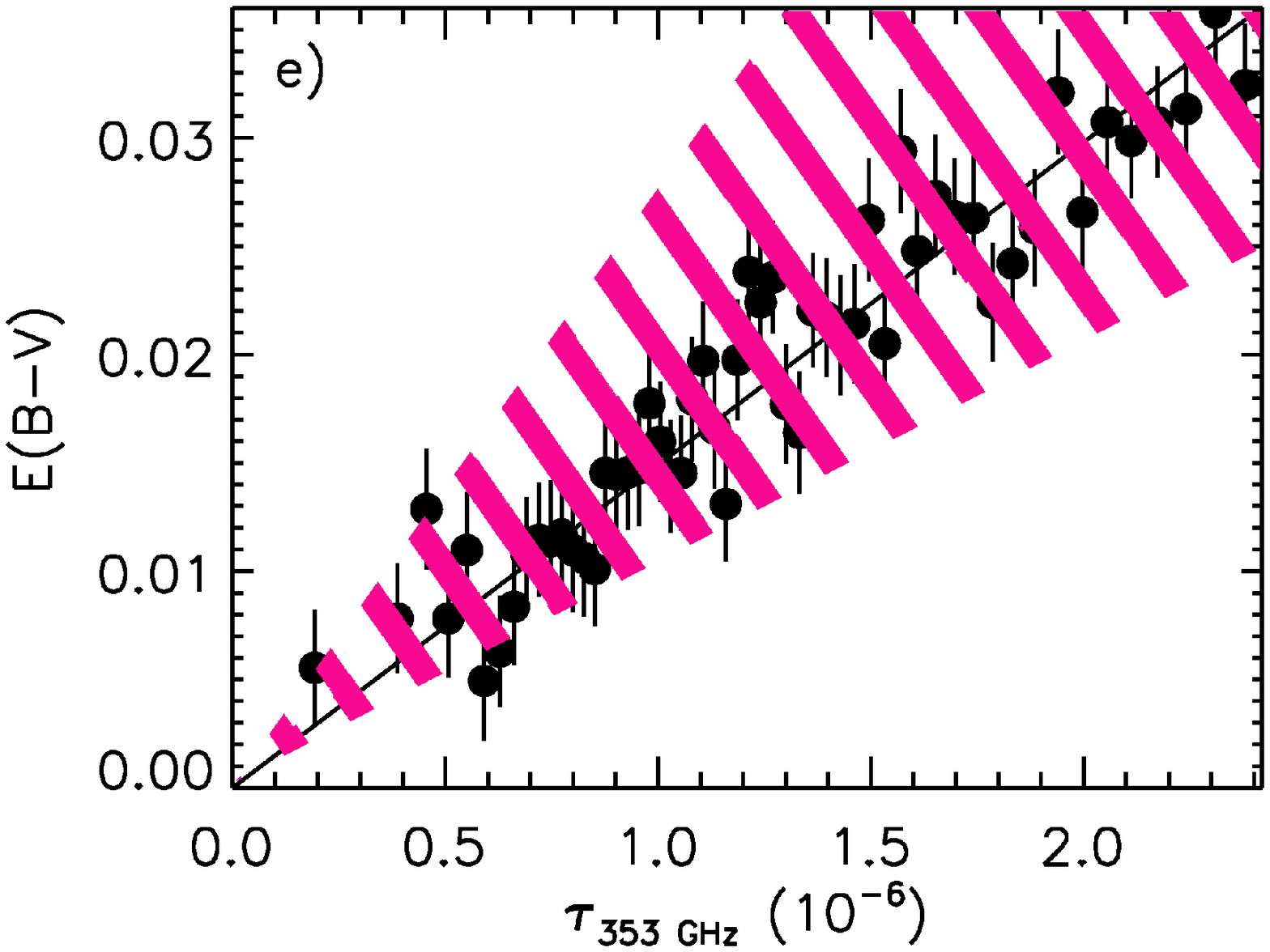} & \includegraphics[width=0.34\textwidth]{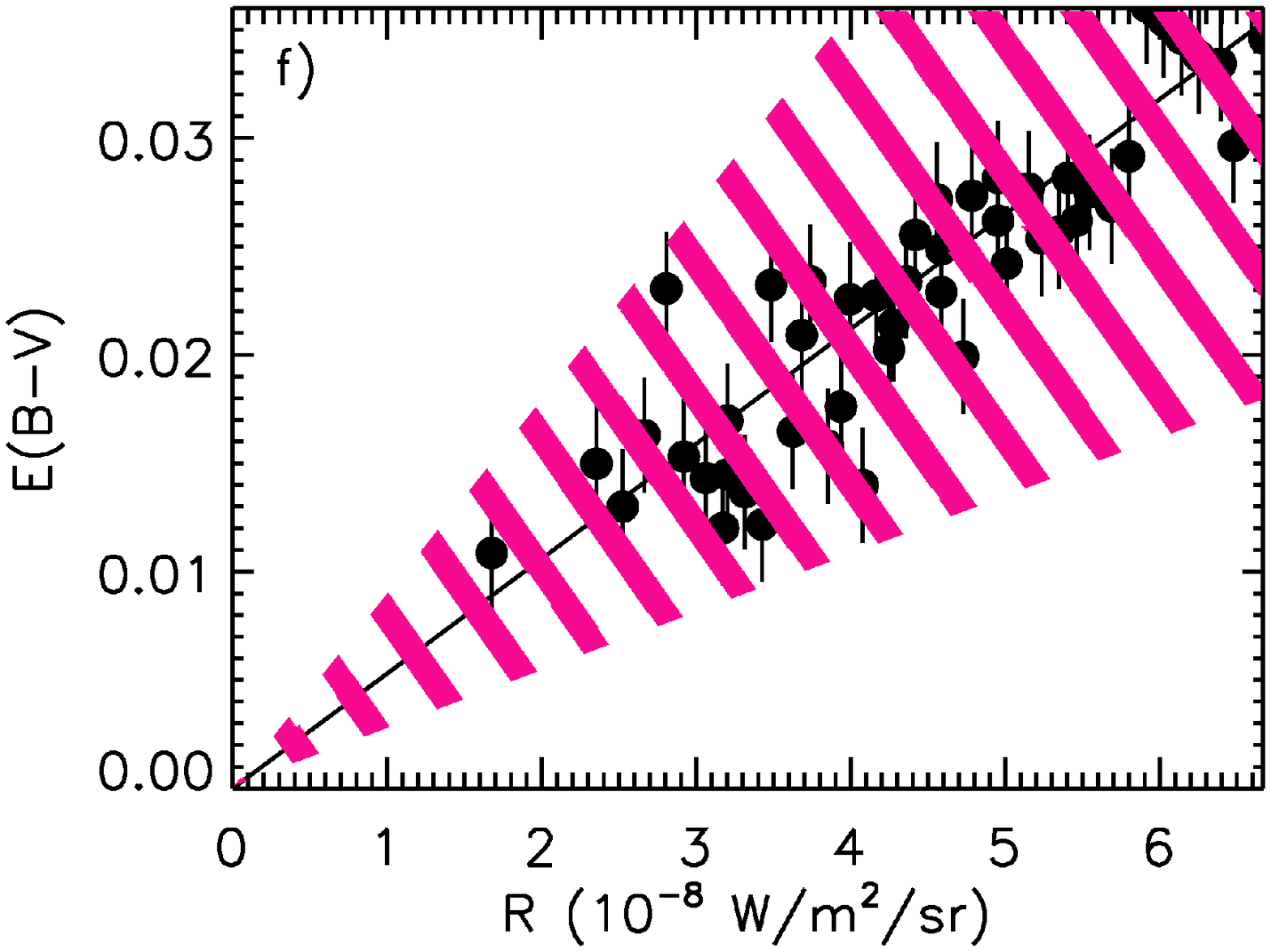}
\end{tabular}}
\caption{Observational data points are the same as in Figs. 1 to 6. The pink dashed areas show all the values of the dust observables that can be produced by the valid dust models presented in Sects.~\ref{environment} and \ref{dust_properties}, i.e. all dust population combinations except 100\% FeS nano-inclusions, silicates with 15 and 20~nm thick aromatic-rich carbon mantles, and aliphatic-rich carbon grains without mantles. For models with increased carbon abundances (see Sect. \ref{carbon_abundance}), we included only those with $N_H < 10^{20}$~H/cm$^2$. Also plotted are models with a decreased carbon abundance. The radiation field is varied with $0.8 \leqslant G_0 \leqslant 1.4$. We further include models for which dust is illuminated by harder radiation fields than the standard ISRF (Sect. \ref{environment}, for black-bodies at 20\,000, 30\,000, and 40\,000~K and $G_0 = 1$, 1.5, 2, 2.5, and 3). See Sect. \ref{methodology} for description of the black and white lines and symbols.}
\label{Fig8} 
\end{figure*}

Using the \citet{Jones2013} core-mantle dust model, we investigated whether the variations in the dust colour temperature, spectral index, and opacity observed by PCXI in the most diffuse areas of the Milky Way ($10^{19} \leqslant N_{\rm H} \leqslant 2.5 \times 10^{20}$~H/cm$^2$) could be explained by varying the dust properties. We also tested for the consistency with the dust extinction and the luminosity measurements. Our results are summarised in Fig.~\ref{Fig8}, where we gather all the valid dust models that we present (pink dashed areas).

We show that the median observations (black and white circles in Fig.~\ref{Fig8}), as well as most of the dispersion, can be explained with little variations in the radiation field and rather small assumptions about the variation/evolution of the dust properties. These include variations in the aromatic-rich carbon mantle thickness on both silicate and carbon grains and variations in the grain size distributions towards smaller grains than what is usually inferred from the observations of the standard diffuse ISM. Smaller grains can for instance explain higher colour temperatures for constant radiation field. For the mantle thickness, we found an upper limit of $\sim 10-15$~nm for the silicate grains and a lower limit of $\sim 5-7.5$~nm for the carbonaceous grains. Thicker mantles lead to lower $\beta$-values for both types of grains but to higher submm opacities. We only considered aromatic-rich carbon mantles as \citet{Jones2014} showed that small carbonaceous particles and aliphatic-rich carbon mantles would be aromatised in less than $10^6$~yrs in the standard diffuse ISM ($A_{\rm V} \lesssim 0.7$ and $G_0 = 1$). Thus, for the high Galactic latitude diffuse ISM, we favour a dust model with aromatic-rich carbon mantle thicknesses in the range $7.5-20$~nm on carbonaceous grains and $5-10$~nm on silicate grains, with the standard \citet{Jones2013} dust size distributions. Local variations in the carbon abundance inside the grains may also take part in producing some of the observed variations in the far-IR/submm SED. Finally, our dust model can also explain the $\beta-T$ relation measured by PCXI. We show that the scatter in the $\beta$-values produced by variations in the dust intrinsic properties are enough to explain most of the observed scatter within $18 \leqslant T \leqslant 23$~K and $1.3 \leqslant \beta \leqslant 1.8$ (Fig.~\ref{Fig8}a). Our results thus suggest that the $\beta-T$ and $\sigma-T$ anti-correlations observed in the diffuse ISM could simply be due to a scatter in the dust properties.

In summary, we have shown the capacity of the new dust model to explain the observed trends with physically-reasonable dust property variations. We therefore consider the \citet{Jones2013} model to be rather robust. However, along a few lines of sight higher temperatures are found ($T \gtrsim 22.5$~K), which we cannot explain with changes in the dust properties or in the radiation field. These few lines of sight might include HII or highly shocked regions, or even grains with peculiar properties due to a peculiar history. Detailed analyses of well-known lines of sight (radiation field colour and intensity, local density, gas properties, carbon depletion, dynamics...) are needed to understand exactly which mechanisms produce the observed variations in the dust properties, and how much these properties depend on grain histories, ages, and environments.

\acknowledgements{We thank our anonymous referee for his careful reading, which helped to improve the paper. We would also like to thank F. Boulanger, V. Guillet, and L. Verstraete for many interesting discussions about cosmic dust. This research was, in part, made possible through the financial support of the Agence Nationale de la Recherche (ANR) through the programs Cold Dust (ANR-07-BLAN-0364-01) and CIMMES (ANR-11-BS56-029-02).}

\bibliography{biblio}

\end{document}